\documentclass[journal]{IEEEtran}

\ifCLASSINFOpdf
  \usepackage[pdftex]{graphicx}
\else
\fi

\usepackage{array}

\ifCLASSOPTIONcompsoc
  \usepackage[caption=false,font=normalsize,labelfont=sf,textfont=sf]{subfig}
\else
  \usepackage[caption=false,font=footnotesize]{subfig}
\fi

\usepackage{stfloats}
\usepackage{pifont} 
\usepackage{slashbox}
\usepackage{commath}

\begin{document}
\title{Noninvasive Corneal Image-Based \\ Gaze Measurement System}

\author{Eunji~Chong, Christian~Nitschke, Atsushi~Nakazawa, Agata~Rozga, and James~M.~Rehg 
\thanks{E. Chong, A. Rozga and J. M. Rehg are with the College of Computing, Georgia Institute of Technology, Atlanta, GA, 30332, USA. e-mails: eunjichong@gatech.edu, agata@gatech.edu and rehg@gatech.edu.}
\thanks{C. Nitschke and A. Nakazawa are with Kyoto University, Yoshidahonmachi, Sakyo Ward, Kyoto, Kyoto Prefecture 606-8501, Japan. e-mails: christian.nitschke@i.kyoto-u.ac.jp and nakazawa.atsushi@i.kyoto-u.ac.jp}}

\maketitle

\begin{abstract}
Gaze tracking is an important technology as the system can give information about a person from what and where the person is seeing. There have been many attempts to make robust and accurate gaze trackers using either monitor or wearable devices. However, those contraptions often require fine individual calibration per session and/or require a person wearing a device, which may not be suitable for certain situations. In this paper, we propose a robust and a completely noninvasive gaze tracking system that involves neither complex calibrations nor the use of wearable devices. We achieve this via direct eye reflection analysis by building a real-time system that effectively enables it. We also show several interesting applications for our system including experiments with young children.
\end{abstract}


\IEEEpeerreviewmaketitle

\section{Introduction}
\label{intro}

\IEEEPARstart{G}{aze} tracking is a key technology for studies of visual attention and
cognitive processes~\cite{underwood2005cognitive}~\cite{land2009looking} and has commercial applications in areas such as user-interface development and evaluation, and the design and assessment of advertising. 
Also, it has become popular in computer vision research as a means to study 
saliency~\cite{mathe2015actions}~\cite{Borji:2013aa}~\cite{Garcia-Diaz:2012aa}~\cite{Judd:2009aa}, object and activity recognition~\cite{li2015delving}~\cite{Yun:2013aa}~\cite{ogaki2012coupling}, gaze prediction~\cite{Li:aa}, and salient object detection~\cite{borji2015salient}~\cite{liu2011learning}.

Recently, gaze analysis has emerged as a critical tool for assessing and treating children with various medical or developmental conditions, ranging from nutrition levels~\cite{colombo2004maternal} to language~\cite{brooks2005development}~\cite{morales1998following} and motor development~\cite{romski2015early} to Autism Spectrum Disorder (ASD)~\cite{jones2008absence}~\cite{Klin:2002aa}~\cite{leekam2000attention}~\cite{osterling1994early}~\cite{mundy1986defining}. Especially, it has been well-established that children with ASD exhibit atypical patterns of looking at social stimuli (e.g., less frequent eye contact and following gaze than their typically-developing peers)~\cite{jones2008absence}~\cite{Klin:2002aa}~\cite{mundy1986defining}.
As a result, there has been growing need for the use of gaze tracker in these contexts because manual labeling is usually very time-consuming and subjective. However, most existing gaze trackers are in the form of monitor-based or wearable, and each type has its own limitations.
(A) Monitor-based system: As a majority of research and clinical treatments are based on face-to-face social interactions~\cite{duncan2015face}, the monitor-based approach lacks ecological validity (i.e. unable to be generalized to real-life settings) and contingency (i.e. interaction is not mutual). (B) Wearable system: Although wearable trackers provide a more ``real-world'' environment than a monitor-based one, it is invasive (head-worn), which causes many other problems. Moreover, approaches (A) and (B) assume that the exact moment when the user looks at a calibration target can be found, which can be very difficult with young children.
This paper explores the hypothesis that corneal imaging can be used to develop a novel approach to noninvasive gaze tracking avoiding the aforementioned troubles, which is suitable for the assessment of naturalistic gaze behavior. 

\begin{figure}[ht]
\centering
\includegraphics[scale=0.4]{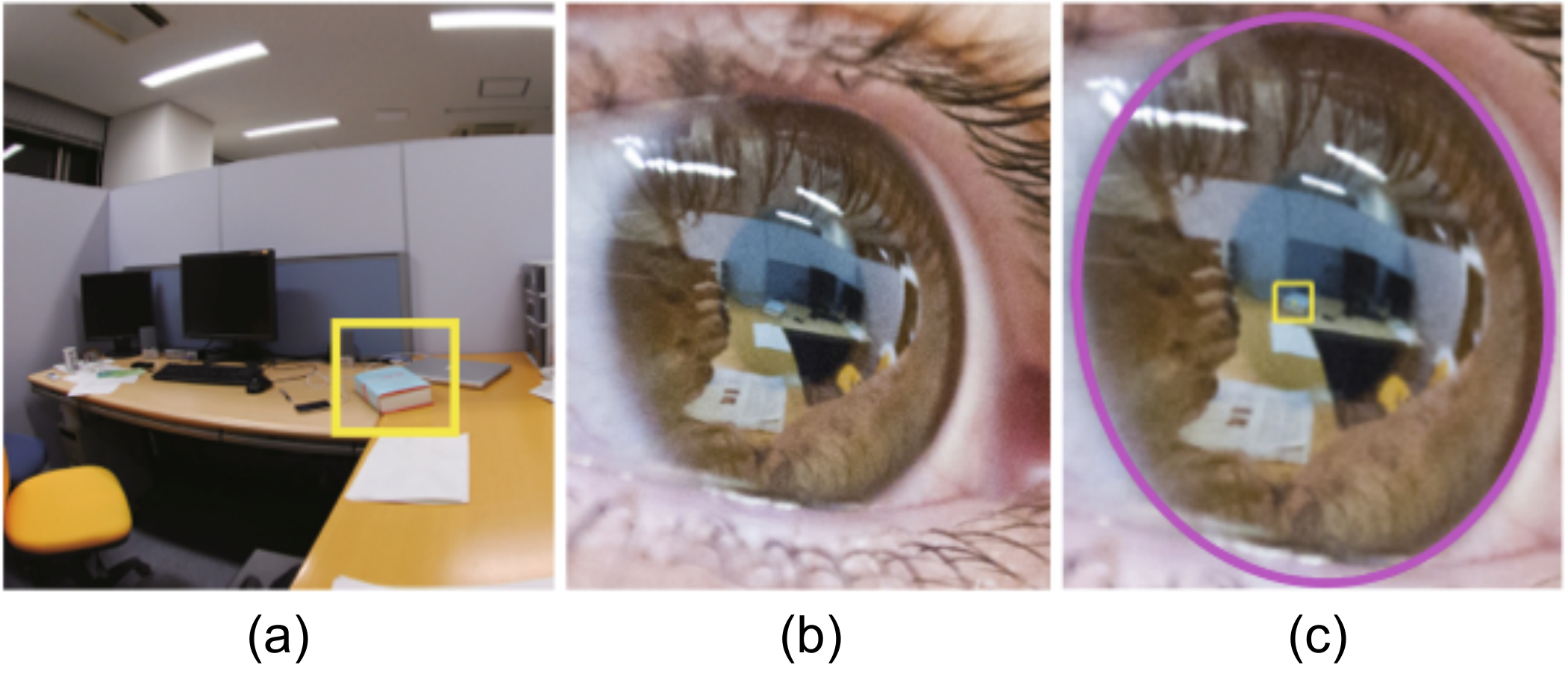}
\caption{Illustration of how a 3D scene is reflected on the corneal image: (a) Scene with an object of interest (a book, in this example) in yellow bounding box, (b) Corneal reflection of the scene, and (c) Estimated corneal boundary in pink ellipse and estimated gaze in yellow square, meaning that the person is looking at the book.}
\label{fig:corneal_concept}
\end{figure}

The starting point for our investigation is the pioneering work of
Nishino and Nayar, who developed the concept of a corneal imaging
system~\cite{nishino2004world} in which a camera captures an image of
the scene reflected in the subject's cornea. This concept is
illustrated in Figure~\ref{fig:corneal_concept}, where (a) is a scene and (b)
is a close-up image of the cornea of a person who is looking at the
scene. By estimating eye pose from the shape of the iris contour, one
can compute the point of gaze on  the corneal image as in
(c). Nishino and Nayar demonstrated that this eye-camera system can be
modeled as a non-central catadioptric imaging
system~\cite{sturm2011camera}, and they showed how to model the
imaging geometry in order to unwarp the corneal image and obtain an
environment map with respect to the subject's eye. This paper extends their foundation and
develops a complete end-to-end real-time gaze measurement system based on corneal
imaging regime. It has two main advantages
over conventional gaze tracking approaches by nature: (1) Calibration needed to obtain gaze measurements 
is minimal to none. (2) The corneal video directly
encodes the relationship between the subject's eye movements and the
content of the scene in an intuitive visual manner. The resulting video is
similar to the output of a wearable gaze tracker that
provides first-person video with overlaid gaze estimates, and the advantage is that corneal video is
obtained noninvasively with a potentially much larger field of
view (FoV). 

This paper makes the following contributions:
\begin{itemize}
\item We develop the first complete real-time gaze
measurement system based on the corneal imaging principle that can overcome many limitations that other forms of (i.e. monitor-based and wearable) gaze trackers possess.
\item We characterize system parameters and
the tradeoffs that govern system design, and conduct careful empirical assessments of
performance to support our design choices.
\item We present the first quantification of the measurement accuracy
of this approach relative to a commercial wearable gaze measurement
system.
\item We present gaze tracking results for a broad range of psychology
experiments, demonstrating the value of the proposed method.
\end{itemize}


\section{Related work}

\subsection{Monitor-Based Gaze Tracking}
A basic property of conventional gaze tracking systems is that they
are based on geometric calculations using the 3D Line of Sight (LoS),
which is the vector in space along which the subject's gaze is
directed (for a single eye). To determine what someone is
looking at, the LoS must be intersected with a 3D representation of
the scene to obtain the point of gaze (PoG). In the case of
monitor-based gaze tracking systems - the most widely used
technology - the intersection test is very simple because the subject
is always looking at a planar monitor screen. In that case, the
calculation of PoG can be performed directly (without explicit
estimation of LoS) by calibrating the relationship between eye
measurements and the PoG (e.g., by asking the user to fixate on a few known
 corners of the screen in a calibration phase).
Many commercial systems use infrared (IR) LEDs to illuminate the eye,
creating specular reflections in the corneal image known as glints. In
the classical pupil-center/corneal-reflection (PCCR)
method~\cite{guestrin2006general}, the vectors from the glints to the
pupil center encode the pose of the eye, and are mapped directly to
the PoG in a monitor-based system. 

Alternatively, 3D stereo-based
gaze estimation methods~\cite{Zhu:2007aa} can directly estimate the 3D
LoS using a geometrically calibrated system with multiple LEDs and
cameras. The problem with this approach is that the intersection test
to determine the PoG still requires knowledge of the 3D scene
geometry, which is problematic in cases other than looking at a
monitor. Thus, in practice, stereo-based methods tend to be used to make
monitor-based tracking more robust to user head movements.

That being said, monitor-based systems are still widely used as they are simple and easy to use,
but they are not adequate for the study of face-to-face social interactions, 
where social partners are engaged in an active back-and-forth
exchange, as opposed to passive viewing of a monitor screen.

\begin{figure*}[t!]
\centering
\includegraphics[width=0.9\linewidth]{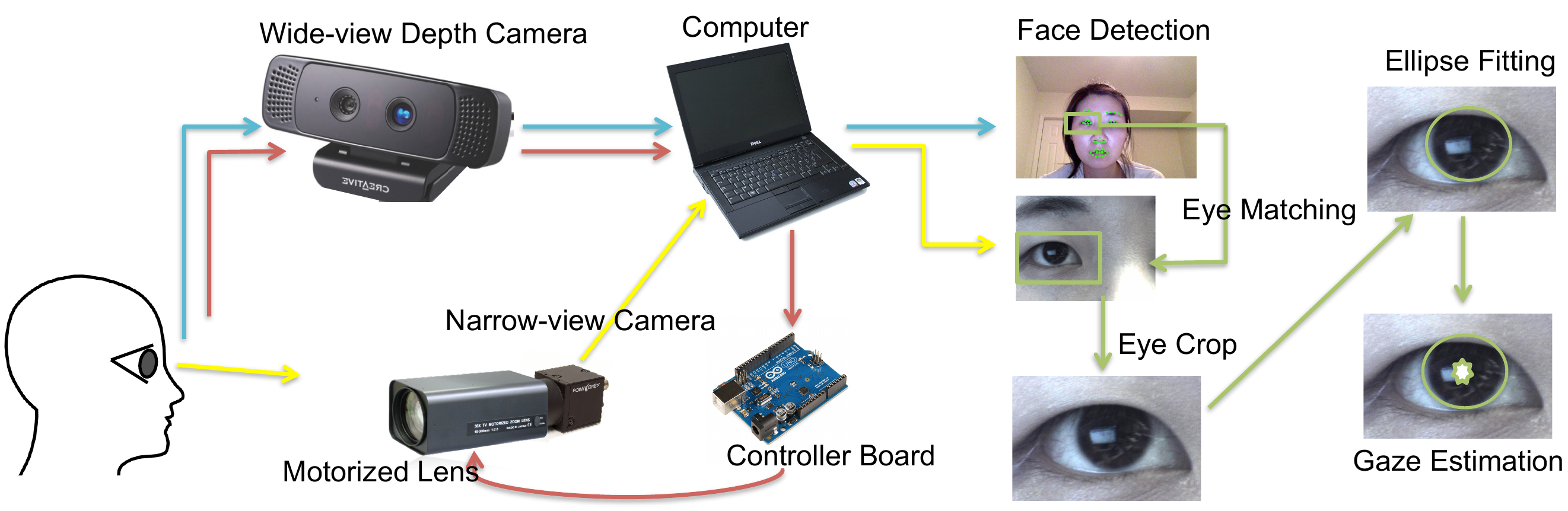}
\caption{Overview of our system. Blue arrows: Wide-view RGB video stream. Red arrows: Wide-view depth video stream and open-loop focus control. Yellow arrows: Narrow-view RGB video stream. Green arrows: Image processing. (1) First, a face is detected in the wide-view camera to extract the location and size of the eyes. (2) Based on the geometric relationship between the wide-view camera and the narrow-view camera, it determines if the eye is in view. (3) Within the narrow-view (thus high-resolution) image, eye region is cropped online. (4) Finally, an ellipse is automatically detected and tracked to estimate the point of gaze. Autofocus control runs in the background using depth information of the eyes provided by the wide-view camera.} 
\label{fig:overview}
\end{figure*}

\subsection{Wearable Gaze Tracking}
The recent capability to make camera/LED systems small enough to be
worn comfortably by the user has led to the emergence of wearable gaze
tracking systems from vendors such as SMI and Tobii. The company Positive Science has developed a
light-weight system designed to be worn by young
children~\cite{PositiveSci}~\cite{franchak2011head}. These systems
integrate a point of view (PoV) or scene camera with PCCR-based gaze
trackers (one for each eye) into a single pair of glasses. The output
of the system is a PoV (or first-person) video, with the estimated PoG
overlaid in each frame. These systems require a subject-specific
calibration phase for each recording session; if the glasses shift on the
user's face during the recording, the calibration must be repeated.
Because of this requirement, it usually takes a considerable amount of effort to ensure that it works properly. 
For example,~\cite{franchak2011head} reported that it took about 15 minutes 
on average to put the device on and to calibrate for the 40 children in their study.

It is also possible to track gaze by learning an appearance model~\cite{noris2011wearable}~\cite{martinez2012gaze}, 
or by tracking the eyeball~\cite{tsukada2011illumination}, 
which can gain some robustness with respect to the movement of the glasses.
However, all of these approaches still require elaborate calibrations between the scene-recording camera and the eye-recording camera
and works under certain assumptions about calibrated moments between the two cameras.

Aside from these issues, although wearable systems provide much greater flexibility than
monitor-based systems, they are significantly more invasive due to the
need to wear the sensing hardware. The current glasses-based systems
are quite large for young children, and poses safety concerns (e.g., if the child fell on the system he or she could injure their eye). Also, these systems are usually tethered, so someone has to follow the child around. More importantly, a significant number of children will not agree to wear the system. For example, a recent study found that 30\% of children didn't comply~\cite{yu2016social}. And even if they do, their behavior may be affected by the technology. A final important concern is the limited FoV of wearable cameras compared to the
FoV of the human subject. Oftentimes, the true PoG is located outside
the FoV of the wearable system, making it impossible to tell what the
subject is looking at during those moments.
\hfill \break

In this paper, we propose that a practical gaze tracking system based on corneal imaging can be a new solution for the assessment of naturalistic gaze behavior. The corneal imaging approach enjoys the following benefits relative to conventional gaze tracking systems, both monitor-based and wearable:

\begin{enumerate}
\item Minimum individual calibration requirement: For each individual, we only need to perform calibration of the kappa angle, which is the subject-specific innate offset between the visual axis and the optical axis. Even without it at all, our method can produce quite accurate ($<2$ degrees of error) and useful results as evaluated in Section~\ref{sec:eval}. In contrast, all other systems require additional calibration (e.g. between the scene-recording camera (or monitor screen) and the eye-recording camera) that is often for every different session, which can be especially problematic for young children.
\item Gaze estimation in 3D scenes without wearables: Our proposed method can accurately estimate gaze in dynamic 3D scenes without requiring the user to wear anything. This is an important advantage when working with children.
\item Higher accuracy in dynamic scenes with significant depth of field: The PCCR approach requires glints to be visible to the scene camera, limiting the maximum standoff distance, and typically uses planar gaze targets such as monitor screens. In contrast, our method can handle dynamic 3D scenes with significant depth variations. Even wearable systems can sometimes have a problem with depth parallax, as the point of view camera is not axis aligned with subject's eye. Our corneal imaging approach avoids all of these issues.
\item Wide field of view: As described in~\cite{nishino2004world}, the corneal image has a wider FoV than the subject's retina itself does and therefore a much wider FoV than any wearable camera currently available. It follows that our system has the ability to collect the maximum scene information.
\end{enumerate}

\hfill \break
The remainder of the paper is organized as follows: Section~\ref{sec:sys} presents design of our system. The system is composed of two parts --- the camera system part and the gaze estimation part. The former corresponds to the left half of Figure~\ref{fig:overview} and its details are explained in Section~\ref{sec:sys}. The latter is the right half of Figure~\ref{fig:overview} and Section~\ref{sec:gaze} describes each step. In Section~\ref{sec:eval}, we present evaluations of our system. Finally in Section~\ref{sec:app}, we demonstrate a variety of real-world applications of the system.

\begin{figure}[h!] 
\centering
\subfloat[Main contstraints]{\includegraphics[width=0.5\linewidth]{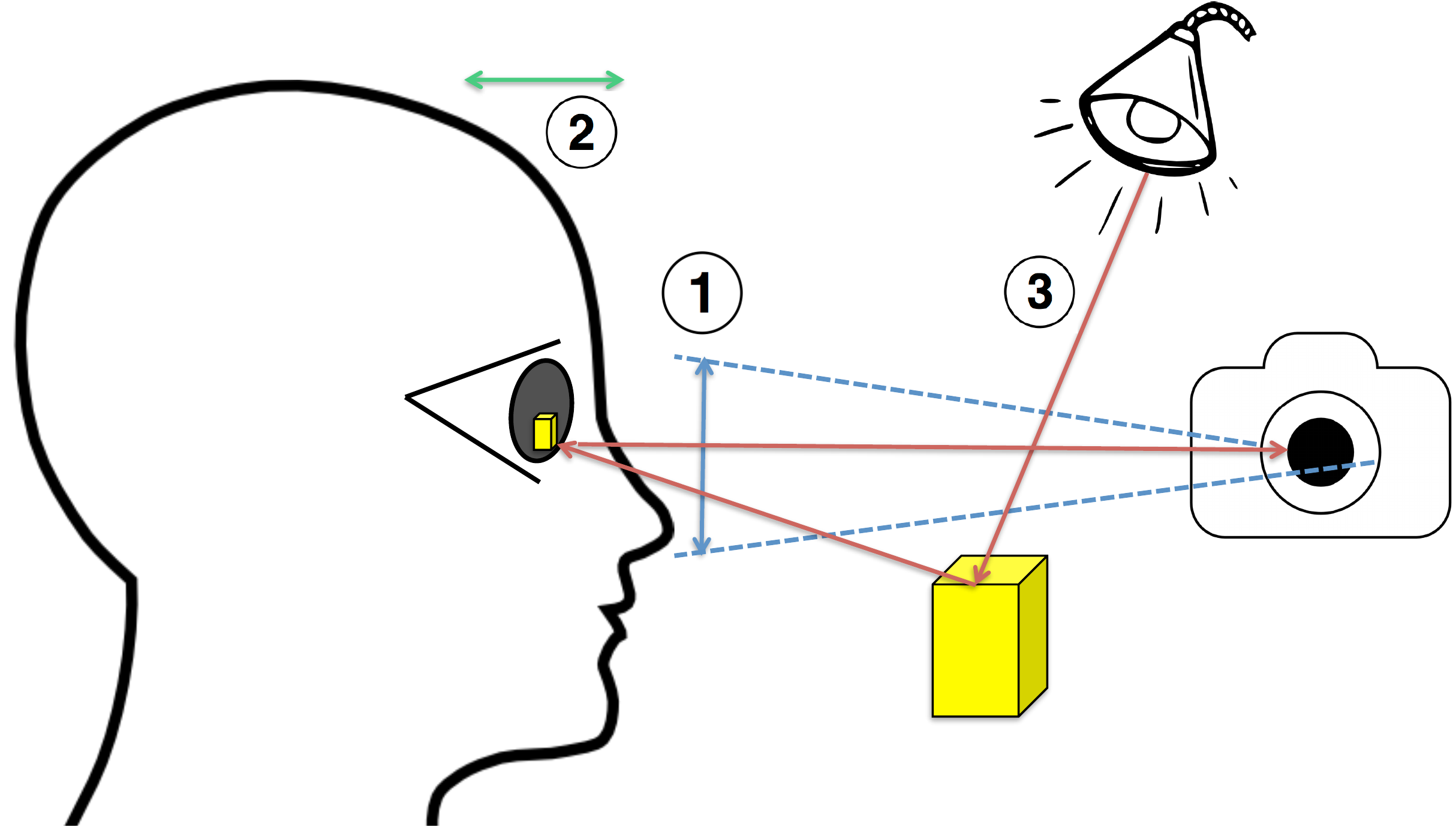}}
\subfloat[Example scenario]{\includegraphics[width=0.45\linewidth]{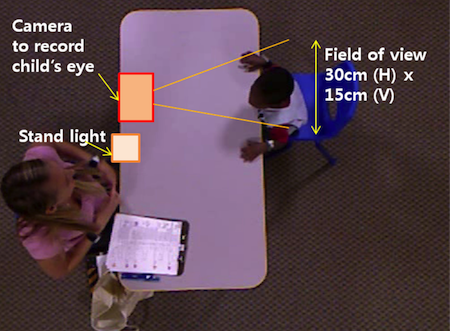}}
\caption{(a) Three constraints are addressed: \ding{192} field of view, \ding{193} depth of field, and \ding{194} the amount of light. (b) One example of how our system can be deployed: A child and an adult sit in chairs across a table. Our camera system can be placed in front of the child and towards the child's face, so it can capture the child's gaze behavior during the interaction (e.g., adult reading a book to the child).  Figure~\ref{fig:child_ball} and Figure~\ref{fig:childmore} show a few samples of corneal images that our system captured in this type of tabletop face-to-face setup.}
\label{fig:csl}
\end{figure}

\section{Camera System}
\label{sec:sys}

\begin{figure}[h!]
\centering
\includegraphics[width=0.95\linewidth]{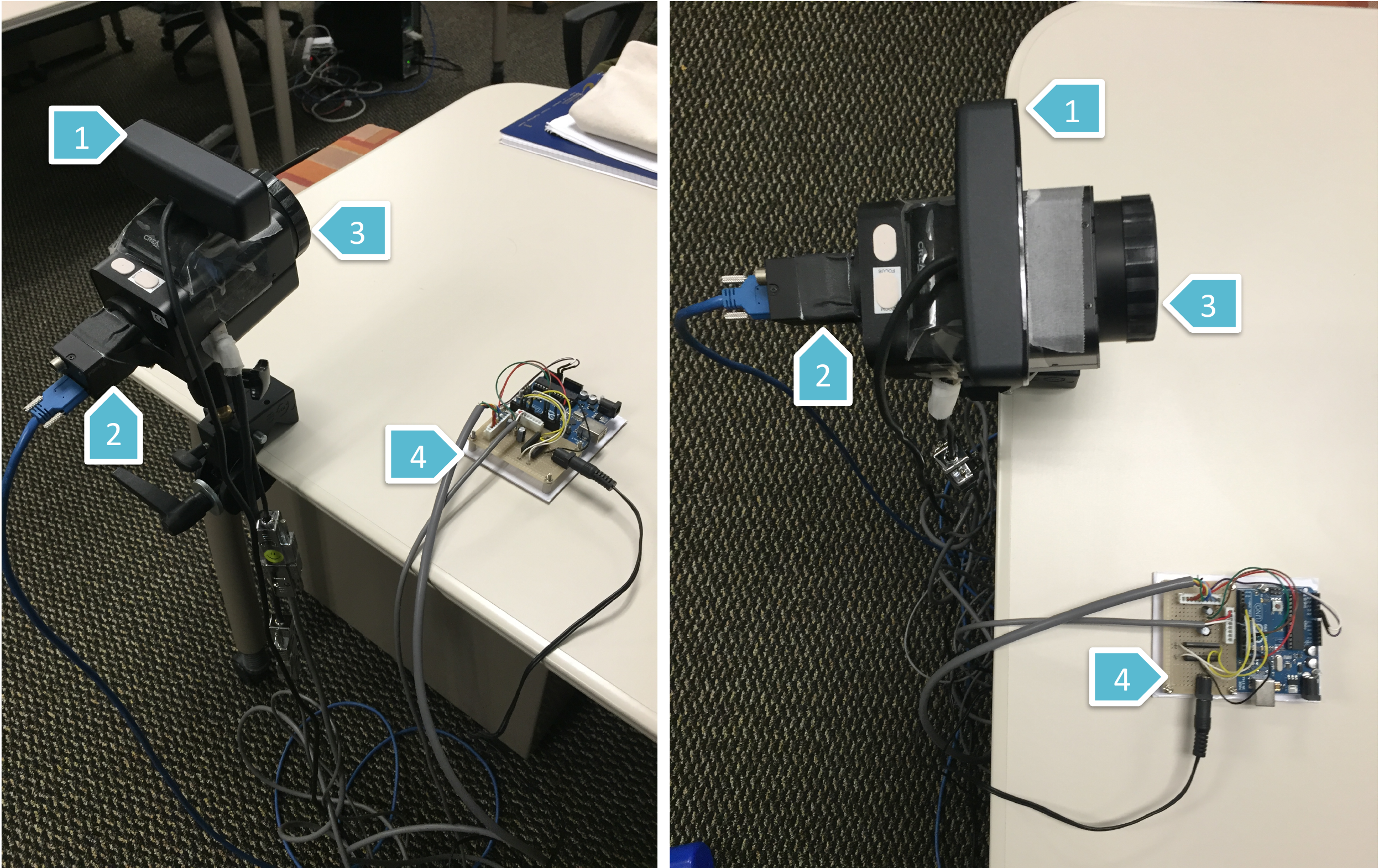}
\caption{The hardware prototype. 1: A wide-view RGBD camera (Intel Creative Senz3D). 2: A narrow-view RGB camera (Point Grey Flea3). 3: Motorized zoom/focus lens (Yamano Optical). 4: Controller board for the motorized lens (custom made along with Arduino Uno). All of these are connected to a single computer.}
\label{fig:hardware}
\end{figure}

Corneal-image based gaze tracking has somewhat contradictory requirements --- it has to obtain a high quality corneal image while minimizing restrictions on the subject's movement. The best quality corneal image is often obtained when the relative pose between a camera and the subject's head remains fixed, but such an arrangement would hinder natural interactions and is not suitable for experiments in which subjects move around. Conversely, in less restrictive settings, movements during the interaction can cause motion blur, focus blur, or the subject moving out of the camera's field of view.
To balance these requirements we designed our system to support a tabletop interaction paradigm, illustrated in Figure~\ref{fig:csl}, which is widely-used in psychology. We systematically determined the values of imaging variables for this setting (e.g., focal lengths, image resolution, depth of field, etc.) through a combination of mathematical modeling and experimentation. The resulting hardware prototype is shown in Figure~\ref{fig:hardware}.

\subsection{Tabletop Interaction Protocol}
\label{sec:targetSetup}
We now briefly explain the protocol for the tabletop interaction which guided the design of our imaging system. A play protocol that is based on tabletop face-to-face social interactions is a good candidate, as such protocols represent a common way to assess child developmental progress in psychology research and clinical settings. The specific play protocol we utilized, called the Rapid-ABC (RABC)~\cite{rehg2013decoding}, is composed of several semi-structured play activities between an adult examiner and a child, such as rolling a ball back and forth and looking through a picture book. See~\cite{rehg2013decoding} for more details on the RABC. This type of semi-structured interaction is a common way to elicit and assess young children's social-communication skills, including gaze behavior such as eye contact and shifts of gaze between objects and the social partner's face. 
We utilized this protocol, not only for the design of our system, but also for assessing the viability of our prototype. In fact, our system was able to successfully capture several moments of social attentions of young children, which will be shown in Section~\ref{sec:app_child}.

\subsection{Finding Imaging Variables and System Architecture}
\begin{figure}[h!] 
\centering
\includegraphics[width=1\linewidth]{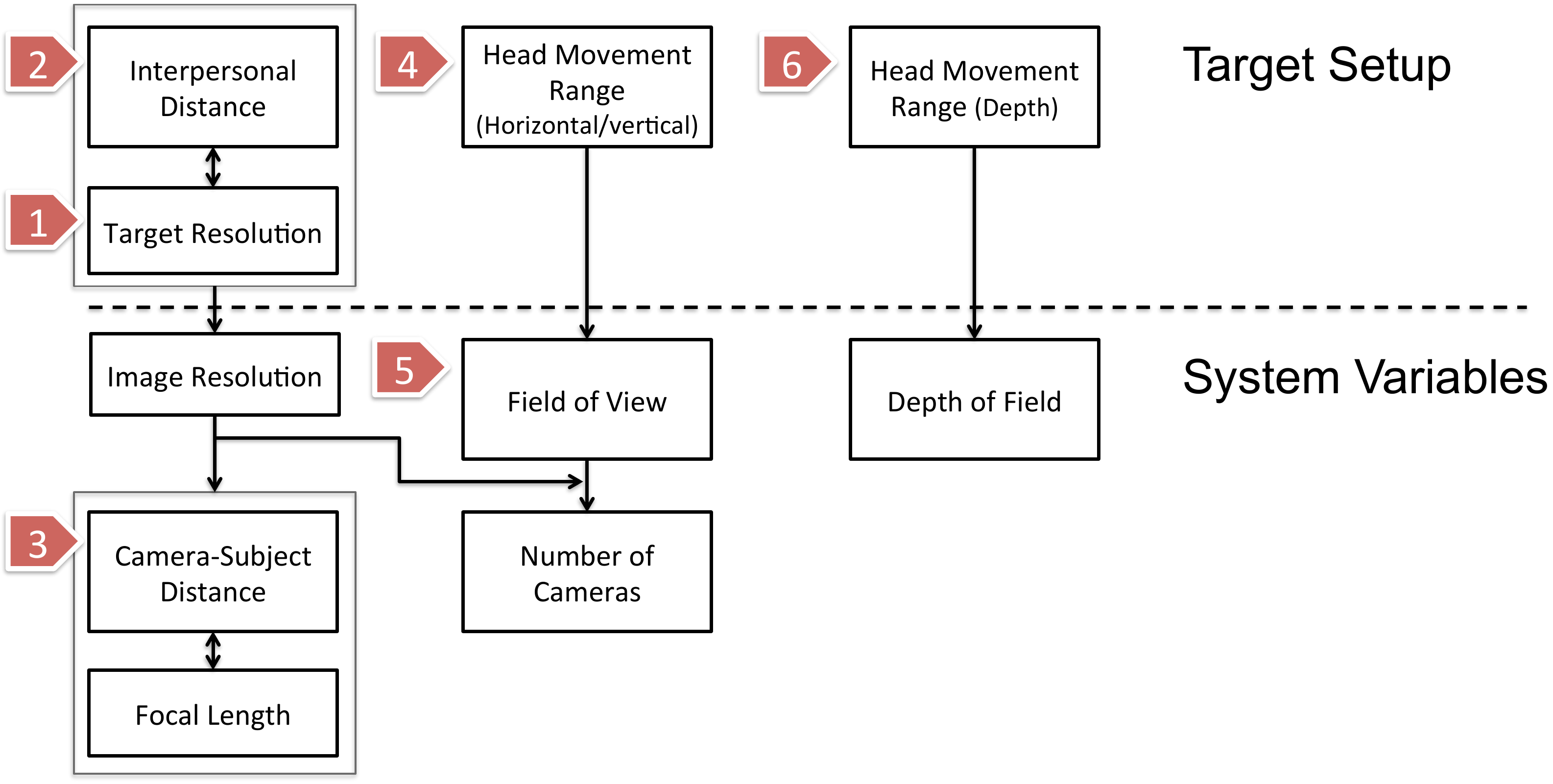}
\includegraphics[width=0.85\linewidth]{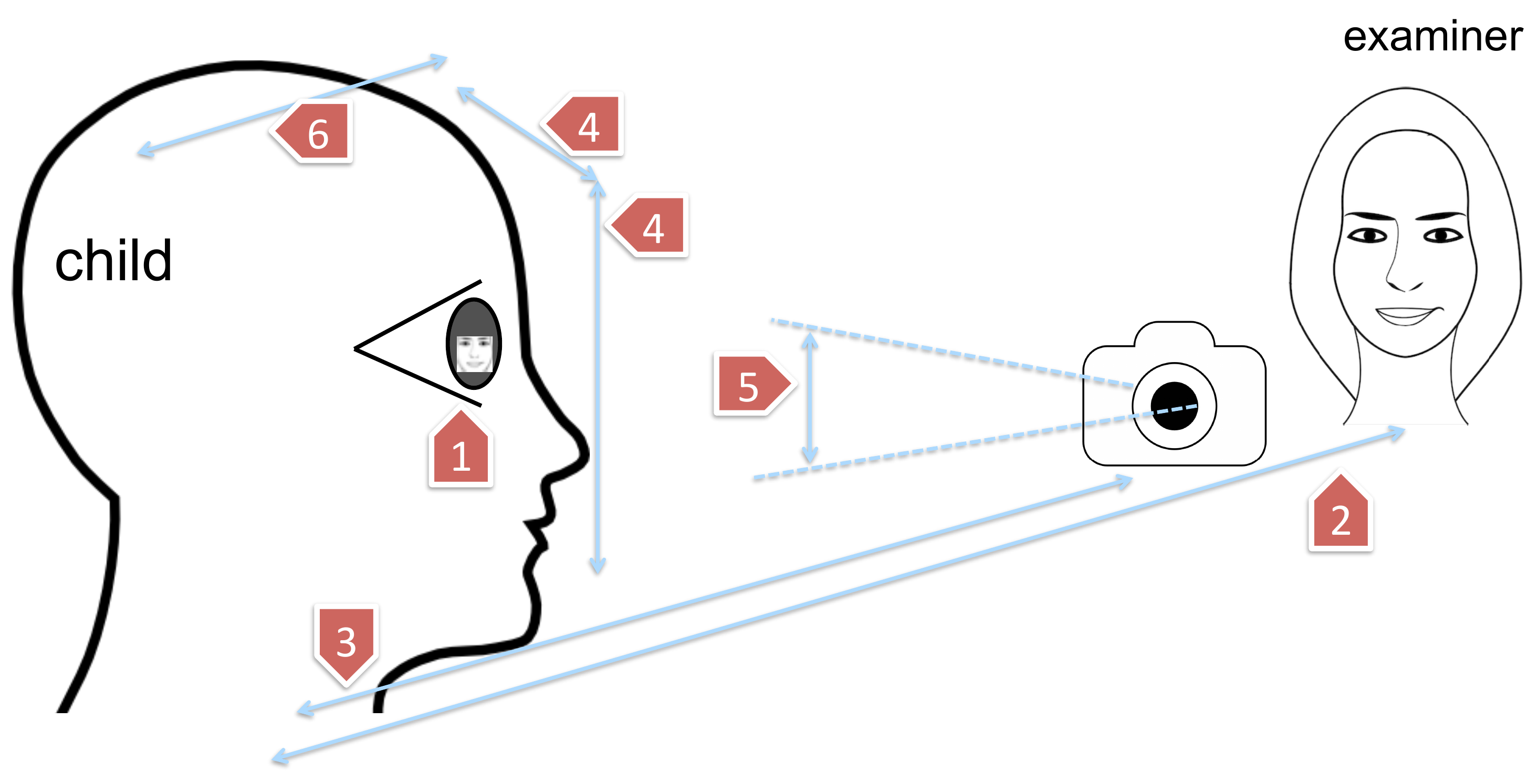}
\caption{Design procedure: we started from measuring the minimum resolution (1) of a face in a corneal image that can be automatically detected as face using~\cite{Xiong:2013aa}. We also considered person-to-person distance (2) to decide the optimal eye size in the image, which we then used to decide on the camera-to-subject distance (3) as well as focal length. Next, we measured the best field of view (5) considering the range of horizontal and vertical head movement (4). This, in conjunction with the image resolution obtained earlier, yielded the optimal number of cameras. Finally, we compensated for depth variation due to head movement (6) using autofocus.}
\label{fig:design_proc}
\end{figure}

In order to ensure the effectiveness of our approach, the imager must be designed to capture the corneal image at sufficient resolution such that important scene elements such as faces and objects are visible. To achieve this goal, we define six interdependent system variables (see Figure~\ref{fig:design_proc}), which are essentially the function of three factors in our target configuration of a tabletop face-to-face interaction (i.e. child's movement range, distance between the two people, and the desired image resolution).

In order to characterize the scene, we randomly selected 10 RABC sessions from~\cite{rehg2013decoding} and analyzed video streams from the overhead camera and the child-facing room camera. We found that the average distance between the examiner and the child (Figure~\ref{fig:design_proc}-(2)) was about 55cm. We also found that the child's eyes were in the child-facing camera's view 70\% of the time within the bounding box of 15cm(V) by 25cm(H) (Figure~\ref{fig:design_proc}-(4)). Finally, we found that the child's depth variation w.r.t. the examiner (Figure~\ref{fig:design_proc}-(6)) was 15cm on average. These numerical values are then used as follows:

When two individuals are 55cm apart and facing each other, the reflection of one's face on the other's eye occupies about 1/5th of diameter of iris boundary (limbus). Considering the average diameter of the limbus is 11.2mm~\cite{snell1998clinical} and the face detector we used~\cite{Xiong:2013aa}~\cite{omron} required 45 pixel x 45 pixel face resolution, we require approximately a resolution of 200 pixels/cm across the surface of the cornea. In our setup, this requires a camera with 35mm-focal-length lens positioned 50cm away from the subject. The effect of varying interpersonal distance will be discussed further in Section~\ref{sec:eval}.

Given an imaging system achieving 200 pixels/cm, there is a question of how to effectively cover the complete volume of the scene (as much as 15cm x 25cm for a child sitting at a tabletop). While a pan-tilt-zoom system could be used, there are two disadvantages with this approach: 1) it could potentially distract subjects with its movements and noise, 2) it could produce focus and motion blur. We believe that a better strategy is to use a motorized focus lens to cope with depth variation and, if necessary, use a camera array to provide the necessary spatial coverage. The lens focus can be controlled using depth measurement from a separate wide-view camera placed directly on top of the imager. This is discussed further in the following section.

\subsection{Autofocus}
\label{sec:autofocus}
Because we can determine the depth of the eyes from the imager using the information from the depth camera, we can easily calculate the focus value that maximizes the sharpness of the image. We can safely assume a thin lens model because the thickness of the lens is negligible compared to the focal length of the lens. Using the thin lens equation, object depth $S$ is related to the focal length $f$ and the back focal distance $s$, as in $f^2 = sS$. As we know $f$ already, and $S$ is obtained at every frame, $s$ is estimated in real time. To relate this $s$ to the focus motor value, we performed focus control calibration, where a linear mapping from $s$ to motor value is estimated with several focused images at known depth $S$.

\section{Gaze Estimation}
\label{sec:gaze}
The process of analyzing an eye image to determine the gaze consists of four steps. First, in order to save resources the image is closely cropped to the eye region. Second, we use a geometric eye model to represent the pose of the eye. Third, we track the limbus with an ellipse to recover the eye parameters. Fourth, we estimate the point of gaze, corrected by a subject-specific offset.

\subsection{Eye Cropping}
\label{subsec:eyecropping}

To save disk space, we perform online eye region cropping and keep only that portion for later use. To increase the speed and the precision of the cropping step, we use the wide-view camera. Because the face is guaranteed to be visible at all times in the wide-view camera, it can be used to select a subset of the image from the narrow-view camera that is recording the corneal image in high resolution. Because the image planes of both cameras are coplanar, we found that simple template matching method works very well.

\begin{figure}[h!]
\centering
\includegraphics[width=1\linewidth]{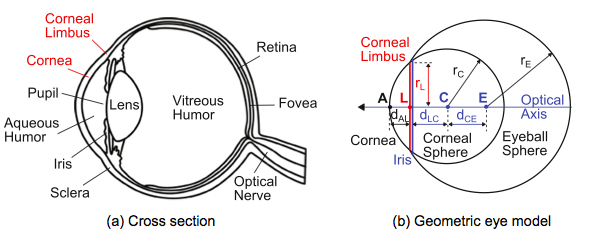}
\caption{(a) Cross section of human eye. (b) Geometric eye model represented as two intersecting spheres.}
\label{fig:eyemodel}
\end{figure}

\begin{figure}[h!]
\centering
\includegraphics[width=0.95\linewidth]{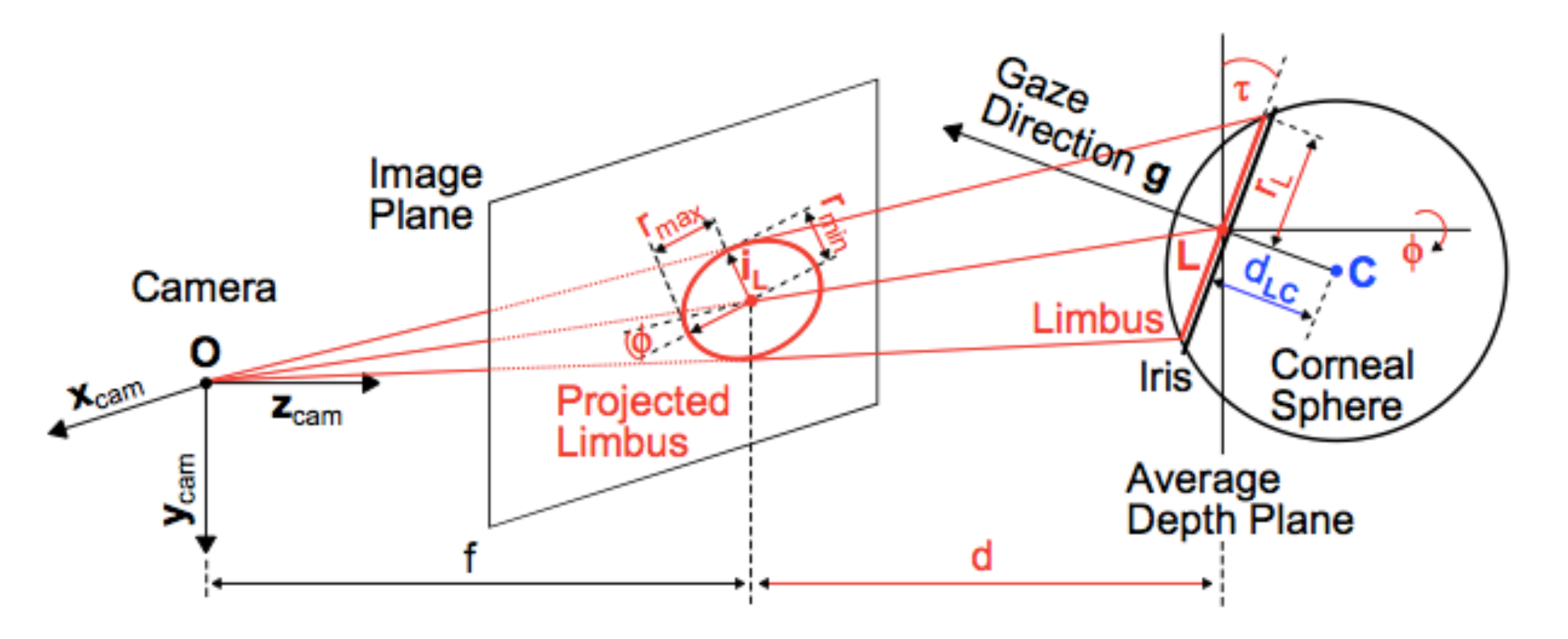}
\caption{Geometrical relationship between eye pose and projected ellipse on image plane}
\label{fig:gazdir}
\end{figure}

\subsection{Eye Model}
Figure~\ref{fig:eyemodel}(a) depicts a cross section of an eye and Figure~\ref{fig:eyemodel}(b) shows an eye model with two spheres, corneal sphere and eyeball sphere. The intersection of the two spheres is a limbus or an iris contour, and we can estimate gaze direction through its pose. The projected shape of the limbus in an image plane is an ellipse. Because the depth of the corneal limbus is much smaller than the eye-camera distance, we can assume a weak perspective projection. Using the model, we can infer the pose of an eye by fitting an ellipse to an image of the limbus.

\subsection{Estimation of the Eye Model Parameters}
Both the limbus and the pupil are visible candidates for ellipse-based tracking in order to determine the eye pose. We investigated pupil tracking but found it to be problematic.\footnote[1]{Many gaze trackers detect the pupil with an IR light source using the dark pupil method. In the dark pupil effect, a pupil appears near black in infrared imagery when an IR light source is not aligned with the camera-pupil axis. However, in our experiments, we found that recording both the infrared and the visible spectrum with a camera resulted in overexposed videos. To avoid this phenomenon, we also tried strobing the IR light source to alternate between visible and infrared streams. However, due to the difference in wavelength between the two streams, this negatively affected focus. Thus, rather than using a separate IR camera, we addressed this problem by improving our ellipse fitting in the visible spectrum.} 
As a consequence, we focused on limbus tracking and improved our method to gain robustness to occlusion by eyelashes and eyelid. 

Now, we explain how we fit an ellipse and how the eye pose is estimated from it. An ellipse is expressed with 5 parameters, $p = (r_{max},r_{min},x,y,\phi)$, where $r_{max}$ and $r_{min}$ are long and short radii respectively, $x$, $y$ are the center coordinate in the image plane, and $\phi$ is the rotation angle (See Figure~\ref{fig:gazdir}). We find these parameters as follows:
\begin{enumerate}
\item Find a set $L = \{l_1, l_2, ..., l_{2m}\}$, possible radii of the ellipse, where $l_{(m+i)} = f \times \frac{RL}{D} + i \ (-m+1 \leq i \leq m)$. Here, $D$ is the depth of the eye w.r.t. the camera, $RL$ is the average radius of the limbus (=5.6mm~\cite{san2010evaluation}), and $f$ is the focal length of the narrow-view camera (35mm). Then, we find a set $C = \{c_1, c_2, ..., c_n\}$ possible center locations of the ellipse, which includes the center of the cropped eye image and several randomly selected points within 10 pixels from the center. Both $m$, $n < 20$ so that the search space is tractable.

\item By Hough transform using $L$ and $C$ from the previous step, we find a candidate set of ellipses $P=\{p_1, p_2, ...\}$ (10$\sim$15 candidates are selected with the highest scores, each p is represented with its five parameters) as well as $E = \{e_1, e_2, ..., e_k\}$, a set of edge points near $P$.

\item Finally we find a single ellipse $p$, using the following rule
\begin{equation} 
\begin{split}
argmax_{p \in P} \Bigg\{ & A \abs{ \frac{I(\text{inside } p)<N}{I(\text{outside } p)<N} } + \\ & B \abs{||E - (p_x,p_y)||<\delta} \Bigg\},
\end{split}
\end{equation}
where $I(\text{inside } p)$ are the pixel values of all the points inside ellipse $p$, $I(\text{outside } p)$are the pixel values of all the points outside ellipse $p$, $N$ is the average pixel value for iris color, $A$ and $B$ are scale factors for each term, $(p_x,p_y)$ is x,y coordinate of ellipse $p$, $\delta$ is empirically chosen threshold, and\abs{...}denotes the number of elements satisfying the inequality.

\item Estimation of 3D eye model parameters:
Given the 2D corneal boundary, we obtain the 3D eye model parameters as depicted in Figure~\ref{fig:gazdir} using the method described in~\cite{nishino2004world}~\cite{nakazawa2012point}. Namely, the angle $\phi$ is known as the rotation of limbus ellipse in the image plane. The angle is given as \mbox{$\tau = \pm \arccos \left( r_{\min } / r_{\max } \right)$}. The center of the corneal sphere $\mathbf{C}$ is located at distance $d_{\mathbf{LC}}$(=~5.6~mm) along $-\mathbf{g}$. Here, there exists an ambiguity in estimating $\tau$, which comes from the ambiguity projecting a 3D circle to the 2D plane.

\item 3D model-based corneal tracking:
Once the initial 3D eye model parameters are solved, we apply particle filter-based tracking for the parameters similar to the approach of~\cite{wu2007tracking}. We generate particles for the corneal center $\mathbf{C}$, two rotation parameters $(\phi, \tau)$ and distance $d$, then obtain a corneal boundary for each particle. We calculate the likelihood of each particle by evaluating the function and obtain the time-series of 3D eye model parameters.
\end{enumerate} 

\subsection{Point of Gaze Estimation and Calibration of Individual Parameter}
\begin{figure}[h!]
\includegraphics[width=1.05\linewidth]{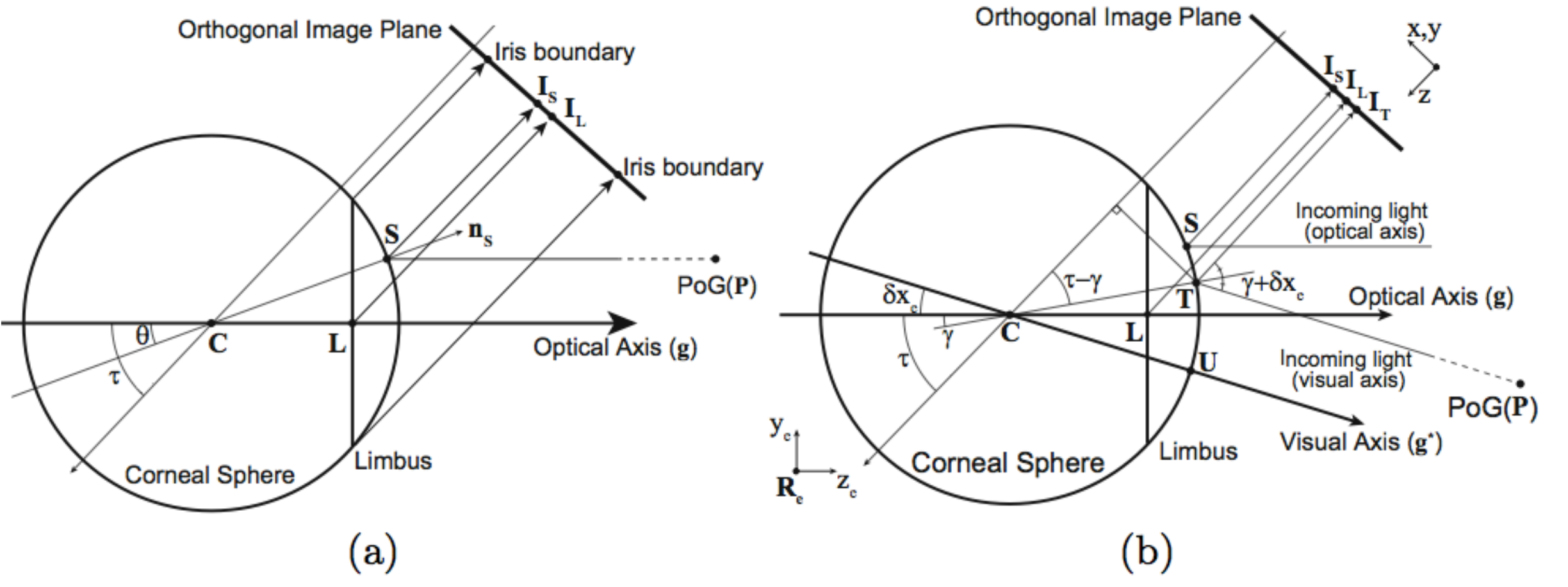}
\caption{(a) The reflection of light-ray parallel to the gaze direction is observed as point of gaze in the image. (b) Illustration of the optical axis $g$ and the visual axis $g*$ of the human eye. The angular difference between these two axes is individual offset, which can be corrected via kappa calibration.}
\label{fig:cornea_val}
\end{figure}

Once an ellipse is found, gaze direction $g$ is obtained from our model in Figure~\ref{fig:eyemodel} and Figure~\ref{fig:gazdir} as
\begin{equation}
g = [sin(\tau) sin(\phi) -sin(\tau) cos(\phi) -cos(\tau)]^{T},
\end{equation}
where $\theta$ is the rotation of ellipse and $\tau=\pm arccos(r_{min}/r_{max})$ is the tilt of the limbus plane w.r.t the image plane.

To find a gaze point in the corneal image, we define a gaze reflection point (GRP) as follows: a light ray parallel to an optical axis of an eye is reflected by its corneal sphere and projected on the image plane. The gaze reflection point is this point in the image. (See Figure~\ref{fig:cornea_val}a.) With the definition, we can find the distance between GRP $I_S$ and $I_L$, a projection of limbus center, as $\left | I_{S} -I_{L} \right |=r_{C}sin(\theta) -d_{LC}sin(\tau)$ where values we used for $r_C$ and $d_{LC}$ are 7.7mm and 5.6mm, respectively~\cite{snell1998clinical}.

\section{Evaluation}
\label{sec:eval}
We evaluated the performance of our system in two ways. First, we conducted experiments to measure the empirical performance. Section~\ref{sec:syseval} describes three evaluations: 1) We quantified the ability of the system to capture in-focus corneal images by measuring face detection performance at a range of depths. 2) We tested the absolute accuracy of PoG estimation using a fixed array of targets and compared our method against a commercial wearable eye tracker. 3) We also characterized the latency of our autofocus control system. Second, we conducted a sensitivity analysis to identify the effect of ellipse measurement error on gaze estimation error and characterized the amount of calibration needed for kappa estimation.

\subsection{System Evaluation}
\label{sec:syseval}
\setlength{\tabcolsep}{1pt}
\begin{table*}[ht]
\label{table:one}
\begin{center}
\begin{tabular}{|c||c|c|c|c|c|c|}
\hline
\backslashbox{Subject}{Depth} & \parbox{1.6cm}{\centering 80cm\\Ours (w. $\kappa$)} &\parbox{1.8cm}{\centering 80cm\\Ours (w.o. $\kappa$)}& \parbox{1.6cm}{\centering 80cm\\SMI Glasses} &\parbox{1.6cm}{\centering 160cm\\Ours (w. $\kappa$)} &\parbox{1.8cm}{\centering 160cm\\Ours (w.o. $\kappa$)}& \parbox{1.6cm}{\centering 160cm\\SMI Glasses}\\
\hline
1 & 1.511$^{\circ}$ & 1.528$^{\circ}$ & 1.911$^{\circ}$ &1.709$^{\circ}$ & 1.913$^{\circ}$ & 1.887$^{\circ}$\\
2 & 1.726$^{\circ}$ & 2.110$^{\circ}$ & 1.182$^{\circ}$ &1.884$^{\circ}$ & 2.110$^{\circ}$ & 1.736$^{\circ}$\\
3 & 1.337$^{\circ}$ & 1.602$^{\circ}$ & 1.024$^{\circ}$ &1.698$^{\circ}$ & 1.732$^{\circ}$ & 2.057$^{\circ}$\\
4 & 1.488$^{\circ}$ & 1.697$^{\circ}$ & 1.406$^{\circ}$ &1.921$^{\circ}$ & 2.055$^{\circ}$ & 1.905$^{\circ}$\\
5 & 1.700$^{\circ}$ & 1.815$^{\circ}$ & 1.533$^{\circ}$ &1.770$^{\circ}$ & 1.926$^{\circ}$ & 1.867$^{\circ}$\\
\hline
Average Error & 1.573$^{\circ}$ & 1.750$^{\circ}$ & 1.411$^{\circ}$ & 1.796$^{\circ}$ & 1.947$^{\circ}$ & 1.892$^{\circ}$\\\hline
\end{tabular}
\end{center}
\caption{Error (in degree) of point of gaze estimation at different depth, with and without kappa calibration (abbreviated as `w. $\kappa$' or `w.o. $\kappa$'), and quantitative comparison with SMI glasses. Note that the SMI glasses always have to be calibrated. The result demonstrates that our system has comparable accuracy with a state-of-the-art wearable gaze tracking system.}
\end{table*} 

\begin{figure}[h]
\subfloat[]{\includegraphics[width=0.55\linewidth]{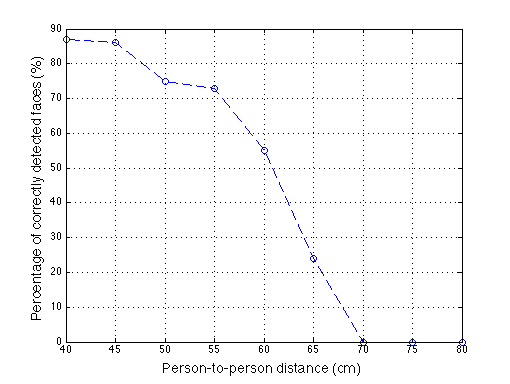}}
\subfloat[]{\includegraphics[width=0.55\linewidth]{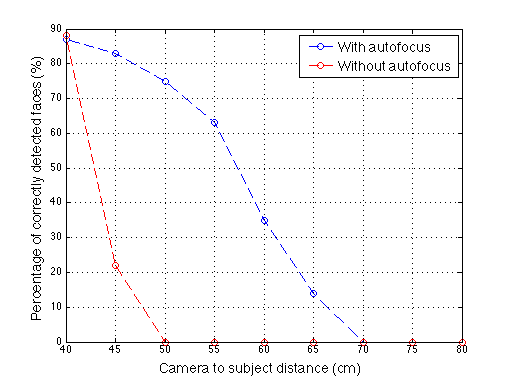}}\\
\centering
\subfloat[]{\includegraphics[width=0.5\linewidth]{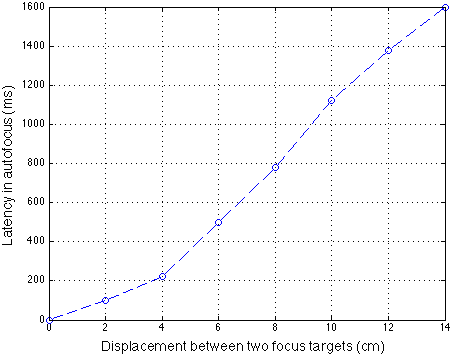}}
\caption{System evaluation with different experimental variables: (a) Face detection results on corneal image under different interpersonal distances. (b) Face detection results on corneal image under different camera-to-subject distances and focus conditions. (c) Latency of autofocus control. The terms ``camera-to-subject distance" and ``interpersonal (person-to-person) distance" are illustrated in Figure~\ref{fig:system_eval_illust}.}
\label{fig:evals}
\end{figure} 

\begin{figure}[h!]
\centering
\includegraphics[width=0.8\linewidth]{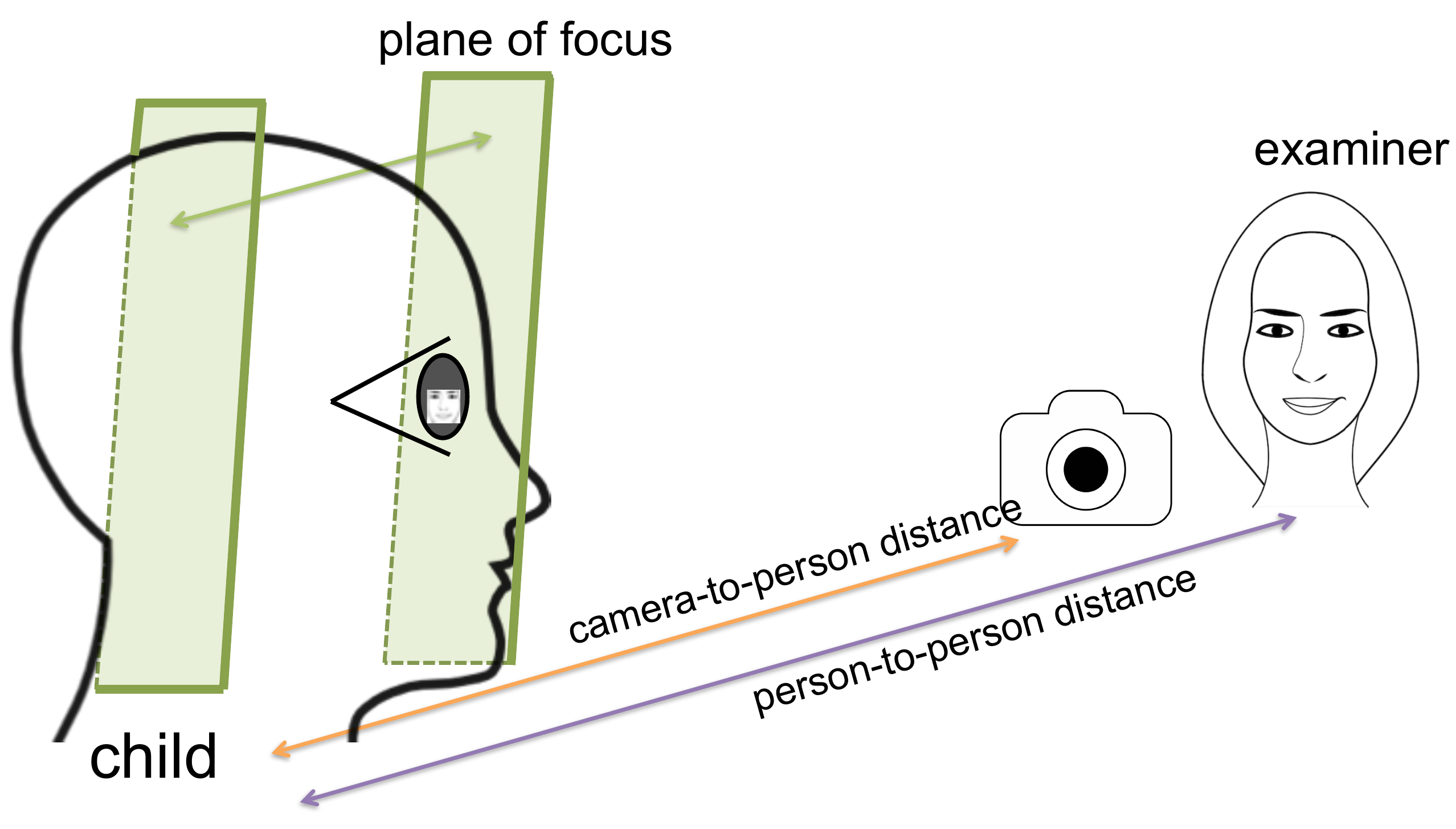}
\caption{Illustration of experimental setup for Figure~\ref{fig:evals}.}
\label{fig:system_eval_illust}
\end{figure} 

\begin{figure}[h!]
\centering
\includegraphics[width=0.9\linewidth]{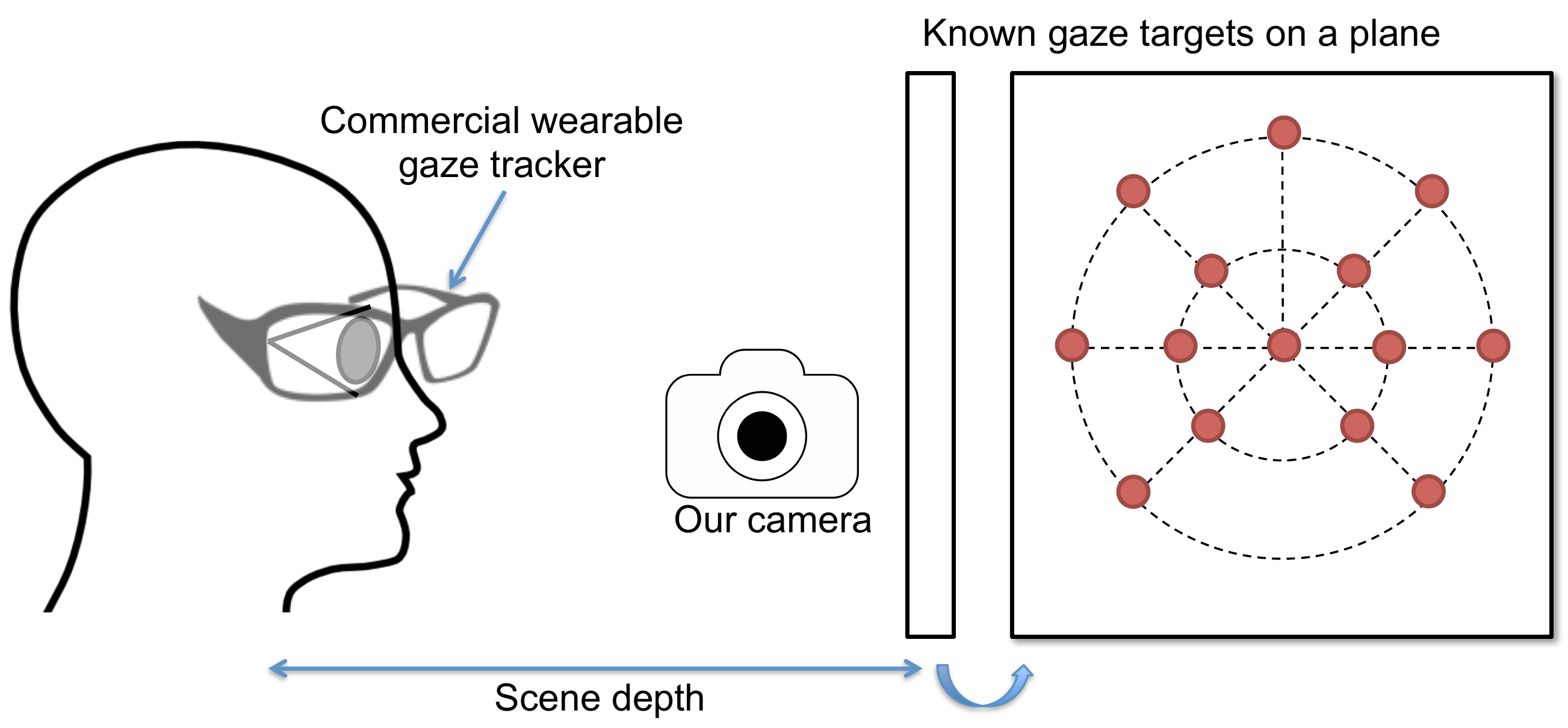}
\caption{Experimental setup for gaze estimation accuracy evaluation}
\label{fig:pog_eval}
\end{figure}

As mentioned in Section~\ref{sec:sys}, one design criterion was to detect faces in the corneal image during natural social interactions. We tested the end-to-end performance of our system using the following protocol: One subject (S1) was facing the camera at a distance of 0.50m while a second subject (S2) was positioned behind the camera and facing S1. (See Figure~\ref{fig:system_eval_illust}). At each iteration, S2 moved away from the camera by 5cm. We sampled 450 frames and measured the number of frames in which the detector was successful at finding S2's face in S1's corneal image. Figure~\ref{fig:evals}(a) shows the frequency of face detection across varying distances between two individuals.  As depicted in Figure~\ref{fig:evals}(a), the detector was mostly successful at detecting face in corneal images when the two individuals were less than 0.55m apart and still working before 0.70m. This validates the usefulness of our system at tabletop interaction distances.

One of the main factors that affected the performance of the face detector was focus blur on the corneal image. In a natural setting where two individuals converse, there are nonneglible head movements. To increase robustness to such movement, we used our online autofocus that could react to both individuals' movements and that was sufficiently fast for optimal face detection without frame loss. Figure~\ref{fig:evals}(b) shows the effect of activating the autofocus module described in Section~\ref{sec:autofocus}. The experiment was conducted similarly as for Figure~\ref{fig:evals}(a) except that only the camera-facing subject moved and the camera and the person behind the camera remained still. We also measured the latency of our autofocus (Figure~\ref{fig:evals}(c)) using the following procedures. We placed two pictures of faces in at different camera-to-face distances, one behind another, and focused the camera on the nearest picture. We then quickly removed this nearest picture and allowed the autofocus to adapt its focus to the other picture. This way we could measure focus latency by counting the number of out-of-focus frames and convert it into milliseconds. In this experiment, we found that the autofocus can reliably adapt to displacement within 6cm in less than 0.5 seconds. 

Once we detect a face in a corneal image we want to measure the subjects' point of gaze. The accuracy of point of gaze estimation is one of the most important criteria for performance of the system, which we measured using 15 ground truth markers  placed in depth planes at 80cm and at 160cm. (See Figure~\ref{fig:pog_eval}). The effect of individual offset (kappa) calibration was measured as well by comparing accuracy with and without this calibration. We also compared the accuracy of our PoG estimation to estimates provided by a pair of commercially available state-of-the art eye tracking glasses (i.e., SMI glasses) by having a subject wear the glasses while simultaneously being recorded by our system. We extracted the frames from both gaze trackers that recorded the same moments to compare PoG estimation accuracy.

Figure~\ref{fig:pog_eval} depicts our experimental setup, and Table~\ref{table:one} shows the estimation errors, measured as the angular difference (degrees) between the position of the ground truth marker, and the PoG position estimated by our method when subjects were looking at the markers at 80cm and 160cm distances.
These results indicate that at a distance of 80cm, even without kappa calibration, our system performance is clearly not worse than SMI glasses. At a distance of 160cm we are comparable to SMI with calibrated kappa. Note that the SMI glasses are always have to be calibrated every time.

\subsection{Error Analysis}

\begin{figure}[h!]
\centering
\subfloat[]{\includegraphics[width=0.32\linewidth]{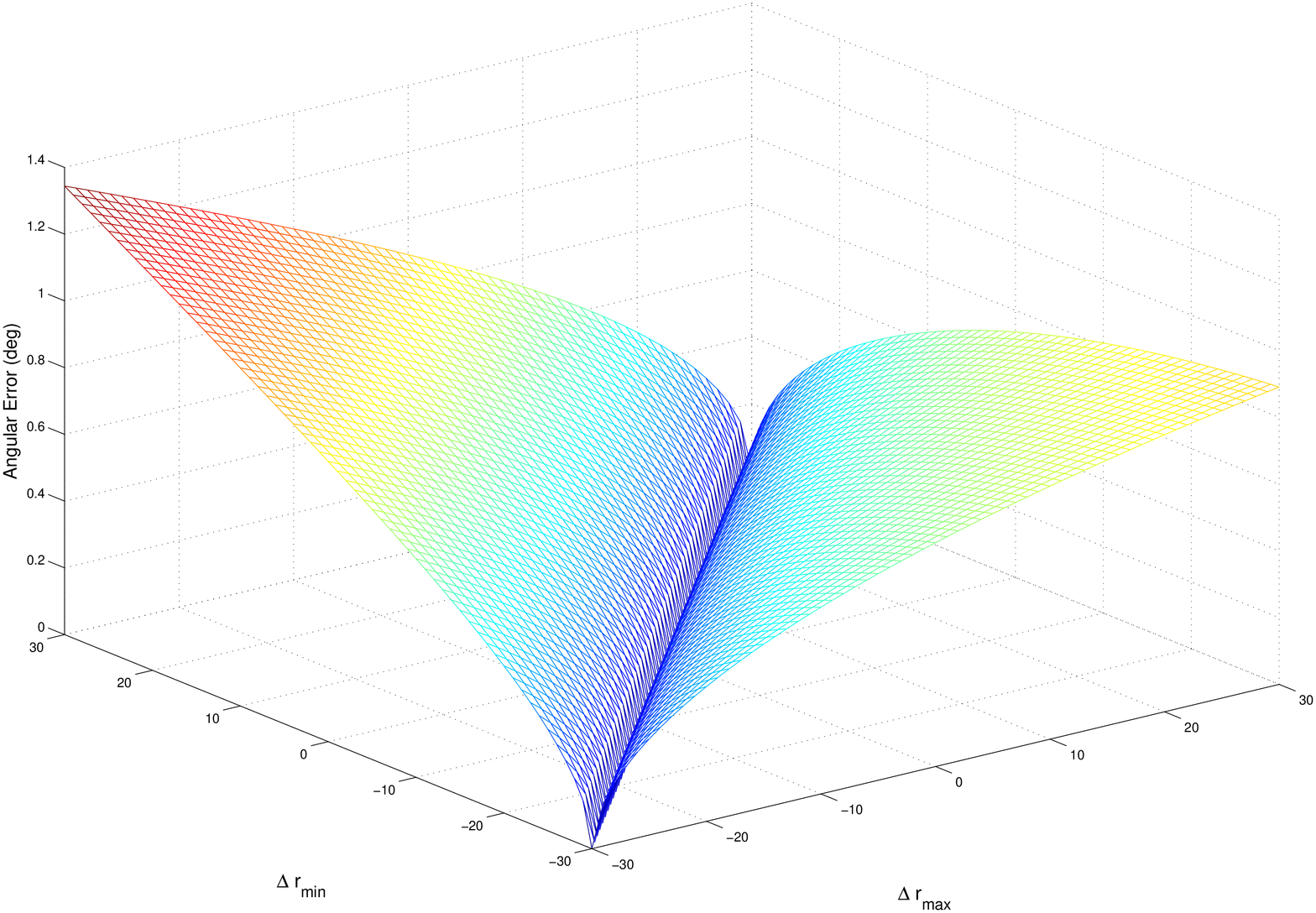}}
\subfloat[]{\includegraphics[width=0.3\linewidth]{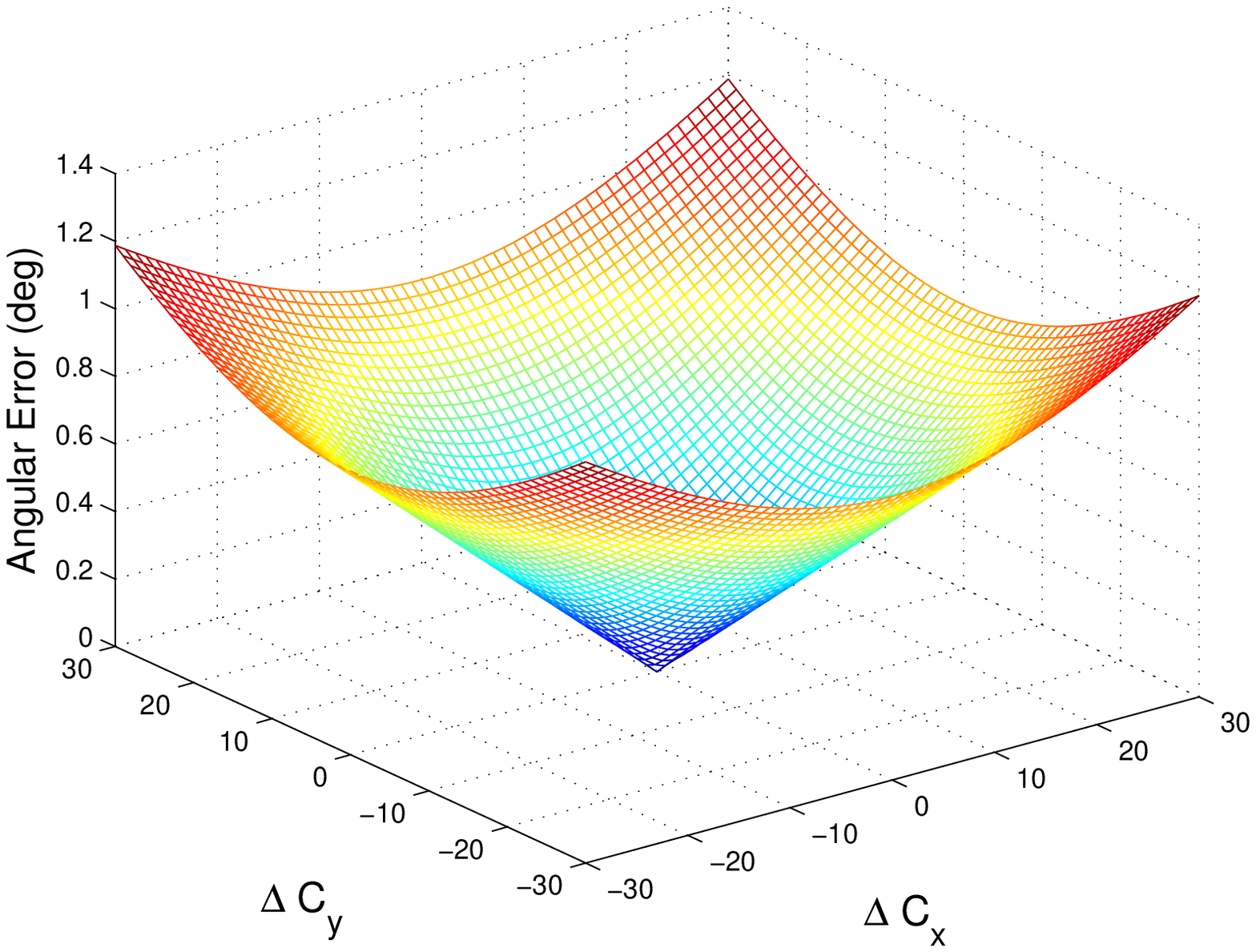}}
\subfloat[]{\includegraphics[width=0.3\linewidth]{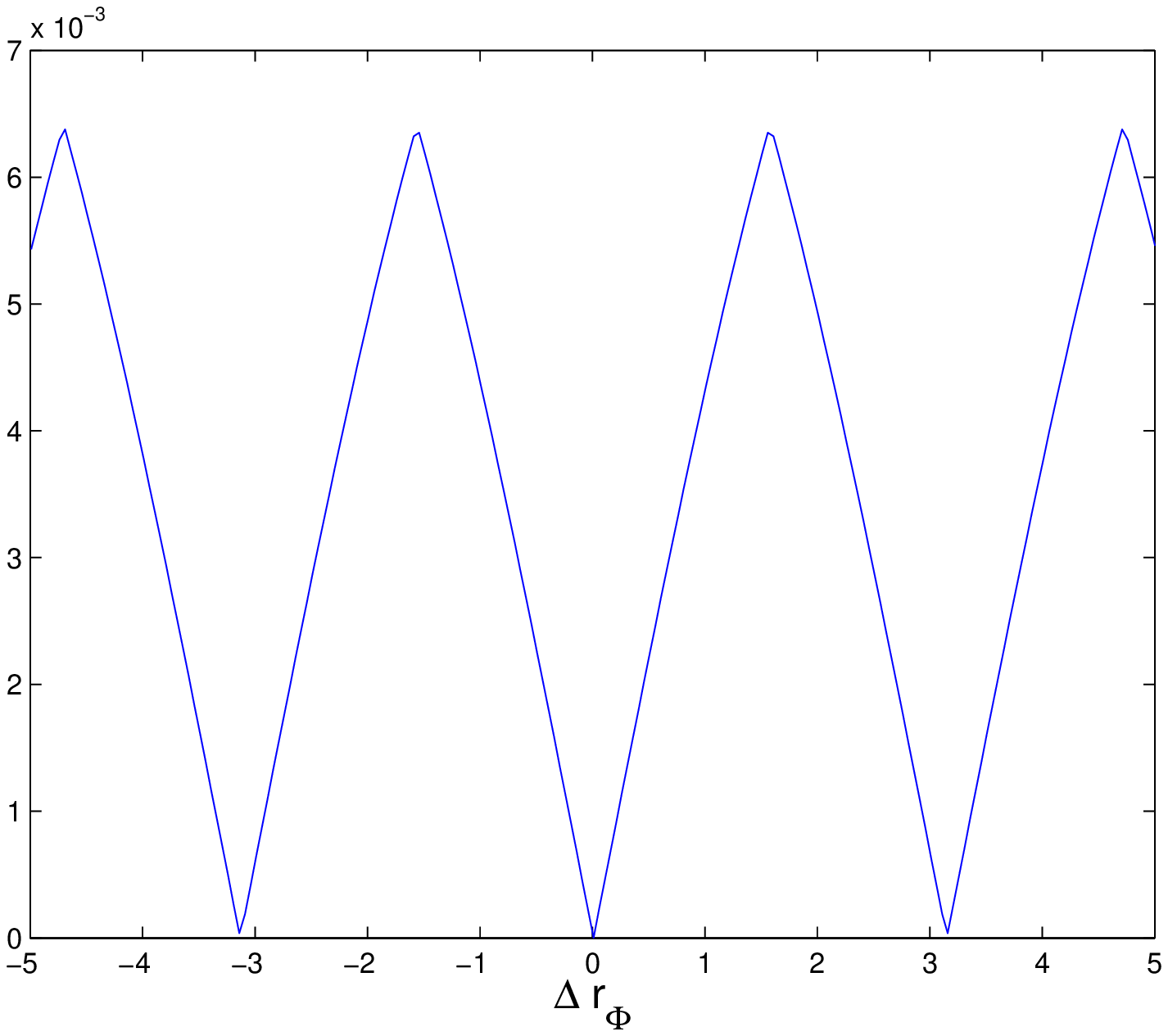}}
\caption{Sensitivity analysis by looking at how gaze estimation error is propagated by changing different parameters in the projected eye model. (a) Gaze angle errors as a function of ellipse axes length estimation error. (b) Gaze angle errors as a function of ellipse center location estimation error. (c) Gaze angle errors as a function of ellipse tilt estimation error.}
\label{fig:sensitivity}
\end{figure} 

\begin{figure}[h!]
\centering
\includegraphics[width=0.7\linewidth]{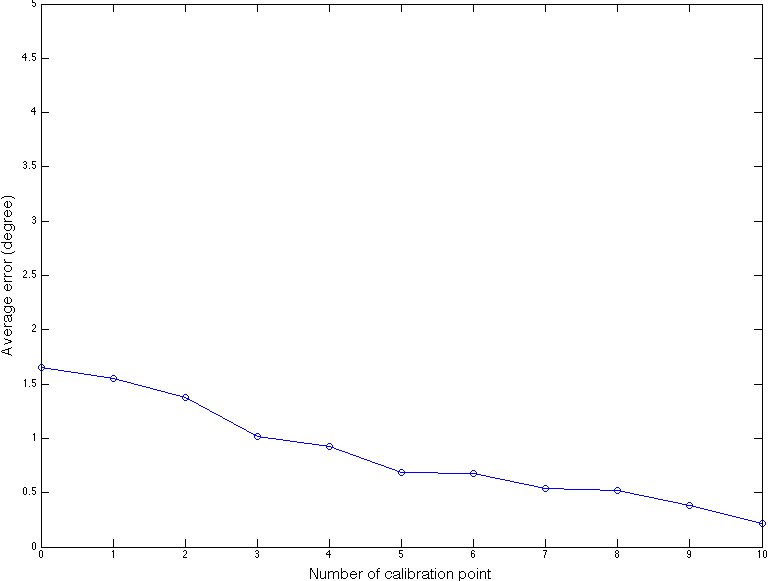}
\caption{Average gaze estimation error vs. number of calibration points. Since it converges within a few number of points, we can rest assured that we would not need many calibration points.}
\label{fig:calib_convergence}
\end{figure}

In order to further quantify the error characteristics of our method, we conducted two additional analyses. 

First, we examined the relationship between errors in the projected eyeball (i.e., ellipse on the image) estimation and errors in 3D gaze angle estimation. Because we use the shape of the ellipse for eye pose estimation (Figure~\ref{fig:gazdir}), gaze angle estimation errors can be viewed as a function of errors of ellipse parameters. As an ellipse is defined with five different parameters (2 for center, 2 for axis length, 1 for tilt), we visualize three separate relations, for center location, axis, and tilt. (See Figure~\ref{fig:sensitivity}). These results suggest that errors in estimating the ellipse center have the biggest impact on gaze angle.

We also investigated the number of calibration points that are needed for kappa calibration that are needed to guarantee a certain level of confidence. Figure~\ref{fig:calib_convergence} shows that gaze estimation error rapidly converges to 0 as the number of calibration points increases. This suggests that 4$\sim$5 points are sufficient. 

\section{Applications}
\label{sec:app}

\begin{figure}[h!]
\centering
\subfloat[]{\includegraphics[width=0.4\linewidth]{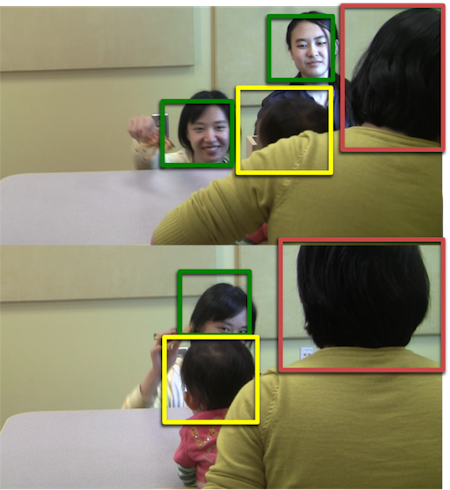}}
\subfloat[]{\includegraphics[width=0.45\linewidth]{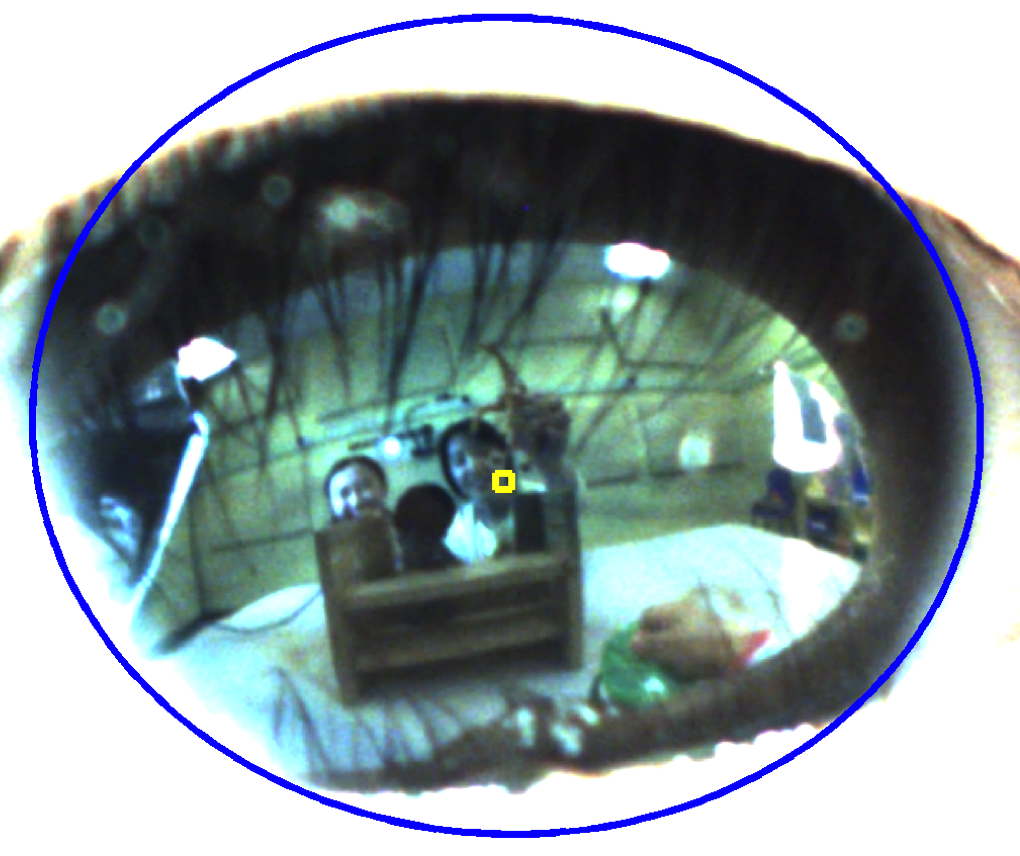}}
\caption{Experimental setup where two adults play with a child. Here, an adult presents a toy to the child. (a) Two snapshots of this session. (b) A corneal image taken by our eye camera.}
\label{fig:giraffe}
\end{figure} 

\begin{figure}[h!]
\setlength\fboxsep{0pt}
\setlength\fboxrule{0.5pt}
\fbox{\includegraphics[width=0.49\linewidth]{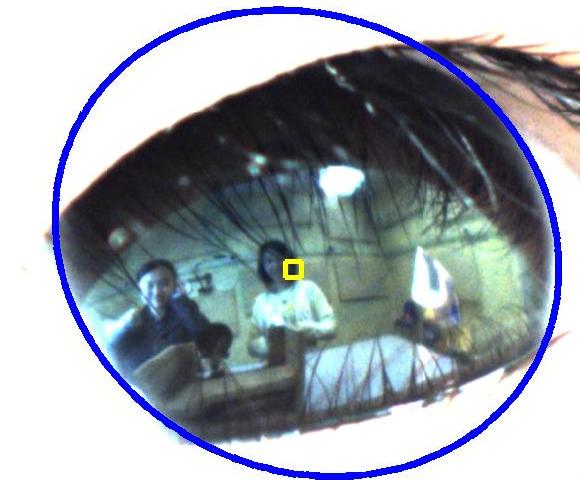}}
\fbox{\includegraphics[width=0.49\linewidth]{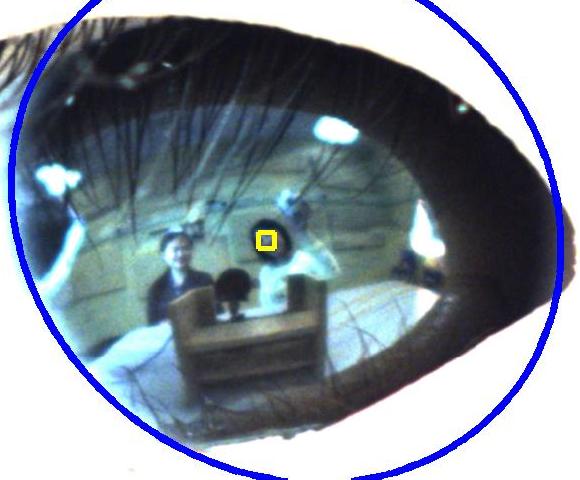}}
\fbox{\includegraphics[width=0.49\linewidth]{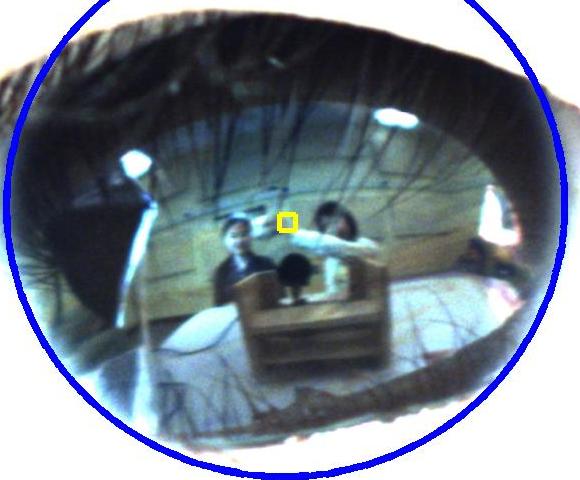}}
\fbox{\includegraphics[width=0.49\linewidth]{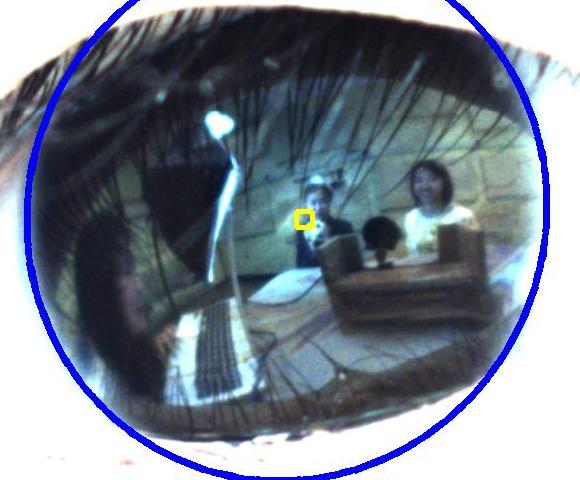}}
\caption{Two adults in front of the viewer (child) are interacting with the child by showing a soccer ball. The yellow square denotes the gaze of the child, and you can see the child's gaze shifts from the right to the left of the scene.}
\label{fig:child_ball}
\end{figure}

In this section, we demonstrate the result of using our system in three relevant application scenarios; interaction between adult and child (Section~\ref{sec:app_child}), testing existing hypothesis in psychology (Section~\ref{sec:app_blind}), and naturalistic viewing (Section~\ref{sec:app_more}). Videos will be provided as Supplementary Materials.

\subsection{Social Interactions with Child}
\label{sec:app_child}
\begin{figure*}[h!]
\centering
\subfloat{
\includegraphics[width=0.2\linewidth]{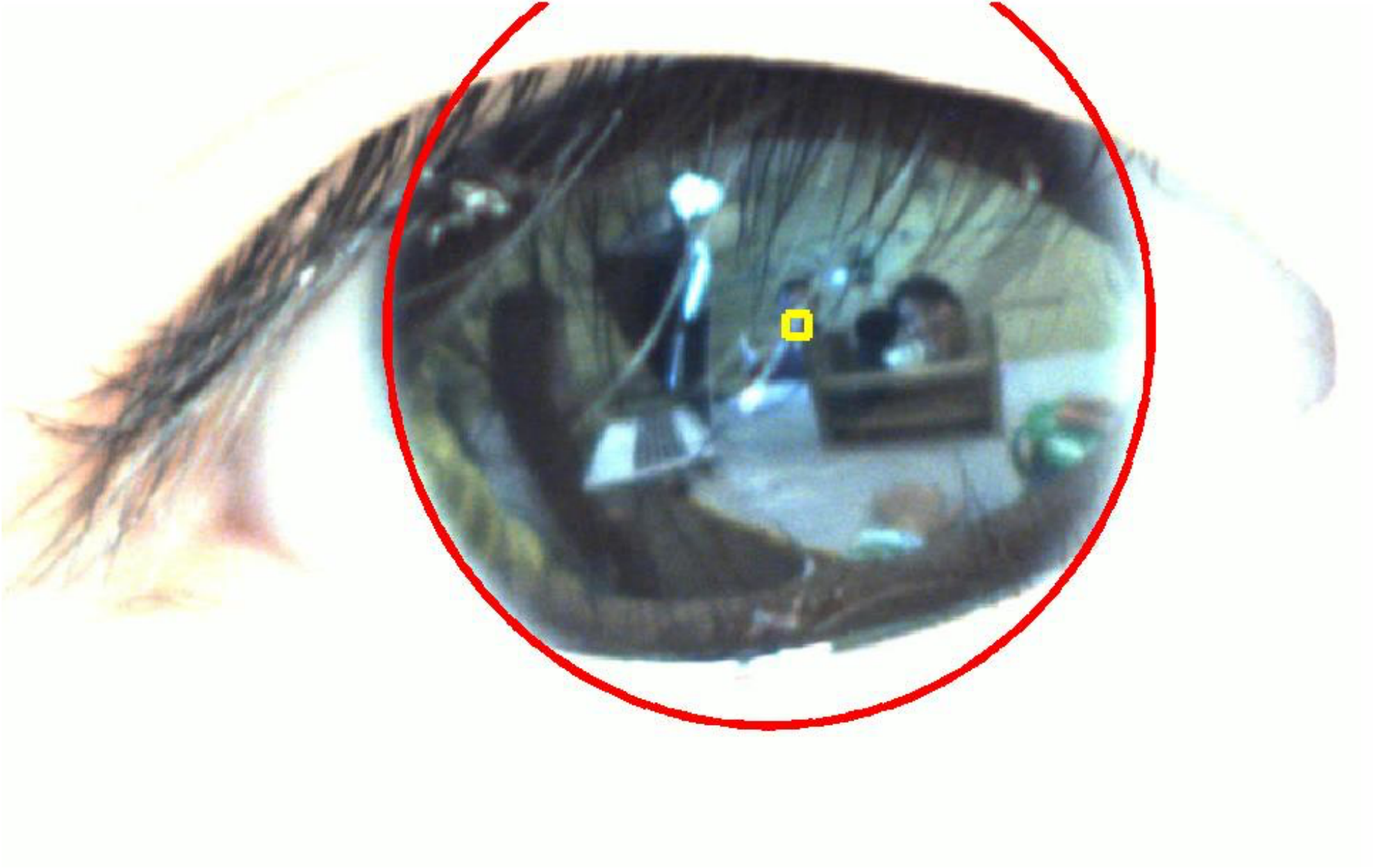}
\includegraphics[width=0.2\linewidth]{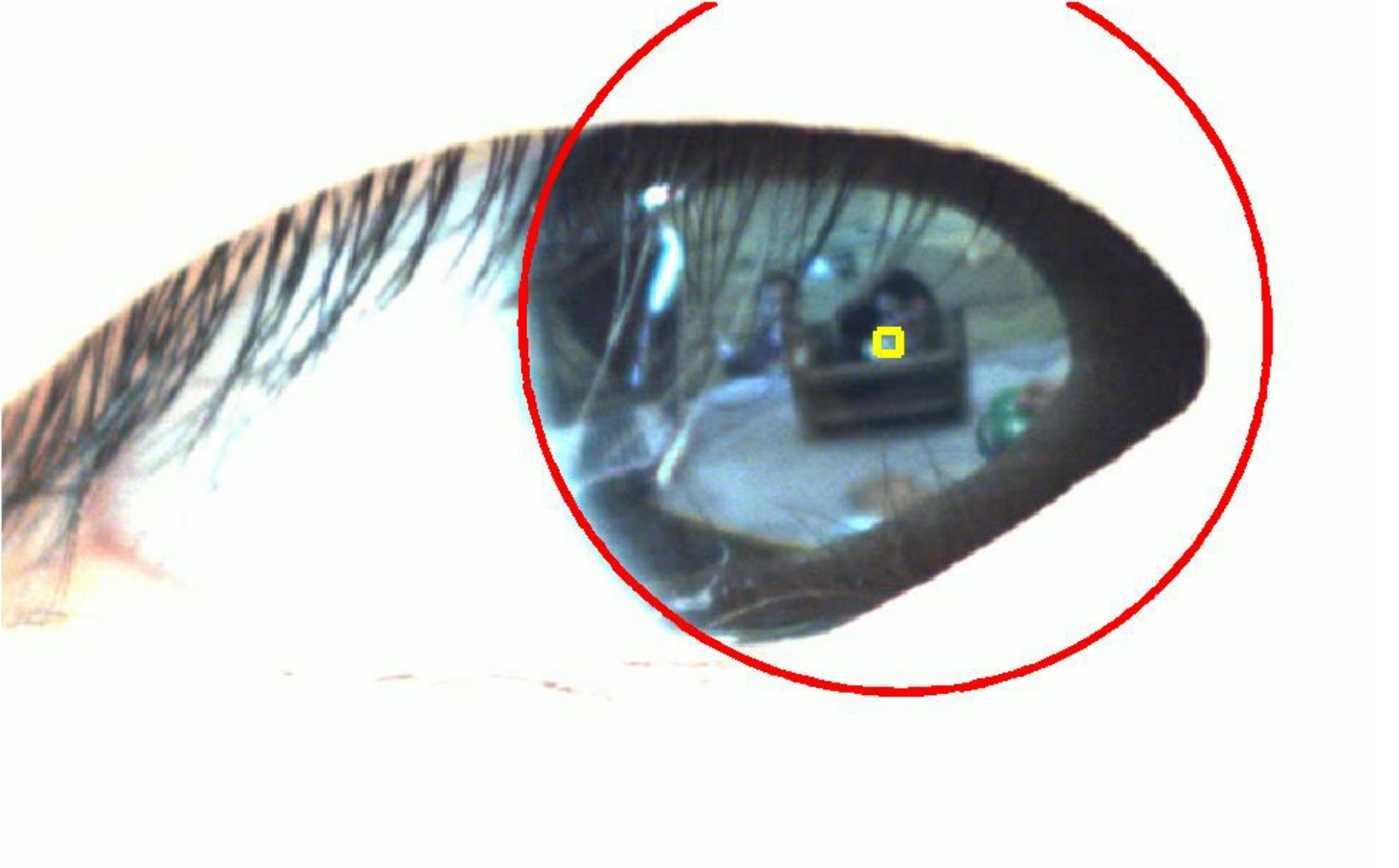}
\includegraphics[width=0.2\linewidth]{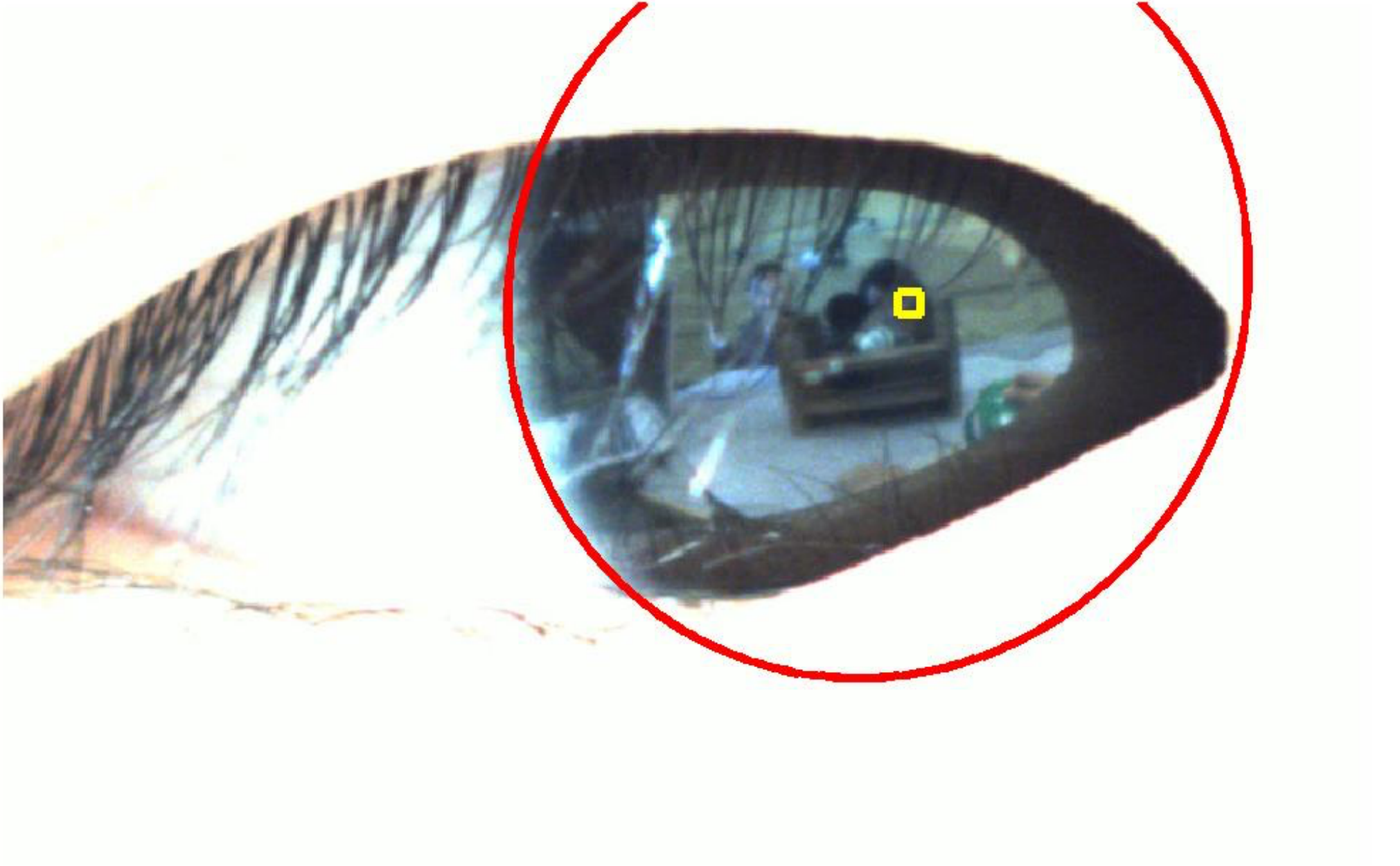}
\includegraphics[width=0.2\linewidth]{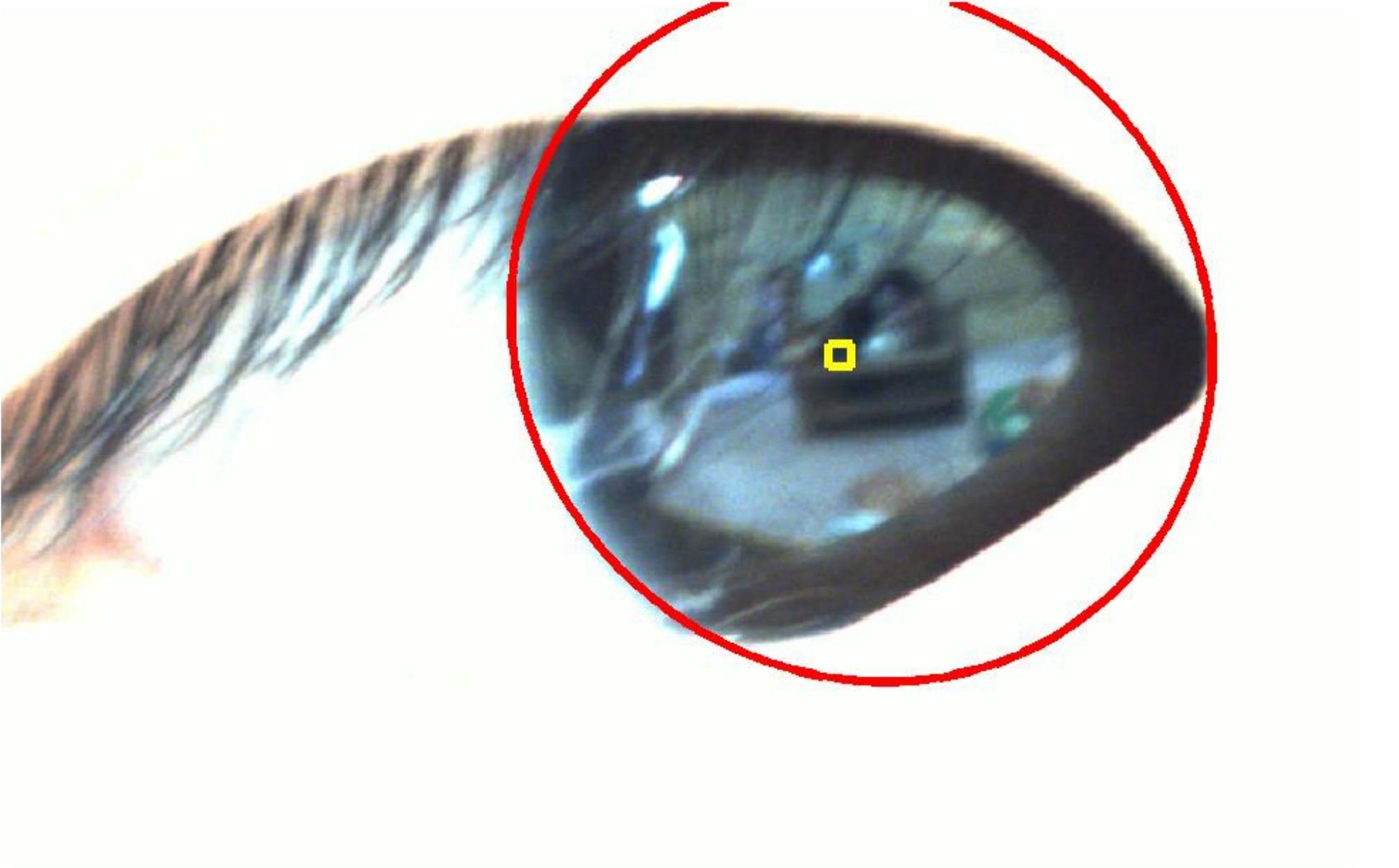}
\includegraphics[width=0.2\linewidth]{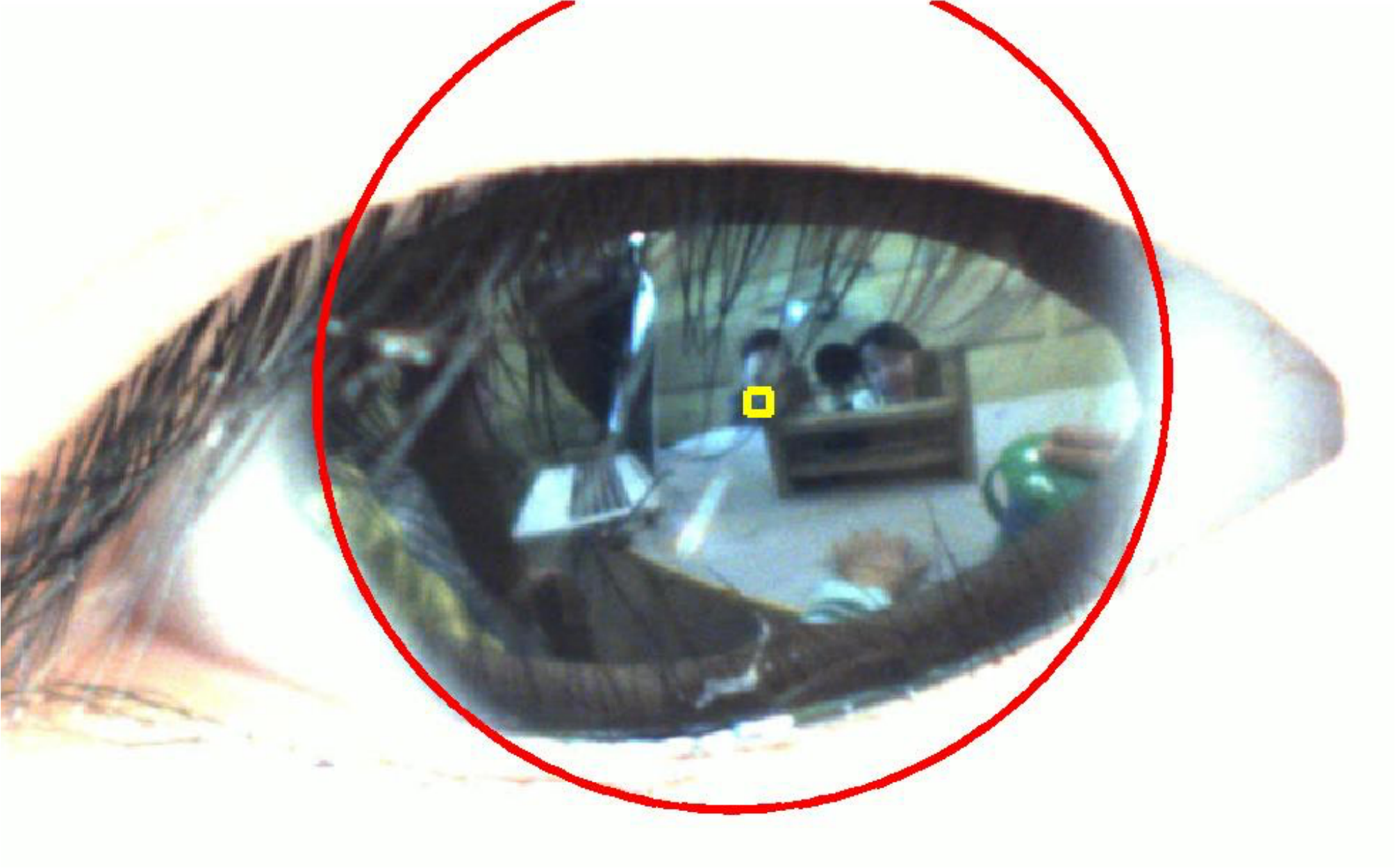}}\\
\subfloat{
\includegraphics[width=0.19\linewidth]{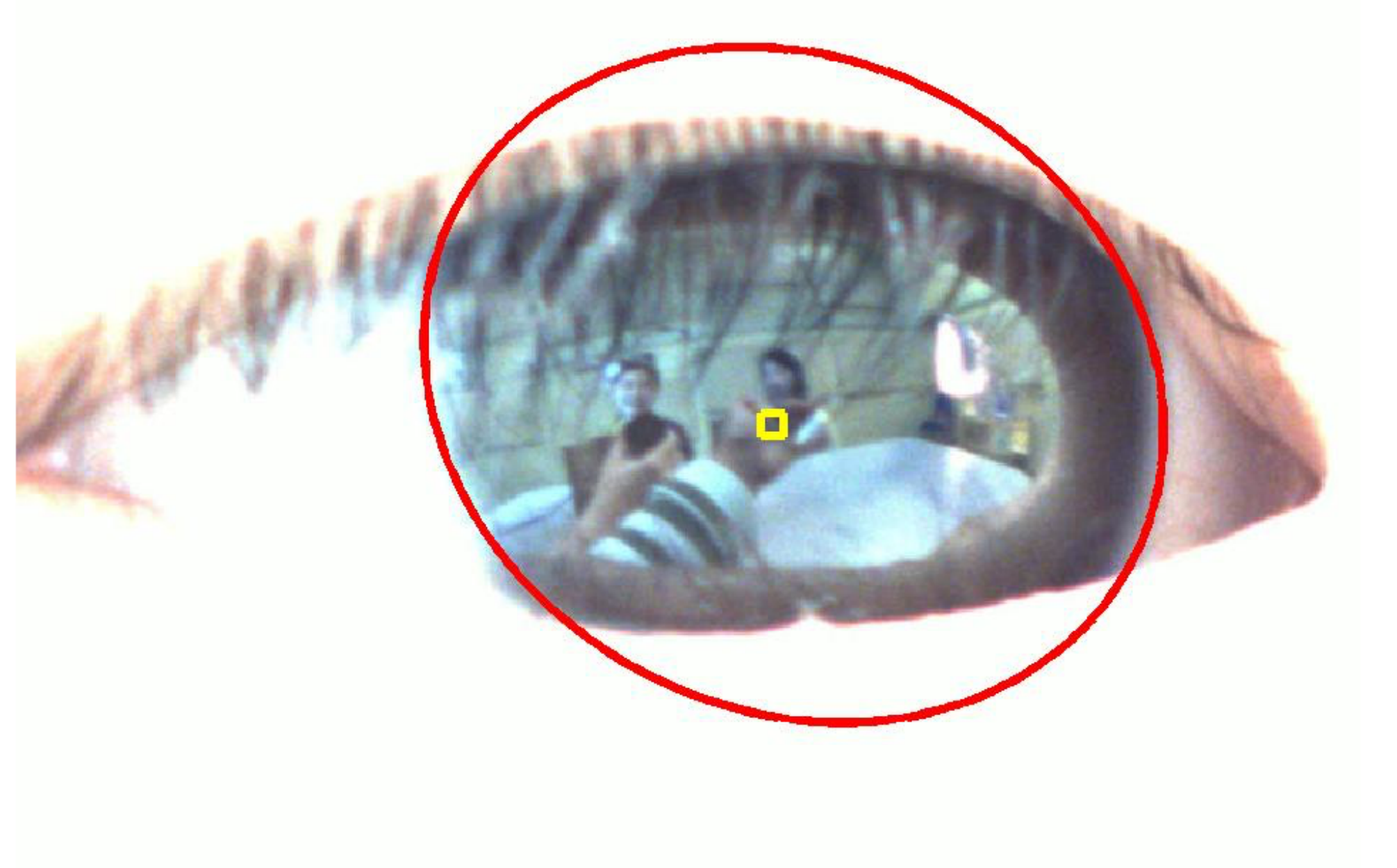}
\includegraphics[width=0.19\linewidth]{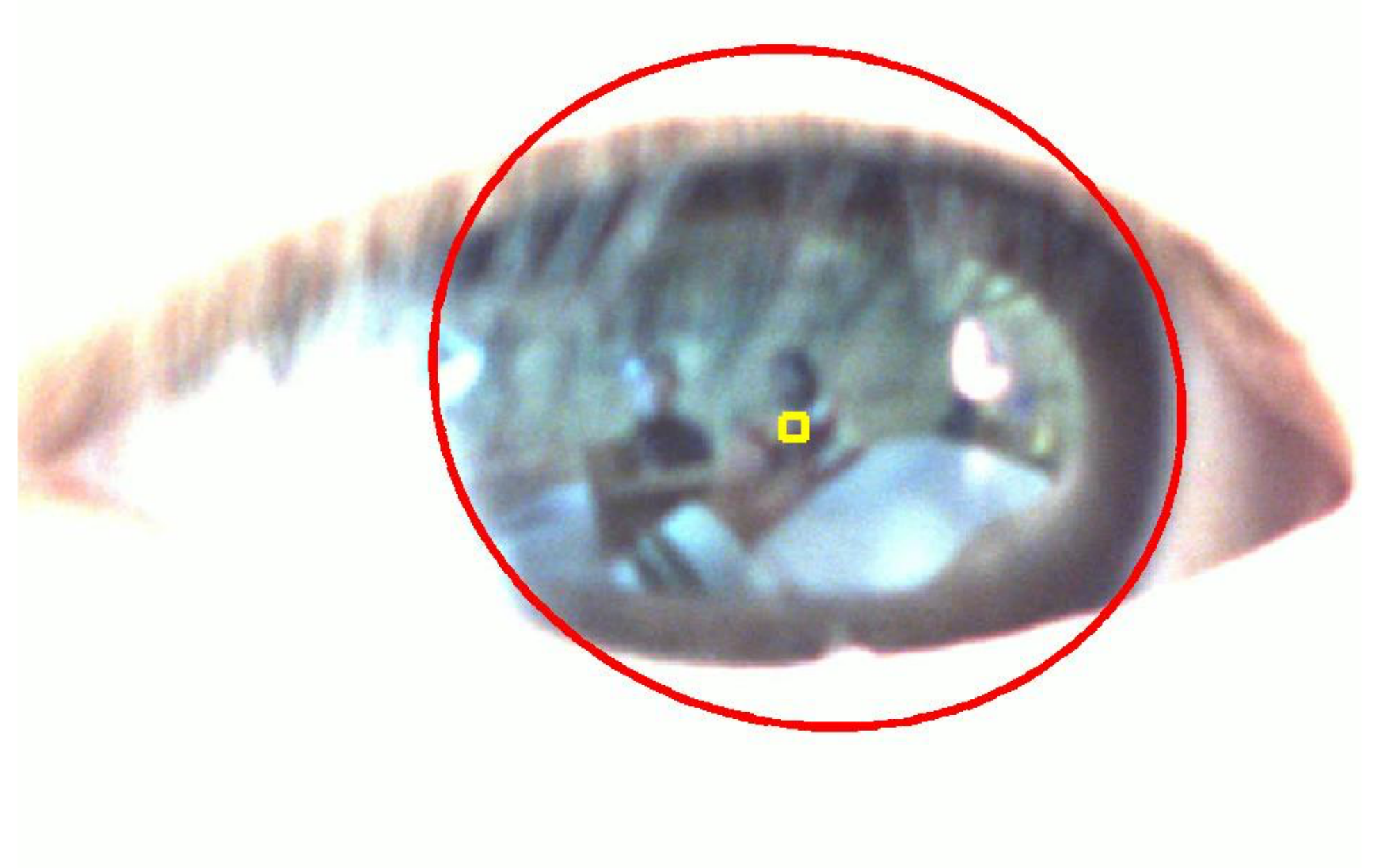}
\includegraphics[width=0.19\linewidth]{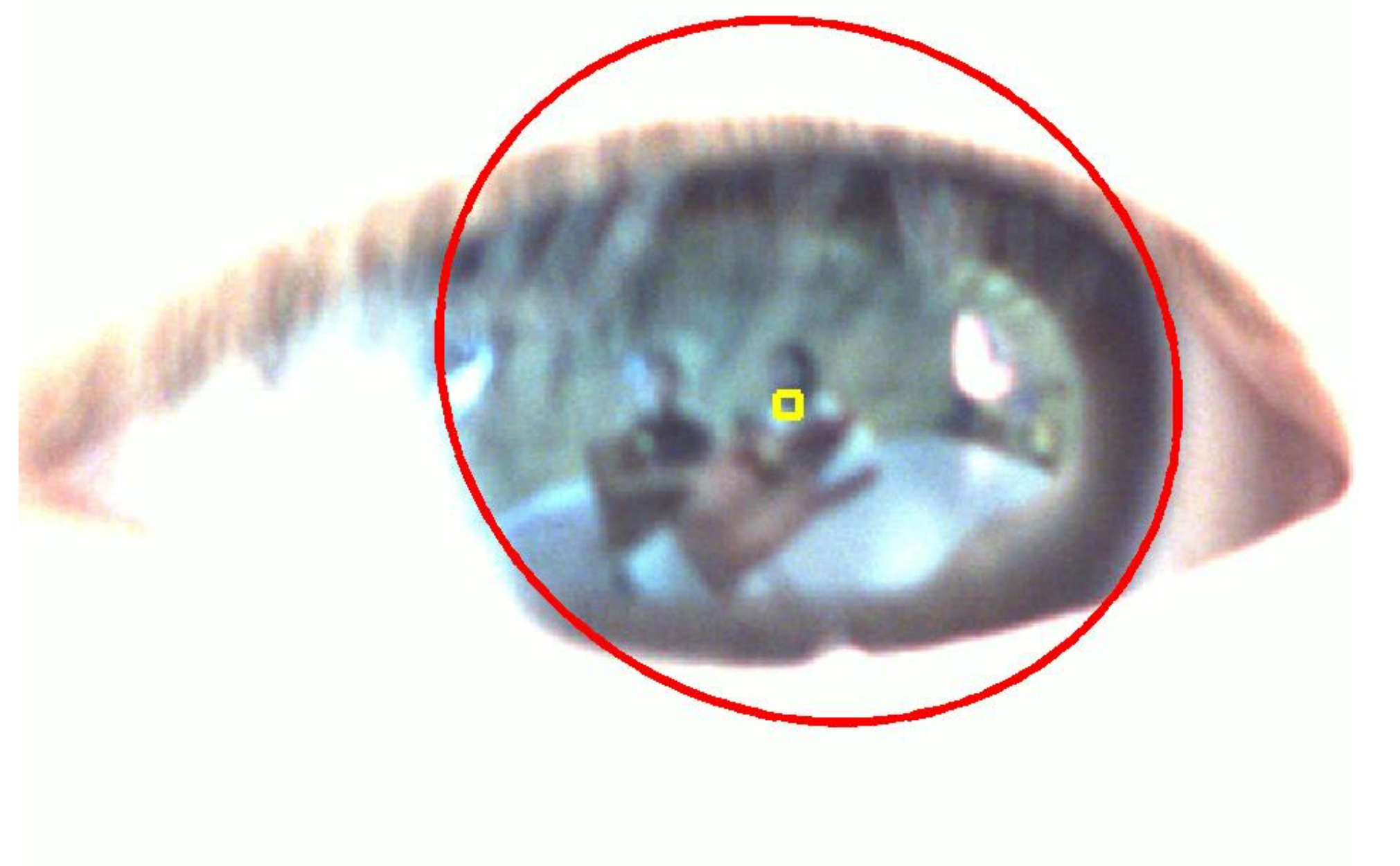}
\includegraphics[width=0.19\linewidth]{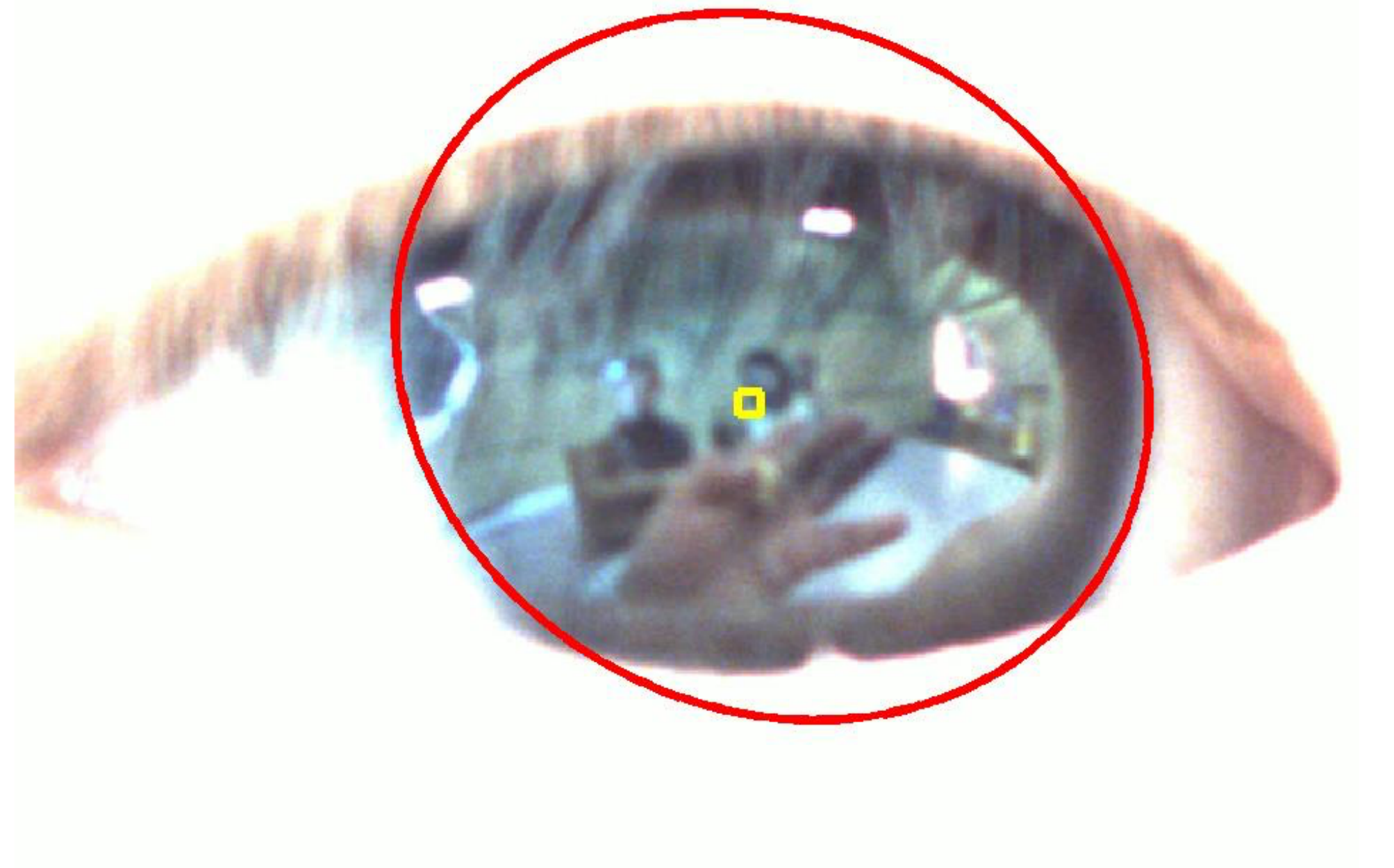}
\includegraphics[width=0.19\linewidth]{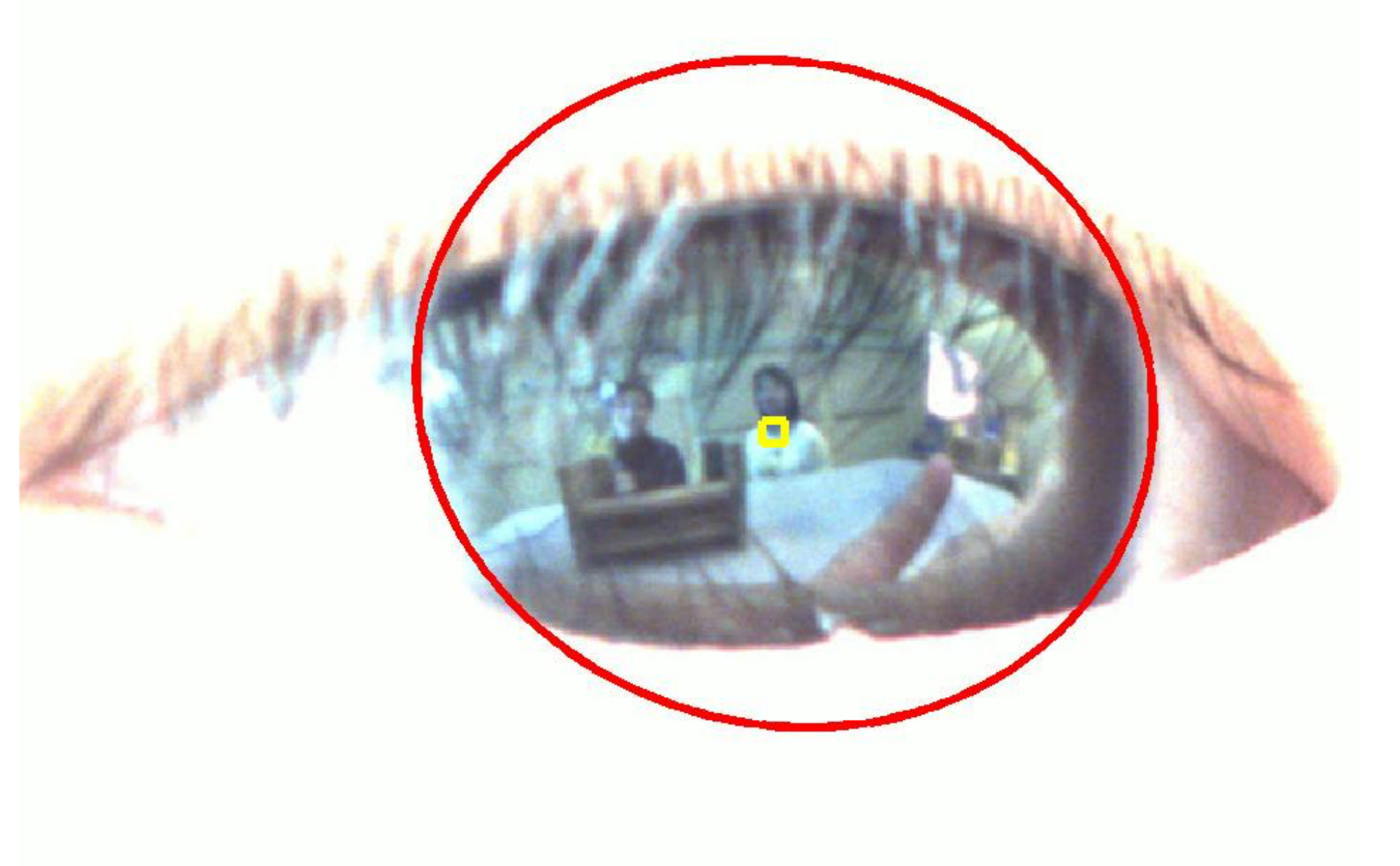}}\\
\subfloat{
\includegraphics[width=0.19\linewidth]{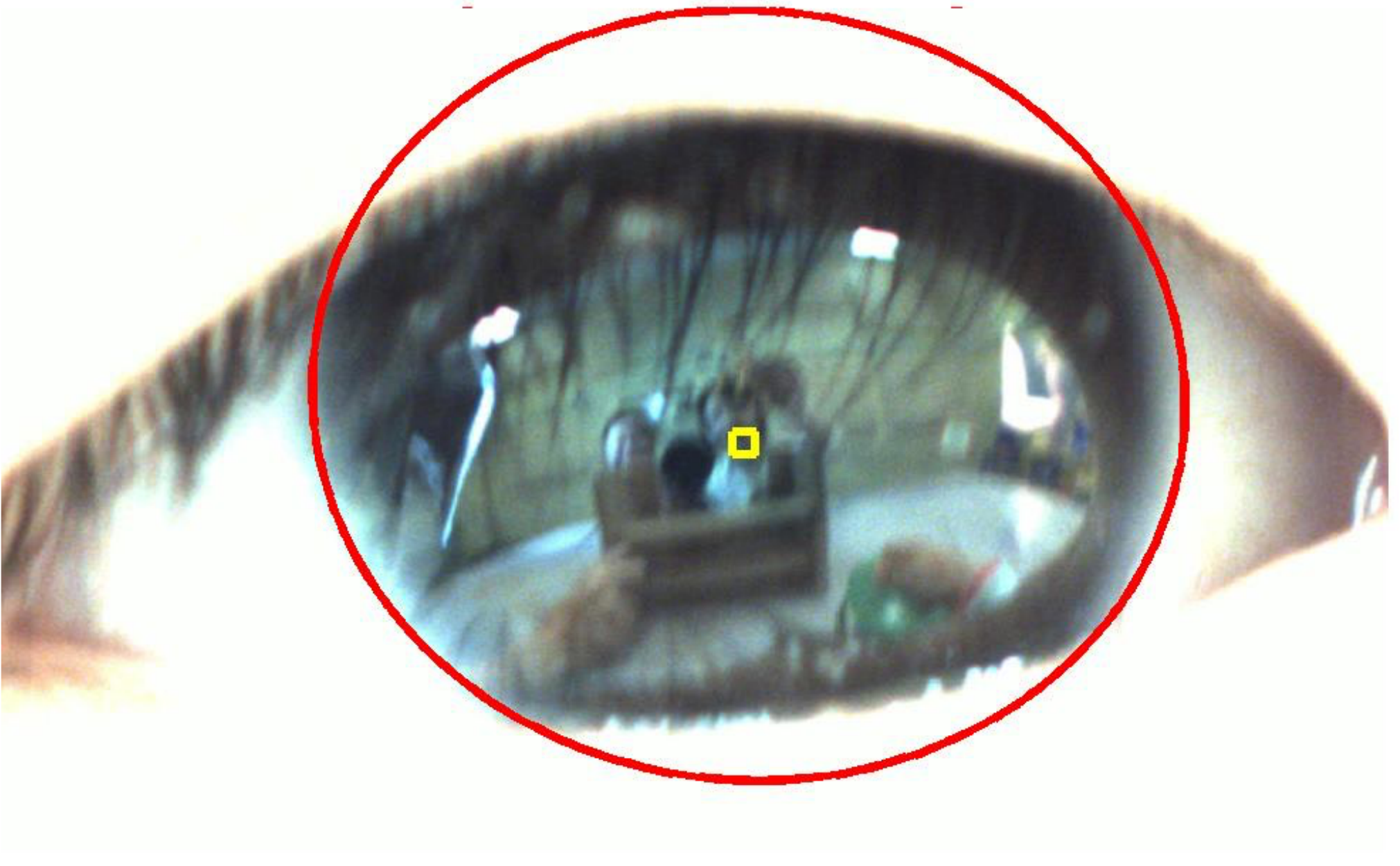}
\includegraphics[width=0.19\linewidth]{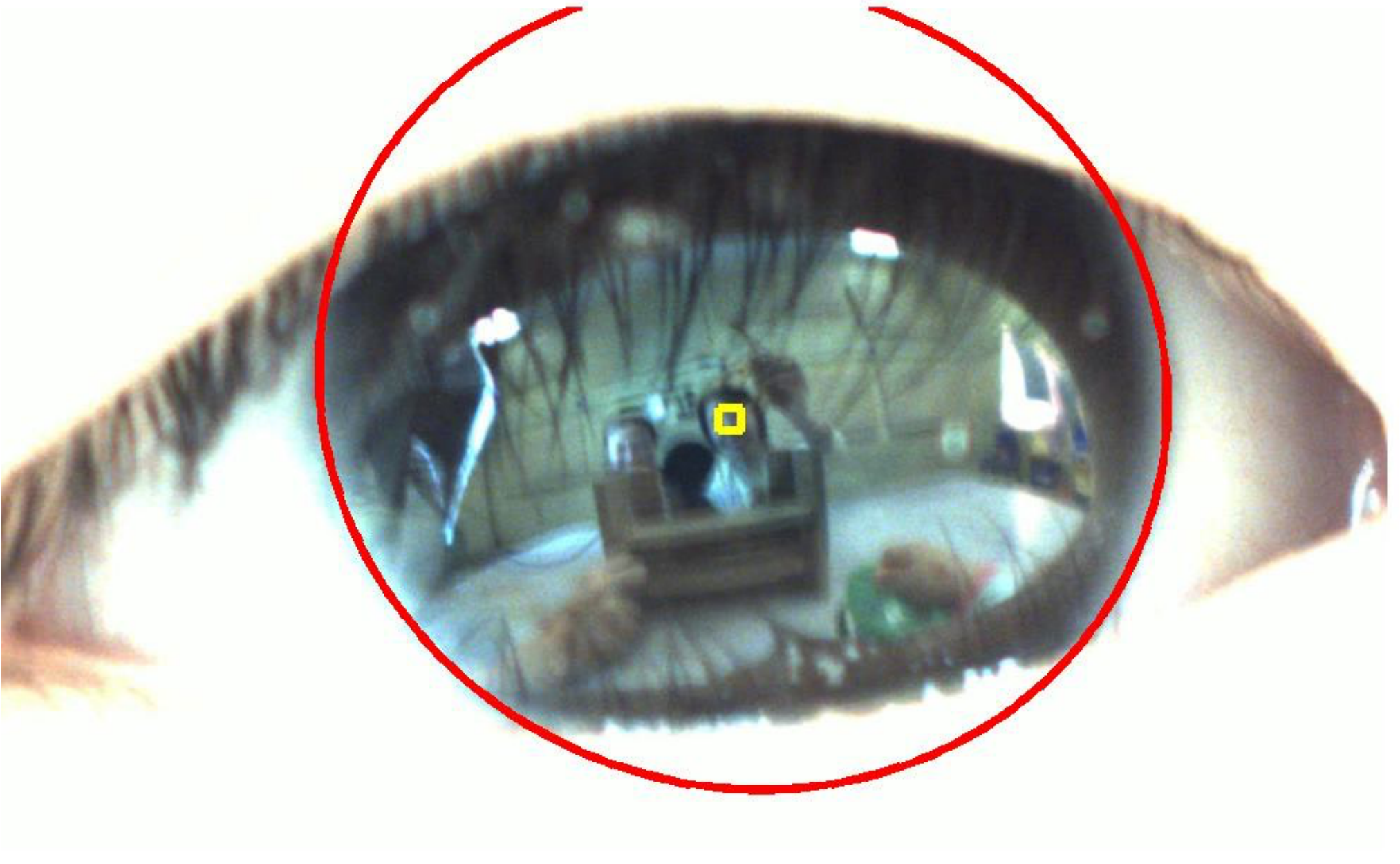}
\includegraphics[width=0.19\linewidth]{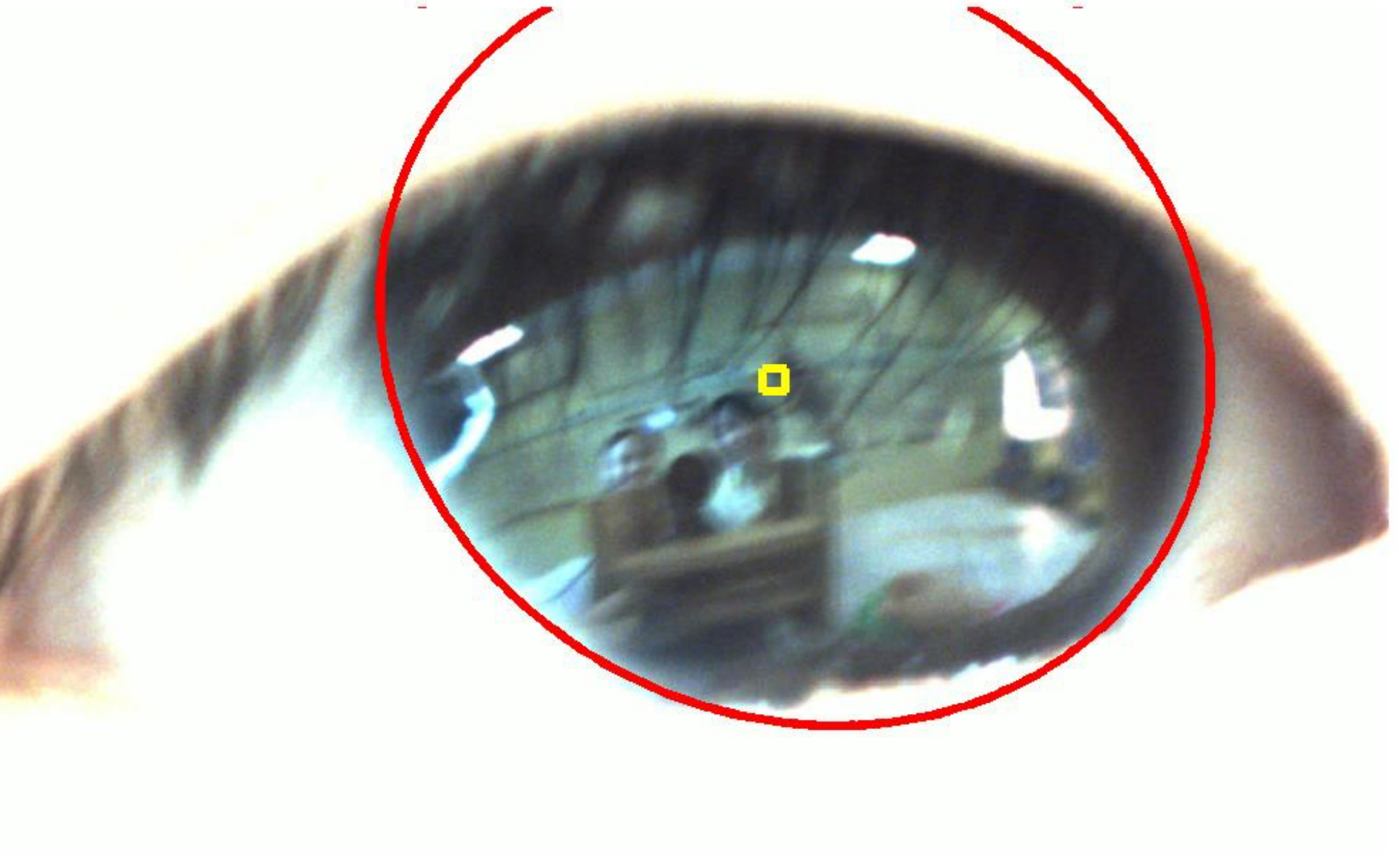}
\includegraphics[width=0.19\linewidth]{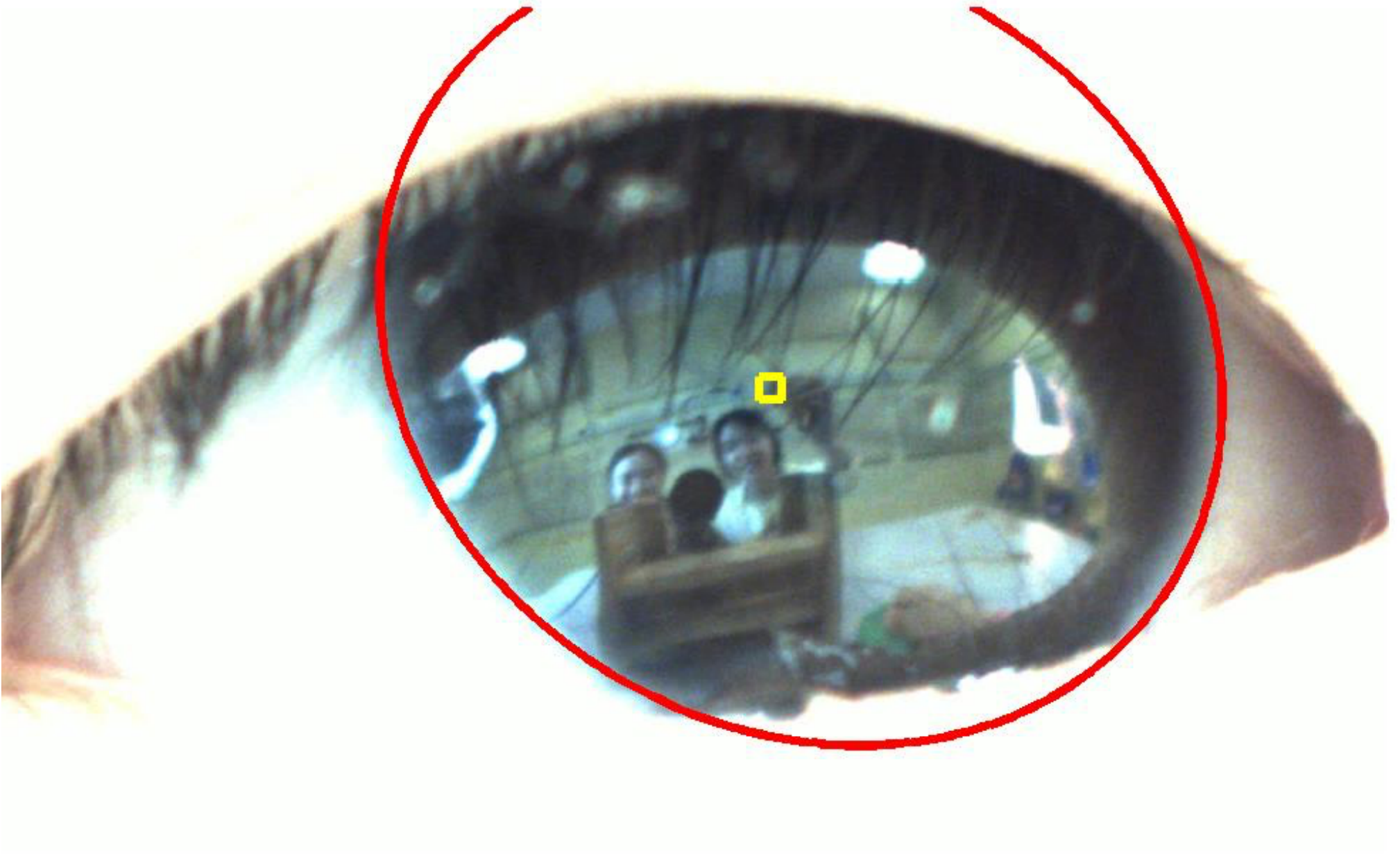}
\includegraphics[width=0.19\linewidth]{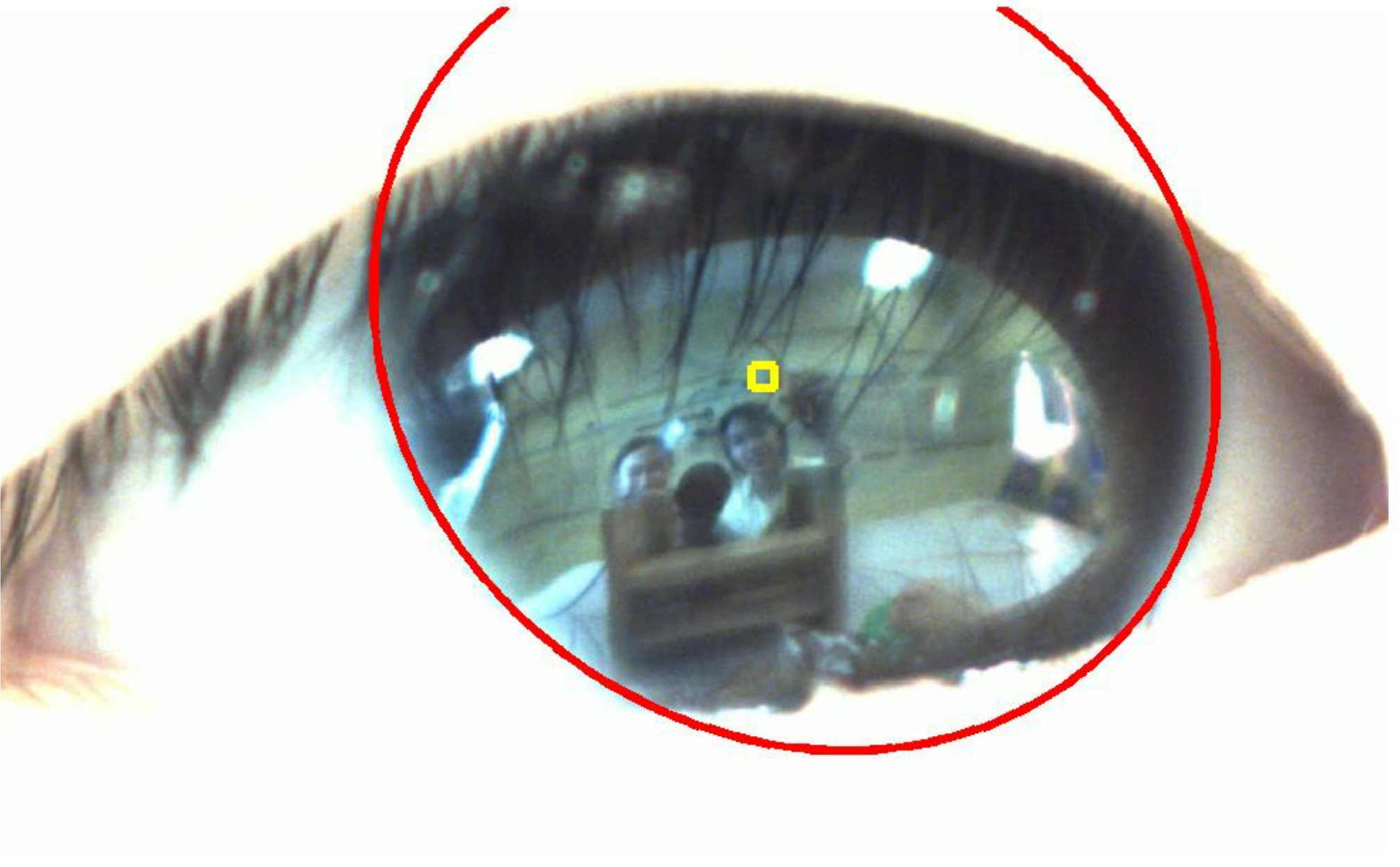}}
\caption{Social attention of young children that was captured by our system. 1st row: The child is shifting gaze between the two people. 2nd row: The child is bringing a snack to the mouth while fixating on a person. 3rd row: The child is attending to a toy that was presented in front of the child.}
\label{fig:childmore}
\end{figure*}

Our experiments with children were performed using the protocol depicted in Section~\ref{sec:targetSetup}; two adults (green box in Figure~\ref{fig:giraffe}(a)) played peekaboo, ball tossing, and dolls with one another in front of a child (yellow box in Figure~\ref{fig:giraffe}(a)) who was either sitting in a chair or in the caregiver (red box in Figure~\ref{fig:giraffe}(a)) 's lap.  See Figure~\ref{fig:giraffe} to get a sense of the layout. Figure~\ref{fig:child_ball} and Figure~\ref{fig:childmore} show a few gaze patterns of children that were captured by our system. The yellow square inside ellipse shows the estimated child's point of gaze. 

We administered the same play protocol with three different young children (age between 7 month and 3 year). We did not restrict the child from moving except that his/her mother kept the baby on her lap to prevent falling down, and since our current prototype had somewhat narrow field of view, the camera captured child's corneal reflection approximately $40\%$ of the time. When the corneal image was acquired, the calculation of PoG was successful $90\%$ of time.

\subsection{Reproducing a Standard Psychological Finding}
\label{sec:app_blind}

\begin{figure*}[h!]
\centering
\subfloat{
\includegraphics[width=0.19\linewidth]{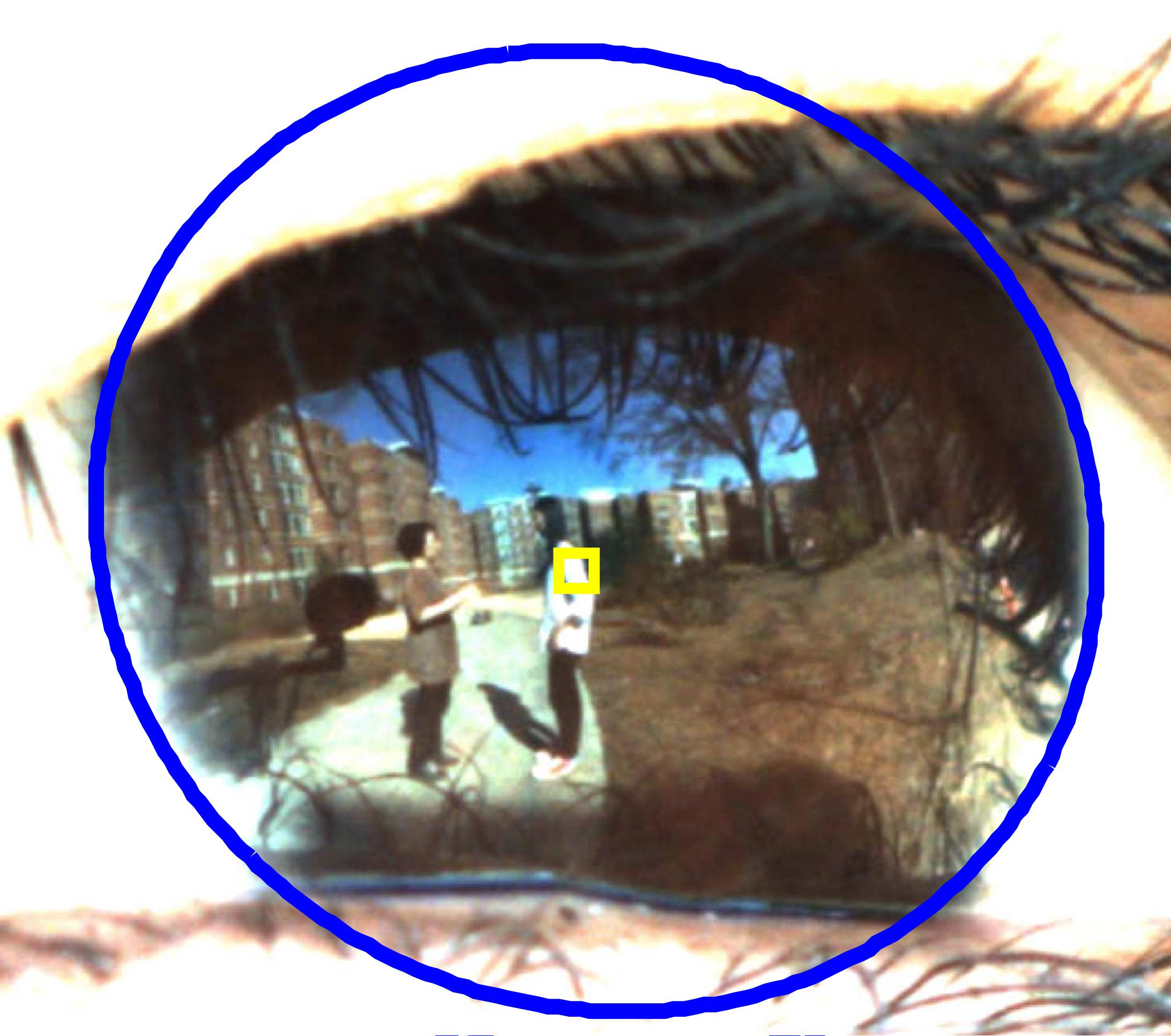}
\includegraphics[width=0.19\linewidth]{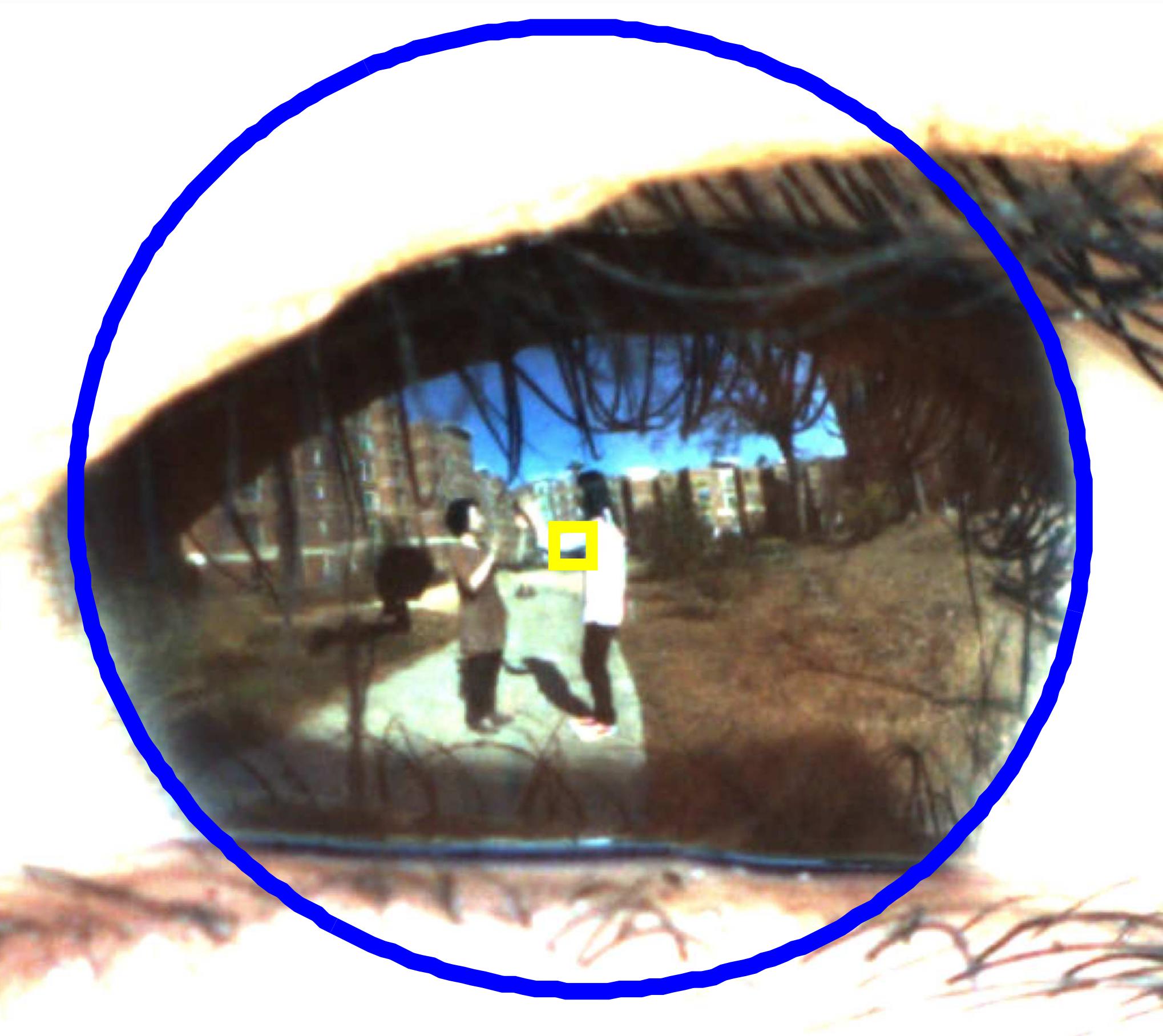}
\includegraphics[width=0.19\linewidth]{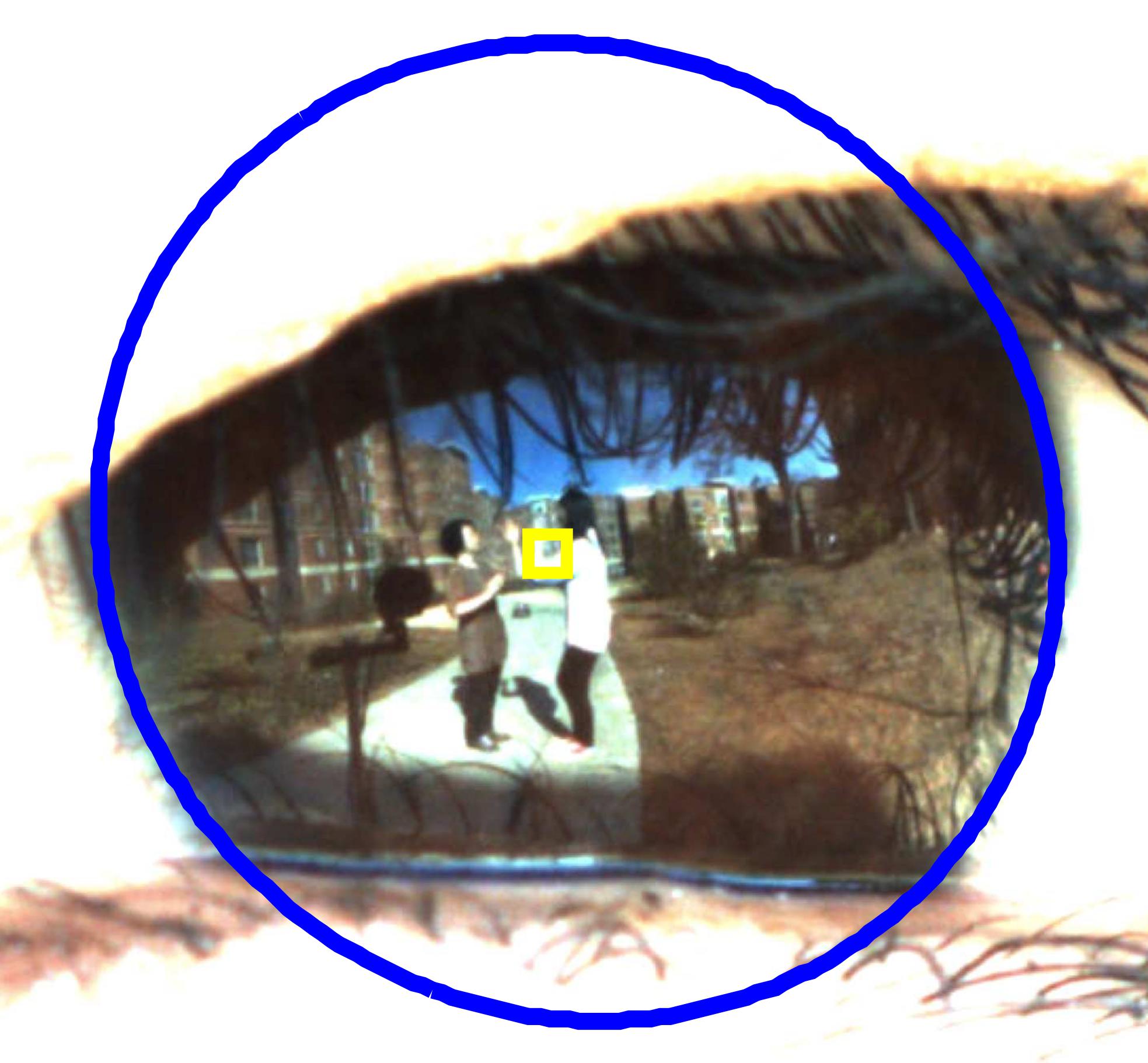}
\includegraphics[width=0.19\linewidth]{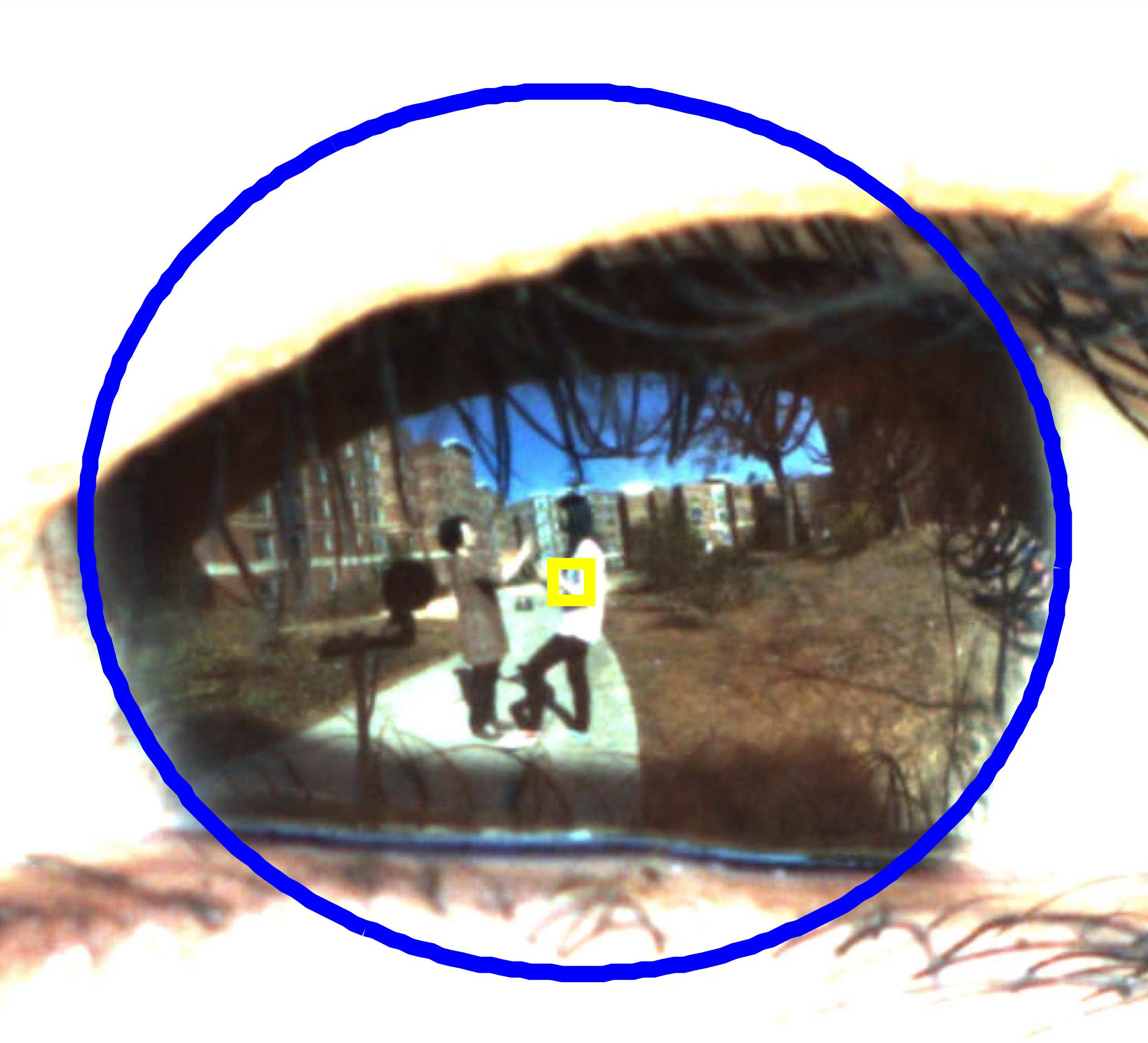}
\includegraphics[width=0.19\linewidth]{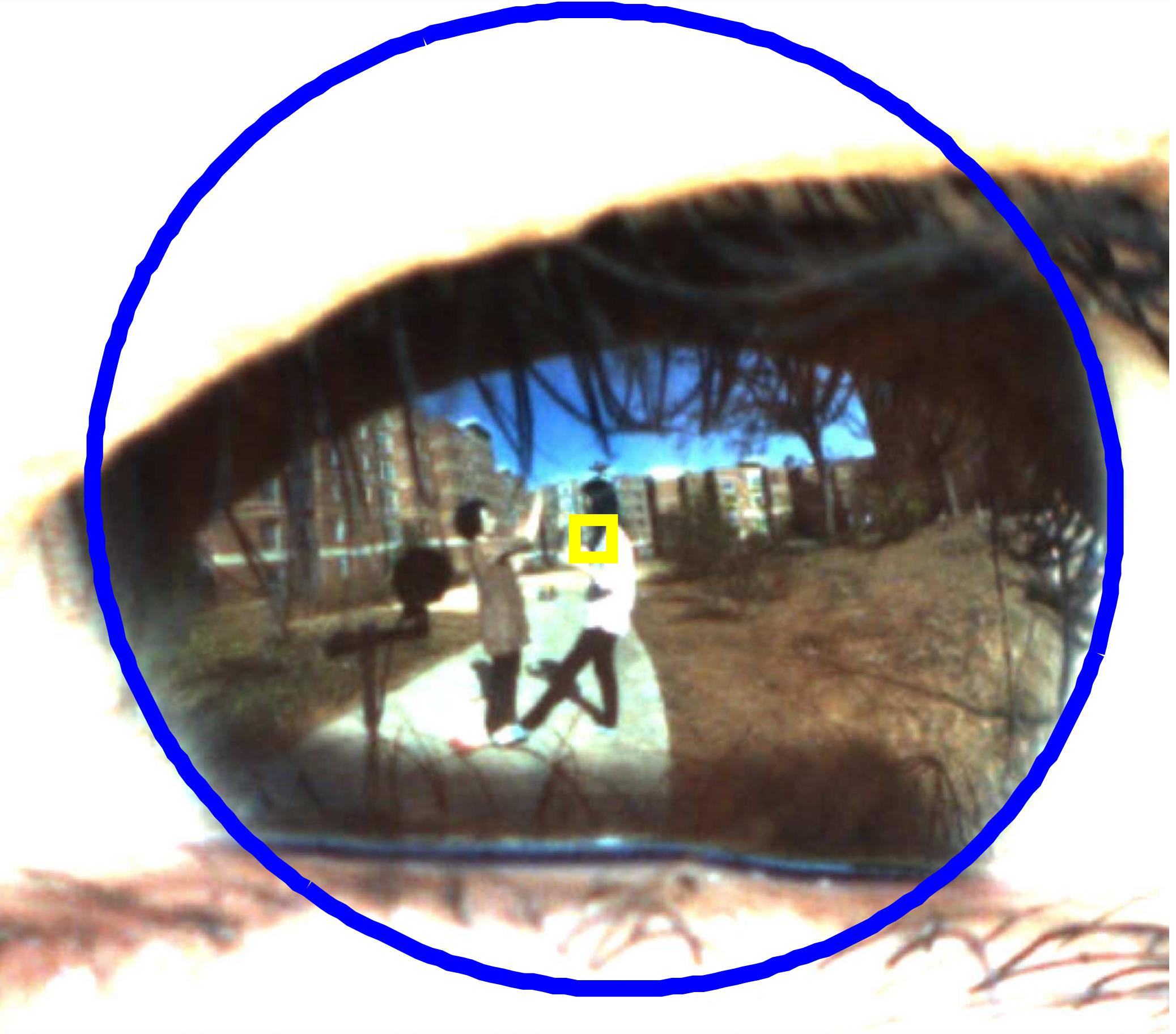}}\\
\subfloat{
\includegraphics[width=0.19\linewidth]{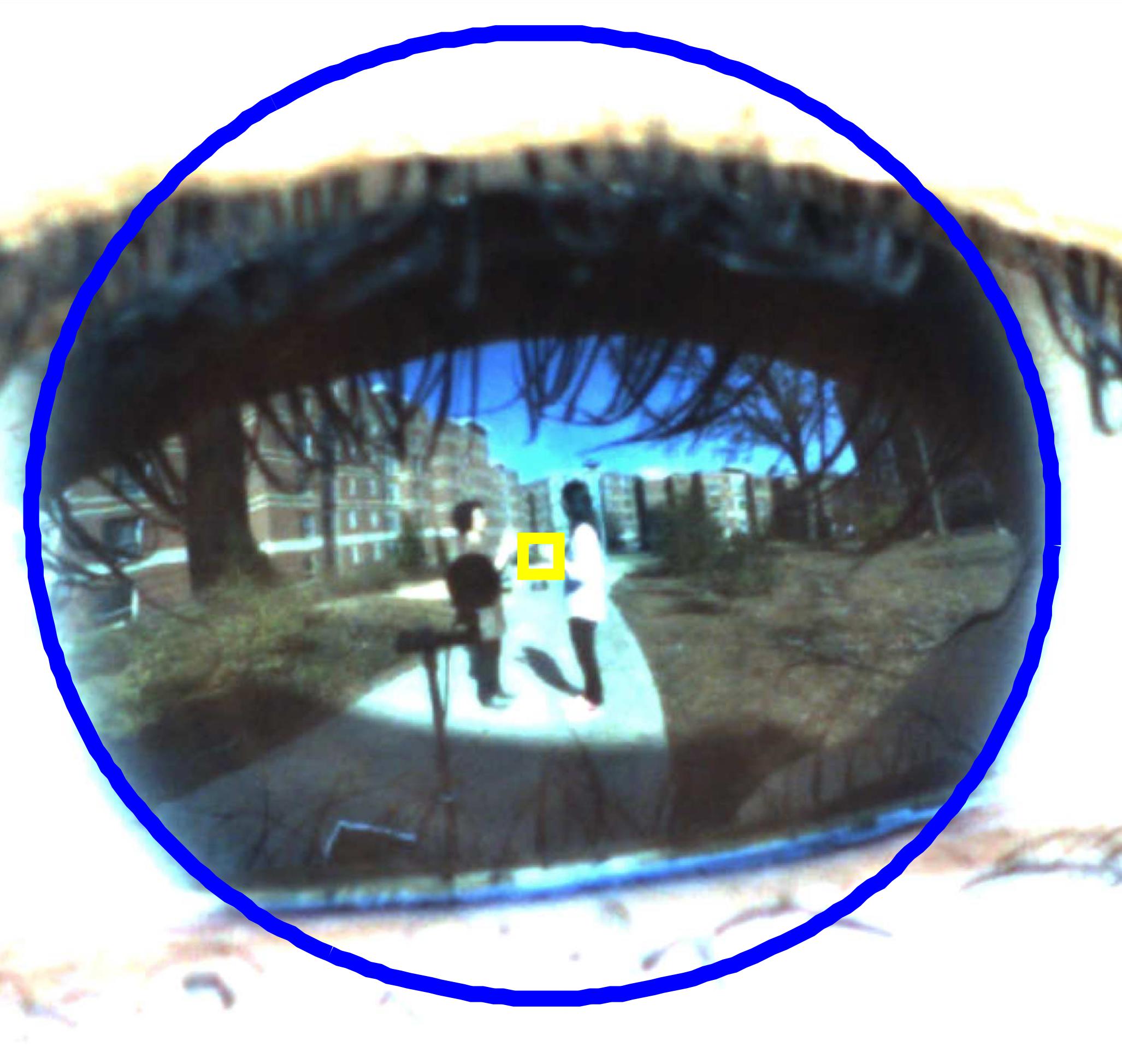}
\includegraphics[width=0.19\linewidth]{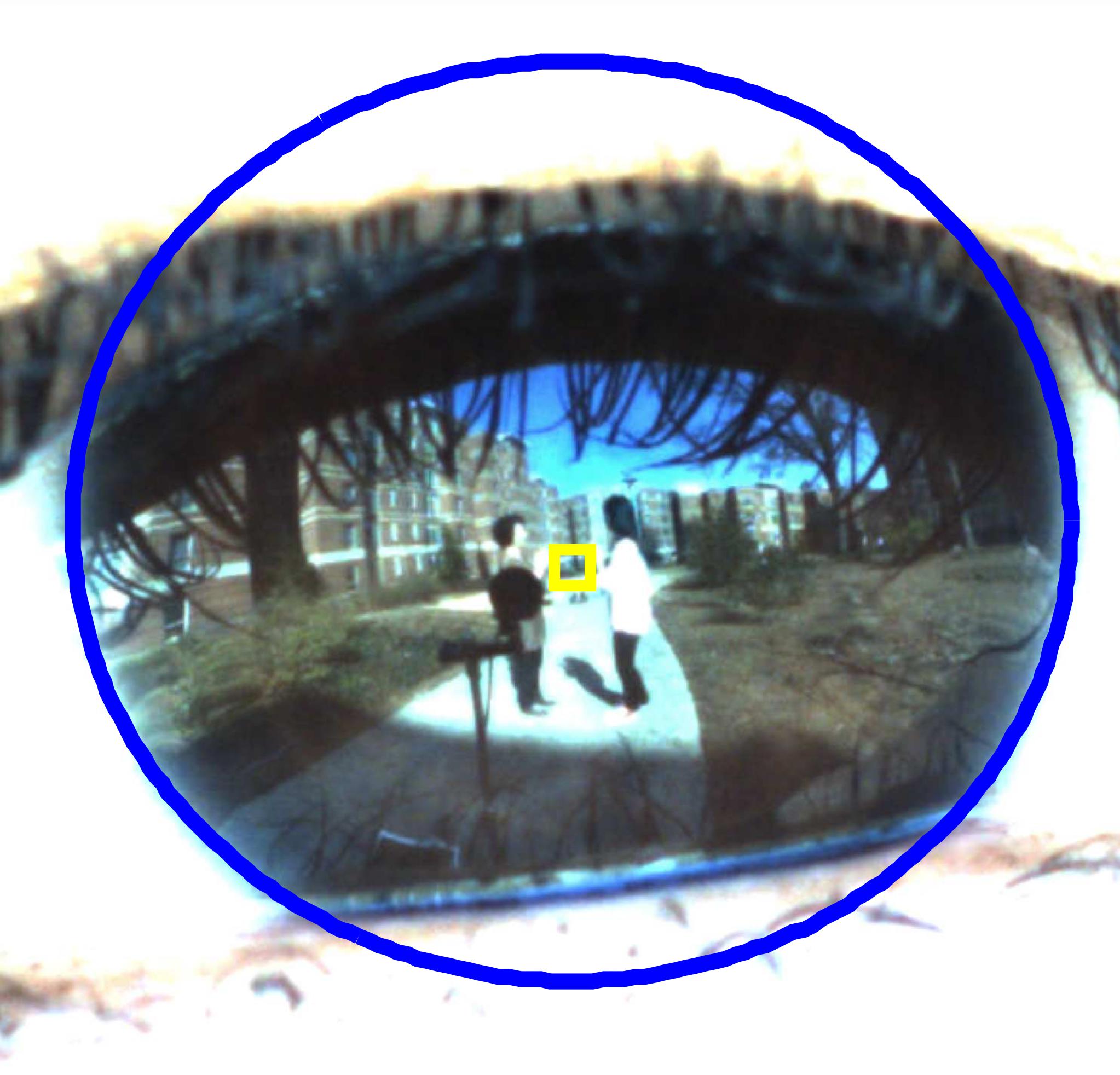}
\includegraphics[width=0.19\linewidth]{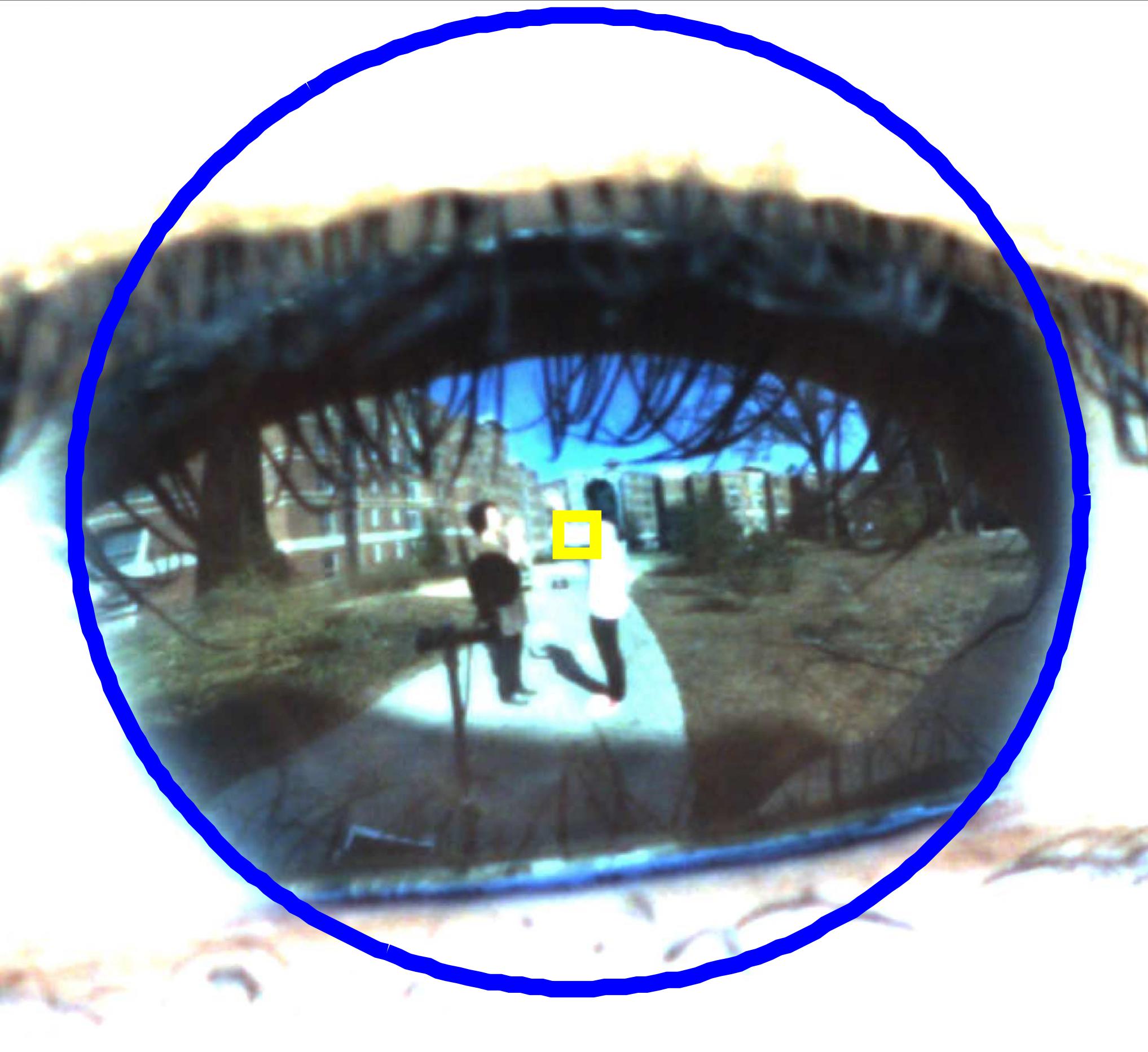}
\includegraphics[width=0.19\linewidth]{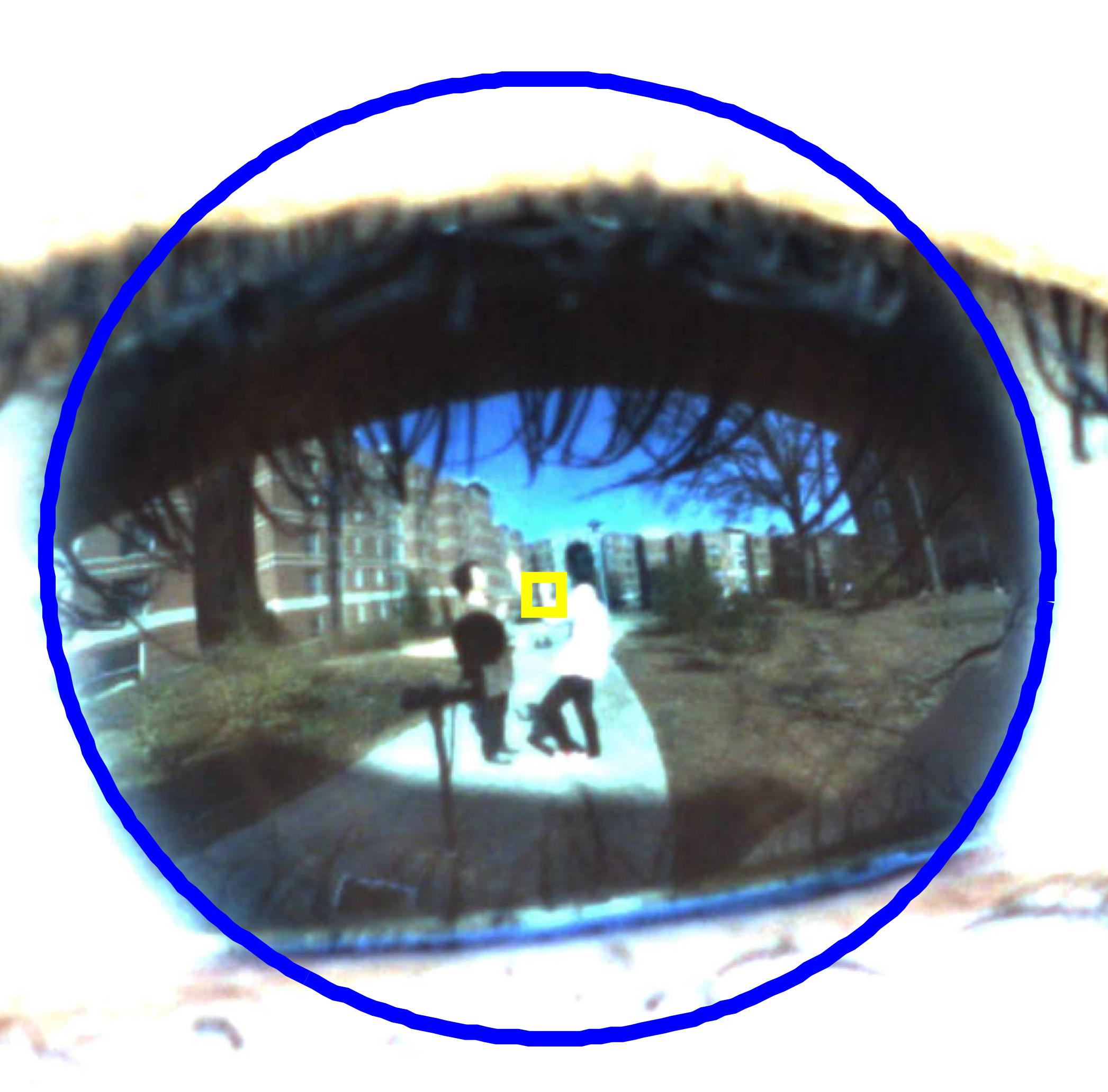}
\includegraphics[width=0.19\linewidth]{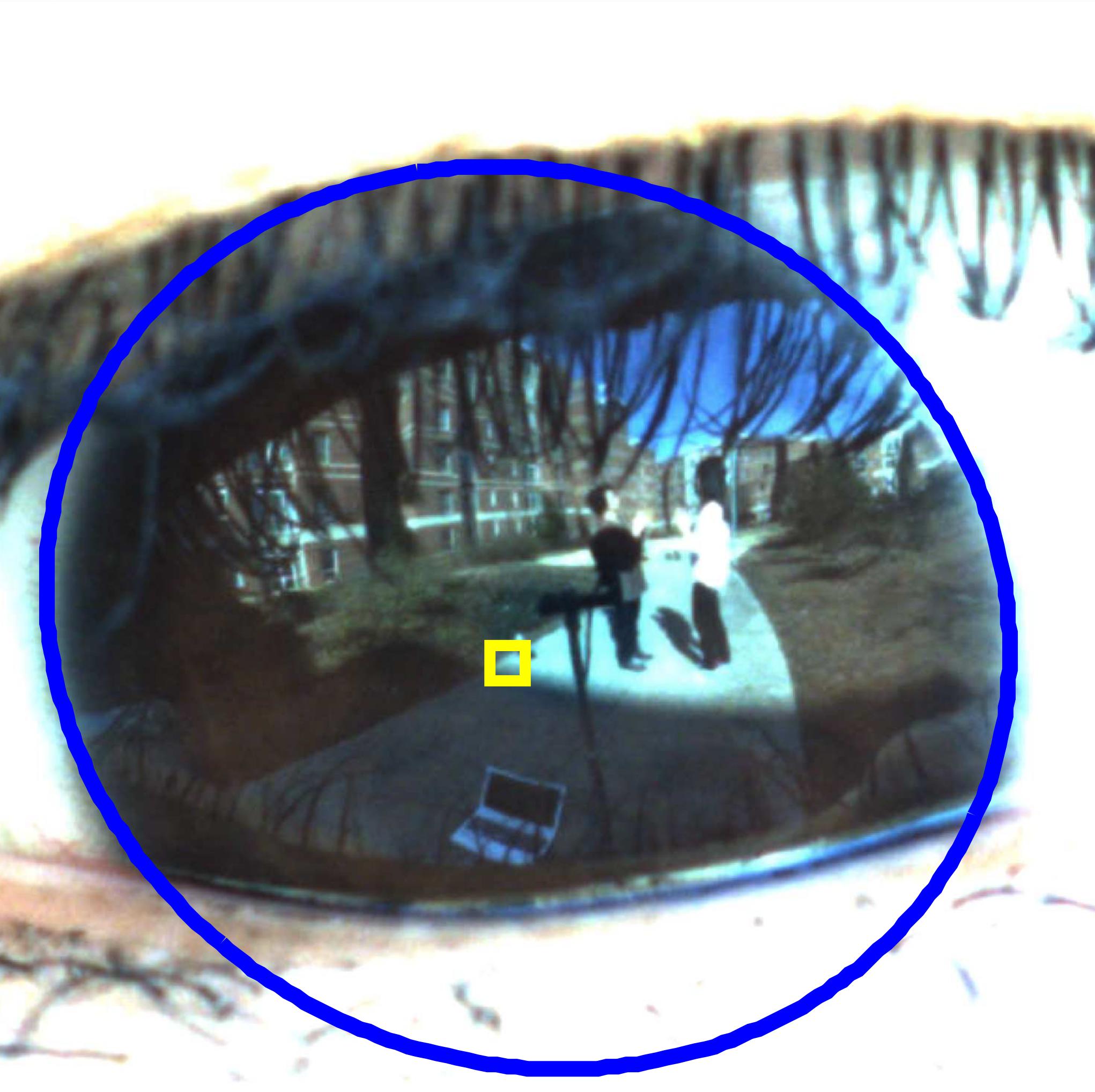}}
\caption{Selective attention test under two conditions; A subject watches two actors doing patty-cake, where an actor surreptitiously removes his or her shoe in midst of experiment. One group was asked to count the number of patty-cakes while another group observed freely. The first row shows one gaze pattern from the first group that did not notice the shoe was taken off. The second row shows gaze pattern from the other group that noticed the shoe removal.}
\label{fig:selective}
\end{figure*}

Gaze tracking is widely utilized in the field of psychology, especially in studies of human mental processes. With that in mind, we devised a simple psychological experiment to demonstrate the potential of our system by replicating an experiment known as the selective attention test.~\cite{Simons:1999aa}. Selective attention is a cognitive process in which a person selectively focuses on a few things and ignores other information when there is too much information to process. Our version of the experiment  was designed in the following way: Subjects observed two actors playing patty-cake. One group of subjects was instructed to count the number of patty-cakes while the second group was instructed to observe freely. While the actors played patty-cake, one actor removed his shoe by kicking it away. Afterward, the subjects were asked whether they noticed anything unusual. In the first group,  3 out of 5 subjects did not notice the shoe removal, and we confirmed that their gaze was near actors' hands the whole time. In contrast, 4 out of 5 subjects in the second group  noticed the removal, and we found that their gaze shifted toward the actors' feet during the shoe removal event. Figure~\ref{fig:selective} show one subject who did not notice it (1st row) and another subject who did not notice it (2nd row).

\subsection{Additional Experiments}
\label{sec:app_more}
\begin{figure*}[h!]
\def\arraystretch{0.185}
\centering
\begin{tabular}{>{\raggedright\arraybackslash}m{.25in} *3{>{\centering\arraybackslash}m{2in}} @{}m{0pt}@{}}
(A) & \includegraphics[width=\linewidth]{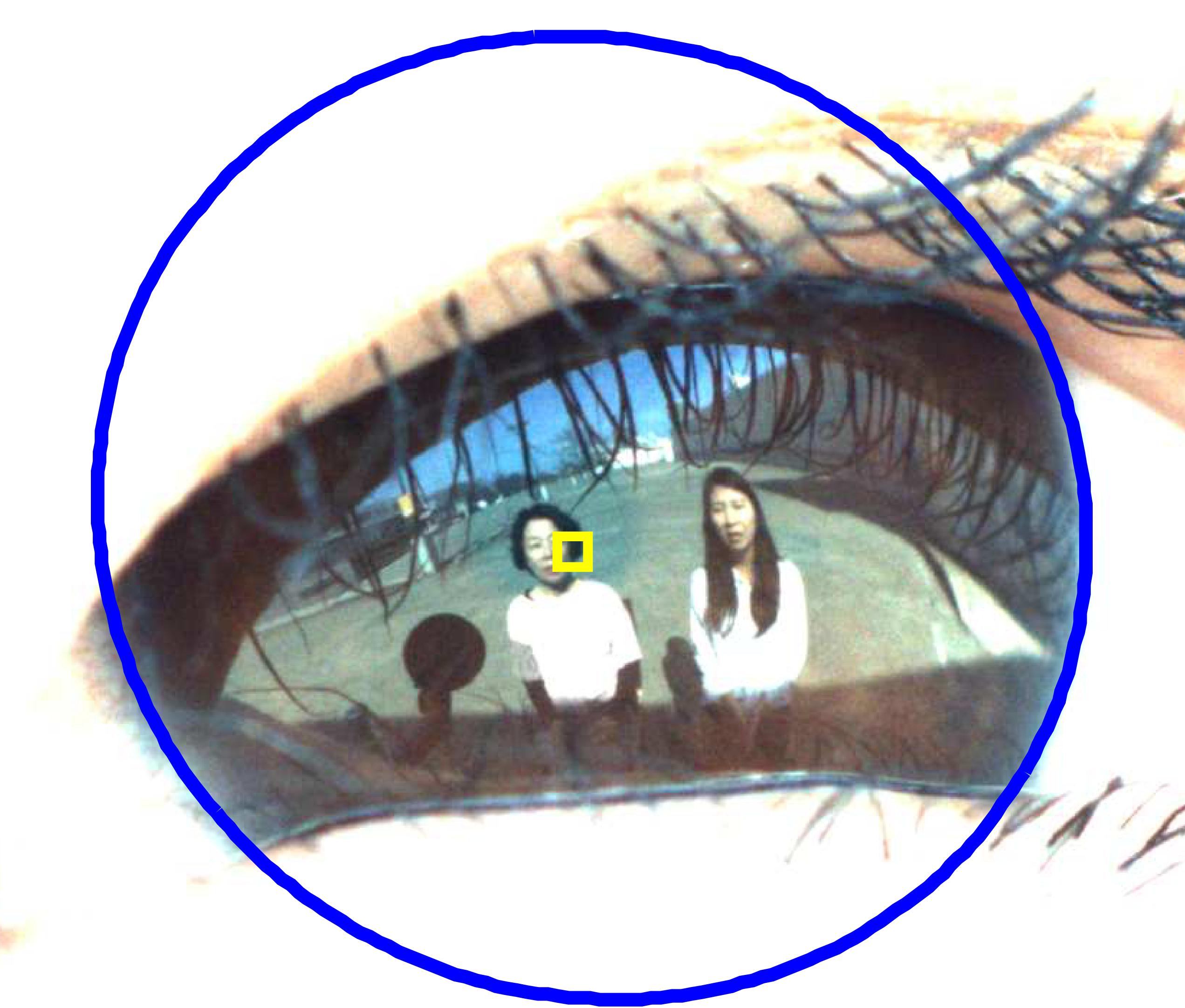} & \includegraphics[width=\linewidth]{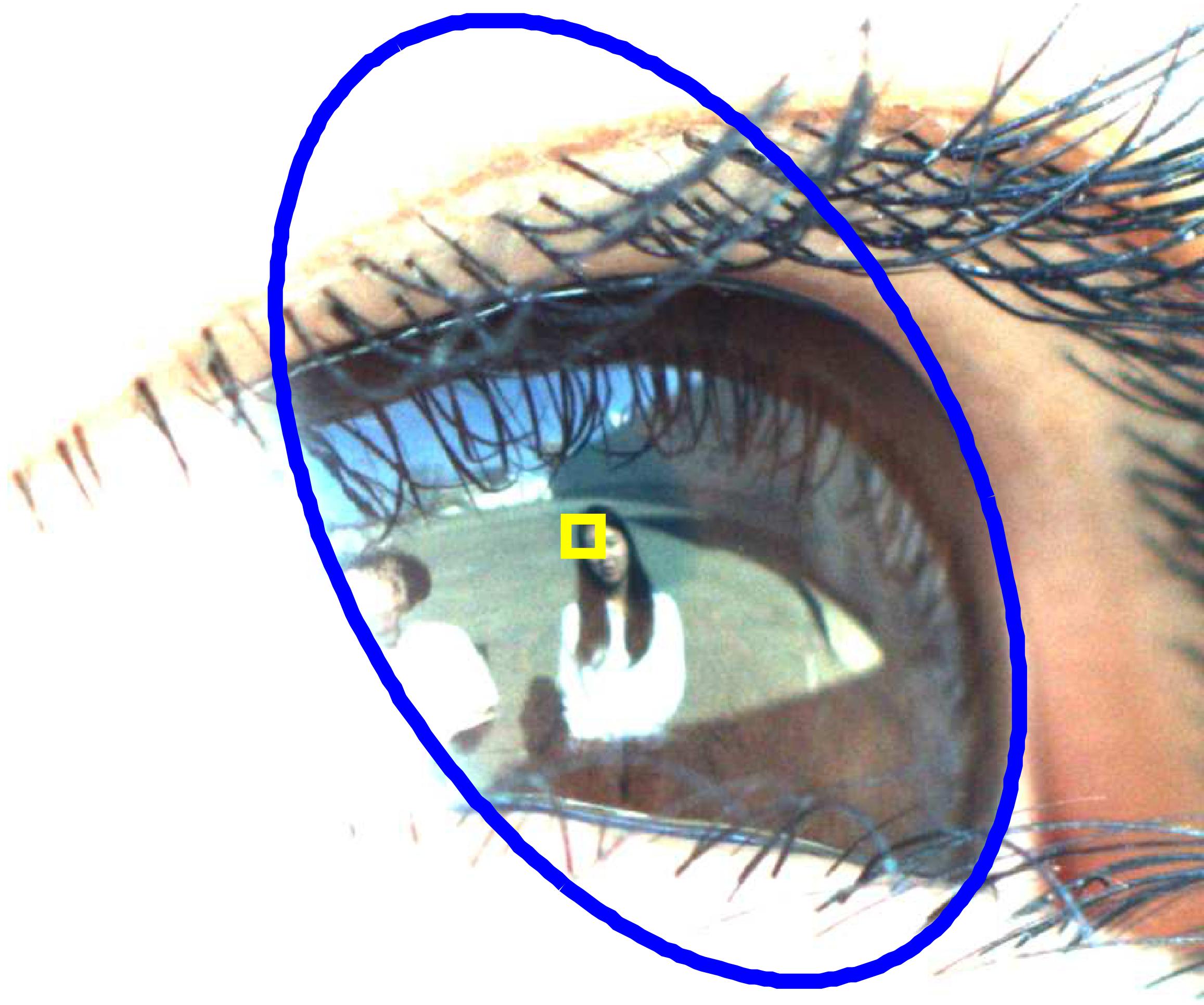} & \includegraphics[width=\linewidth]{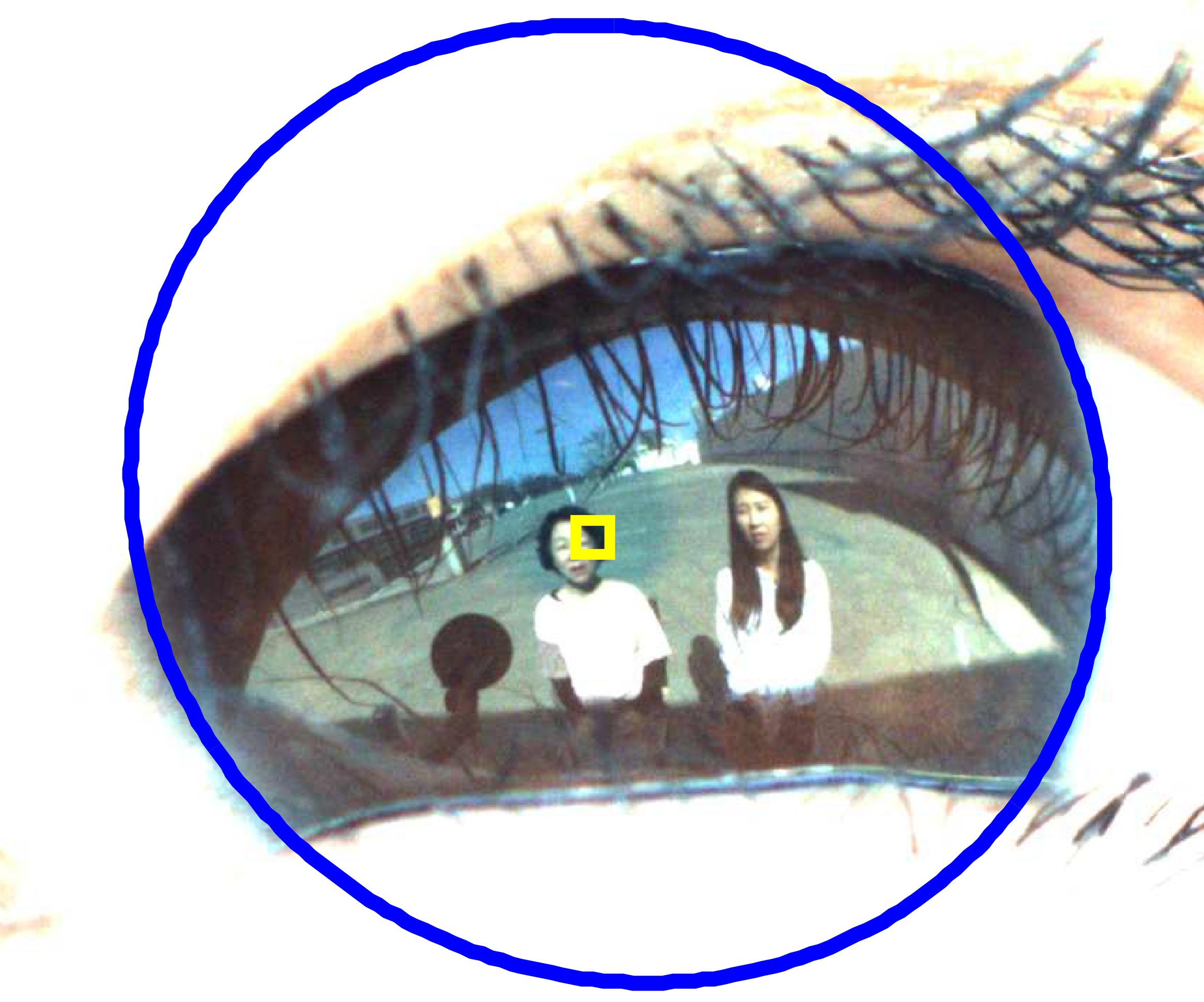}\\
(B) & \includegraphics[width=\linewidth]{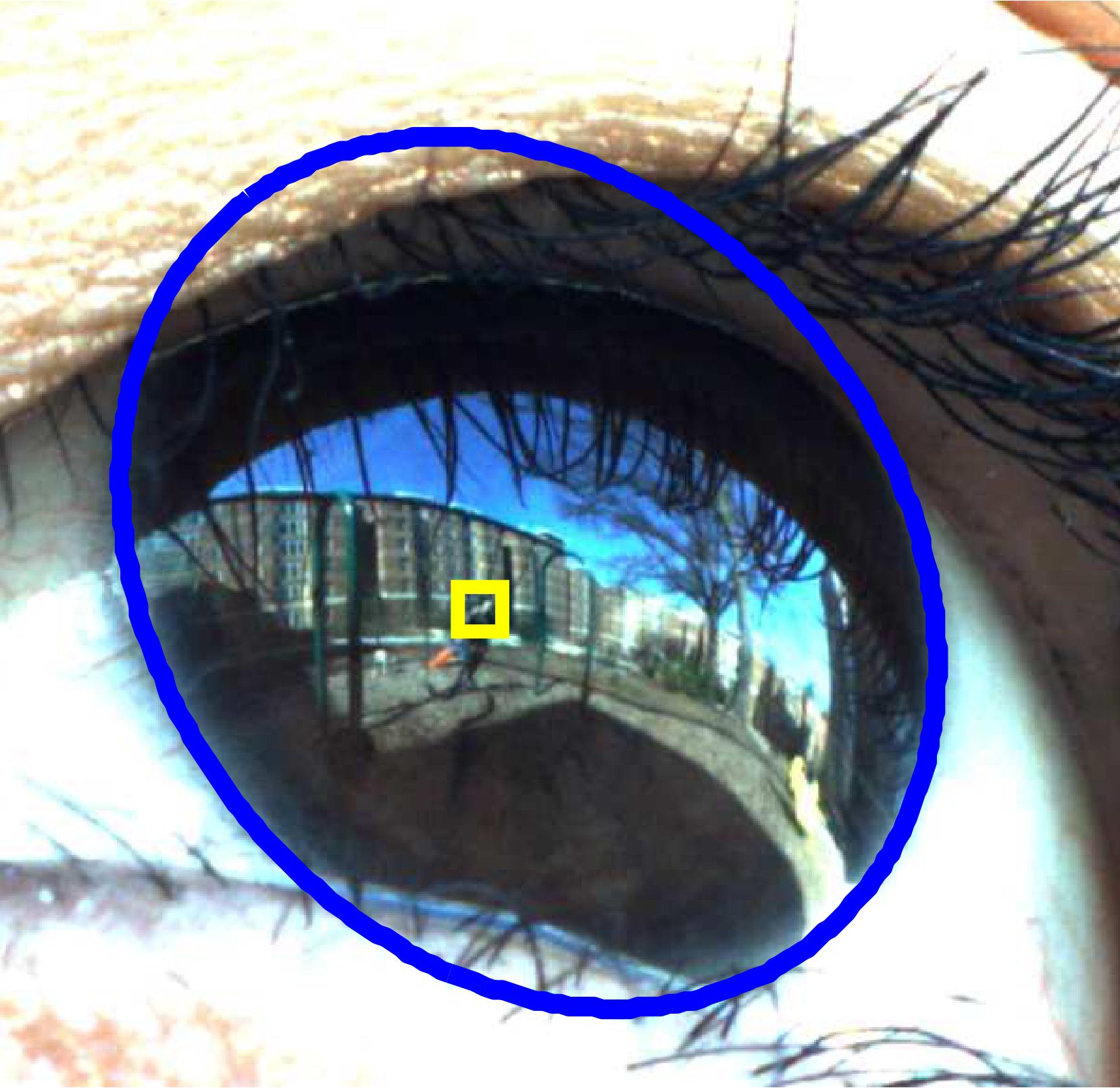} & \includegraphics[width=\linewidth]{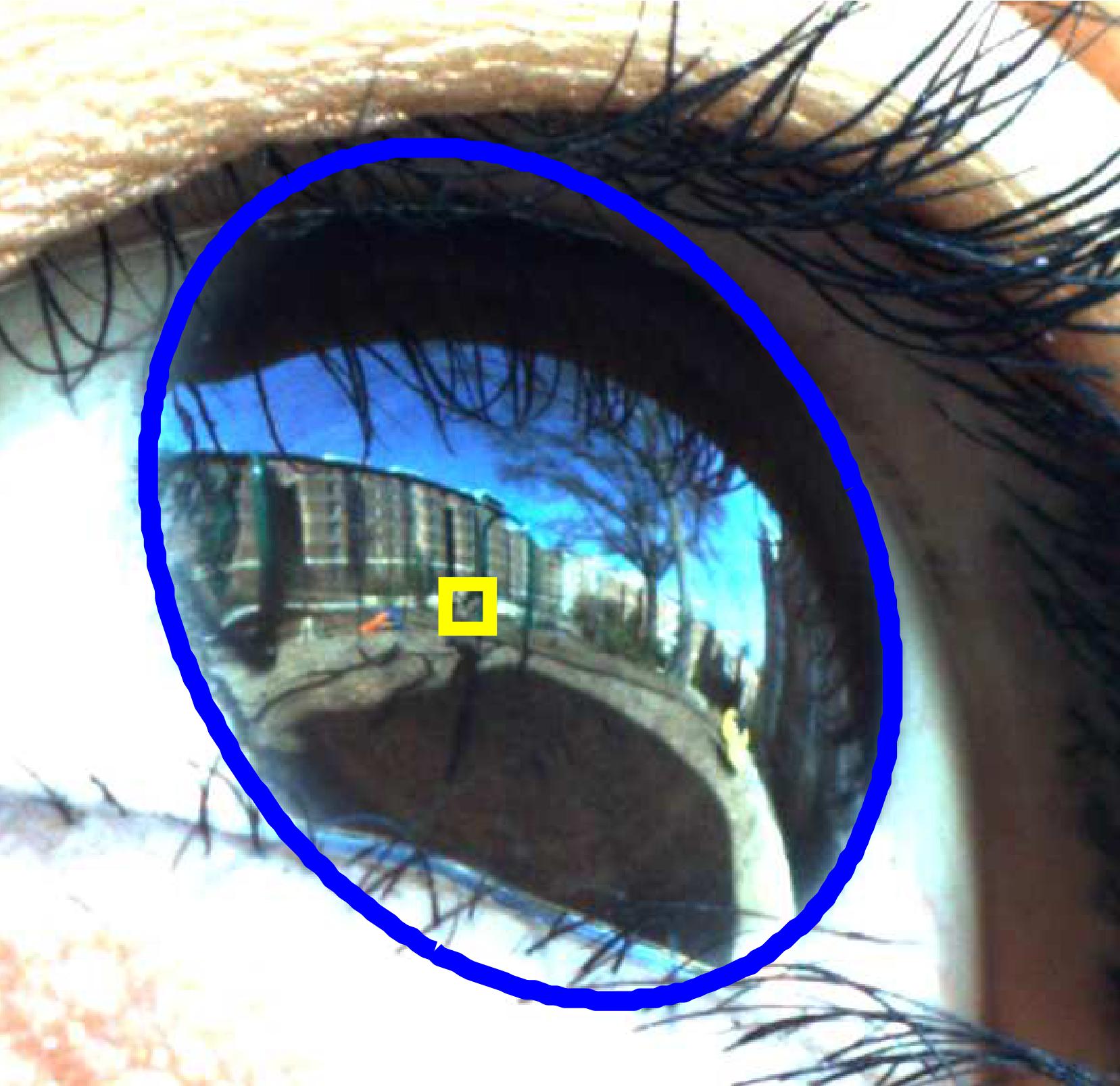} & \includegraphics[width=\linewidth]{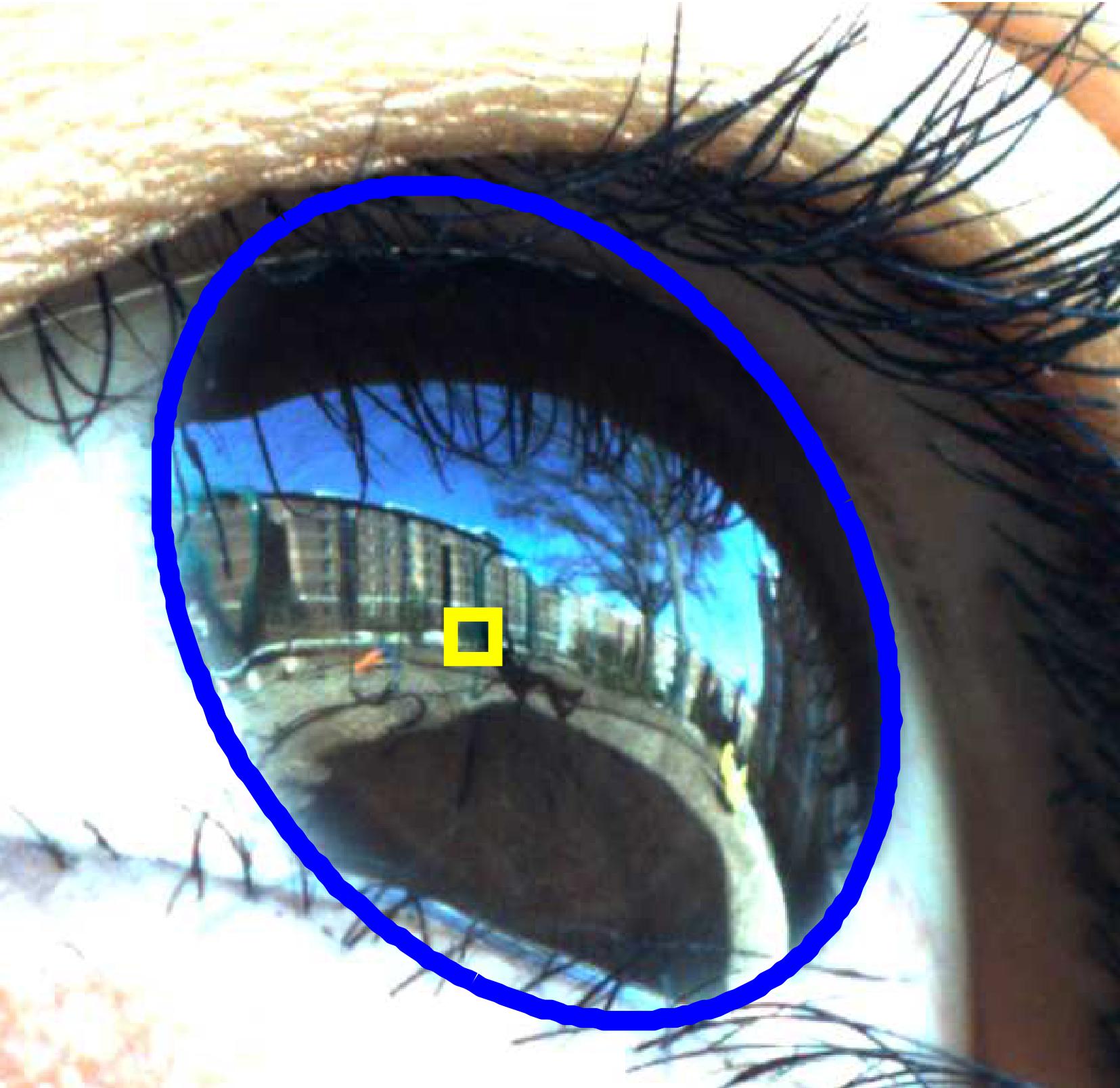}\\
(C) & \includegraphics[width=\linewidth]{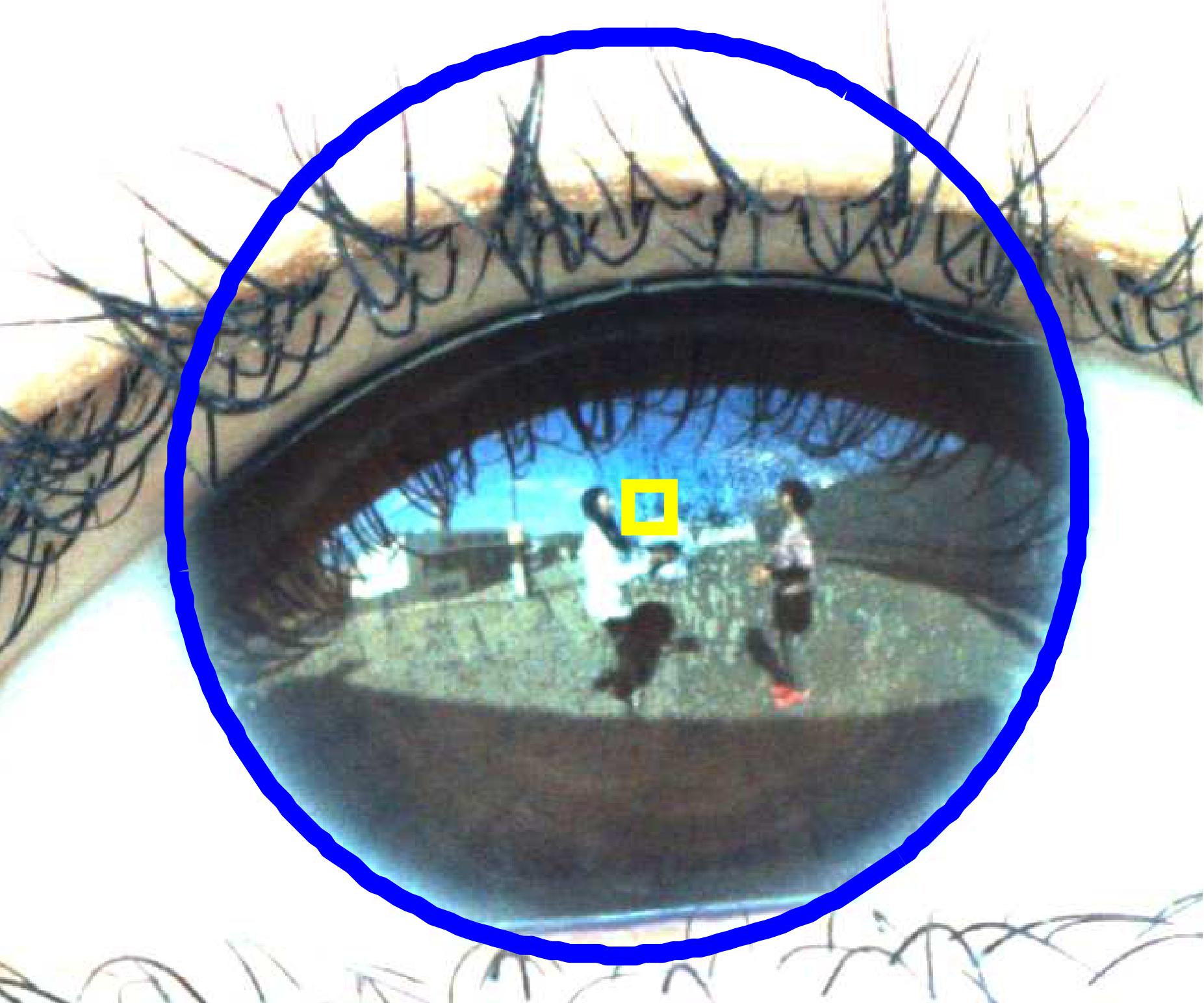} & \includegraphics[width=\linewidth]{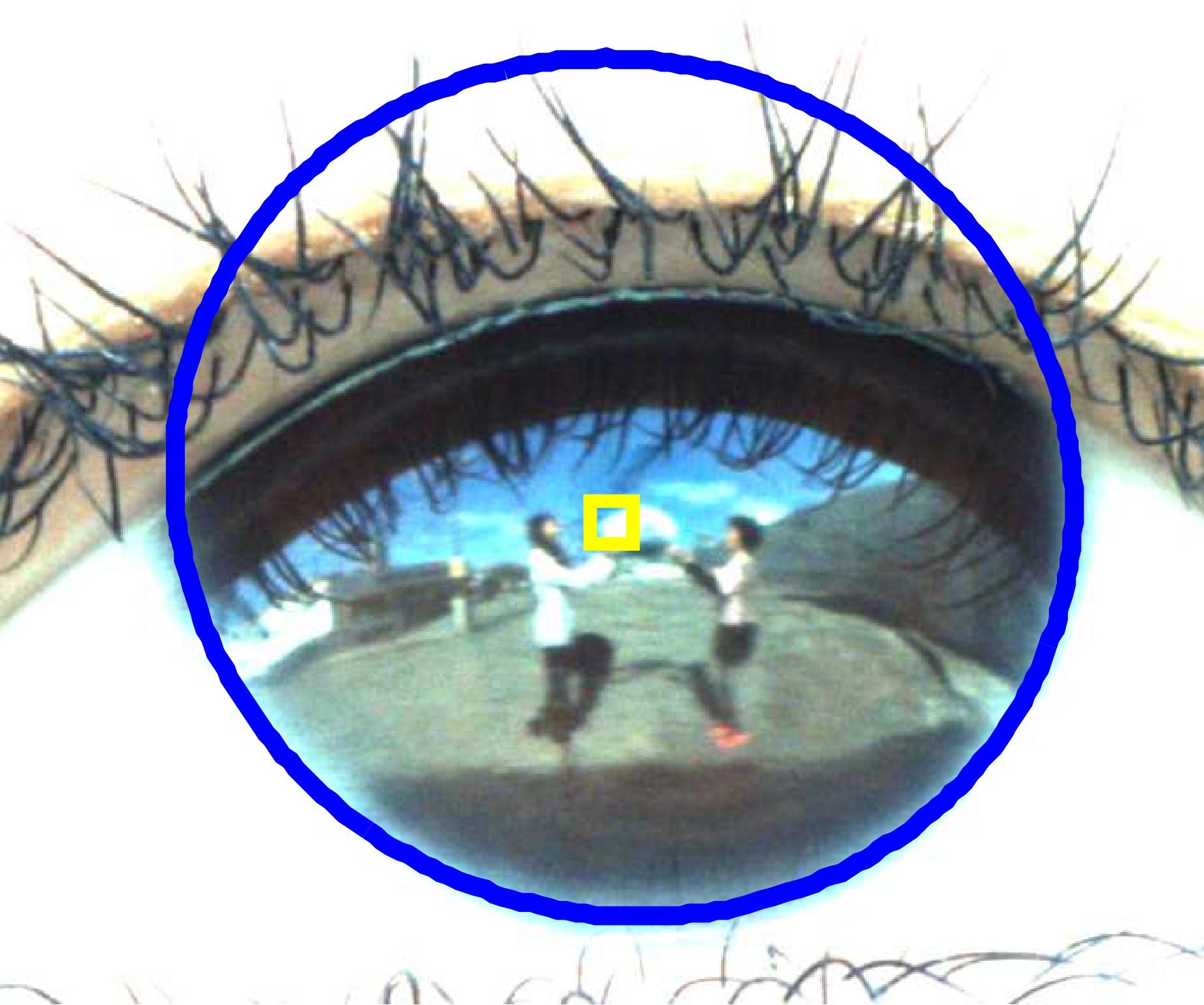} & \includegraphics[width=\linewidth]{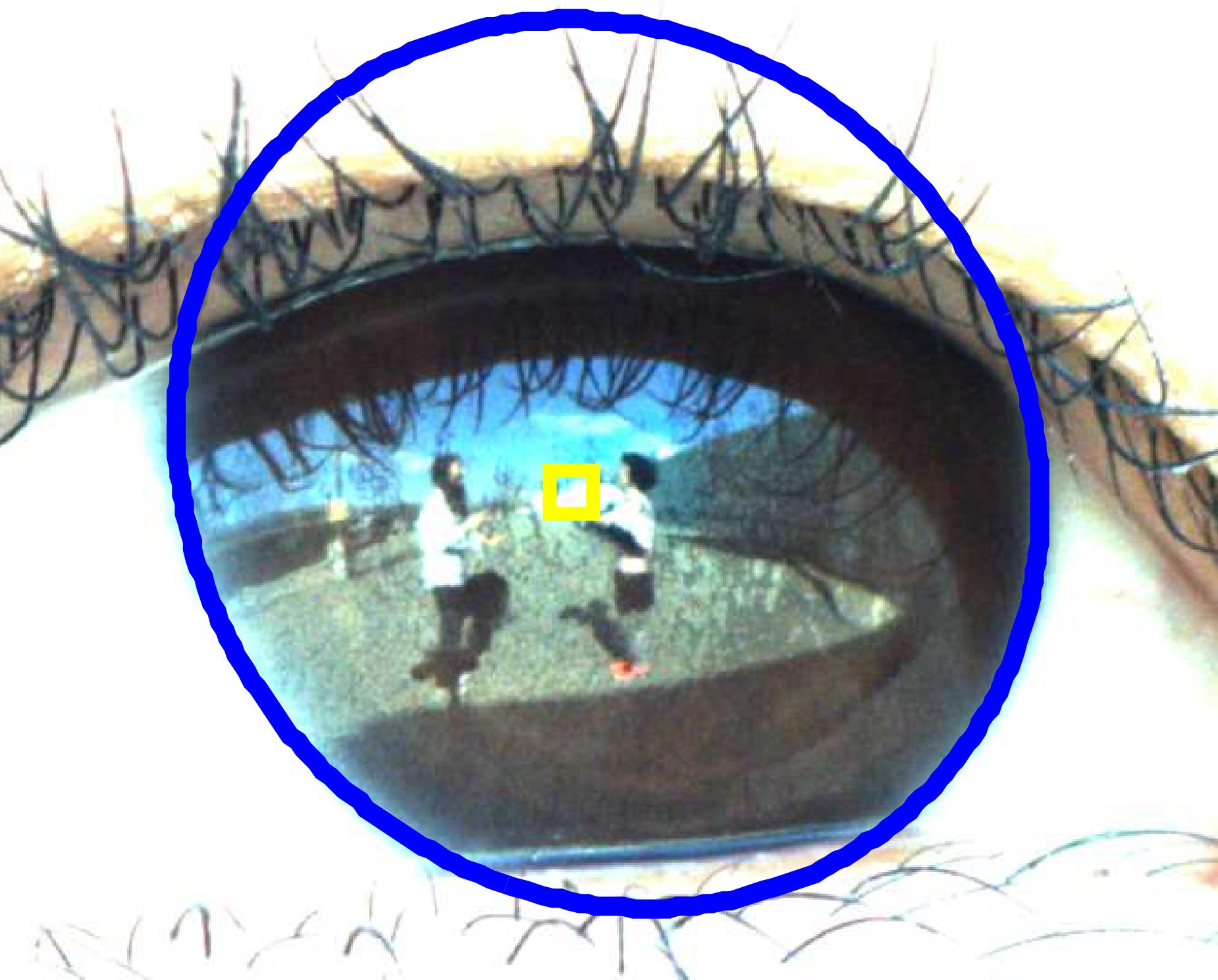}\\
(D) & \includegraphics[width=\linewidth]{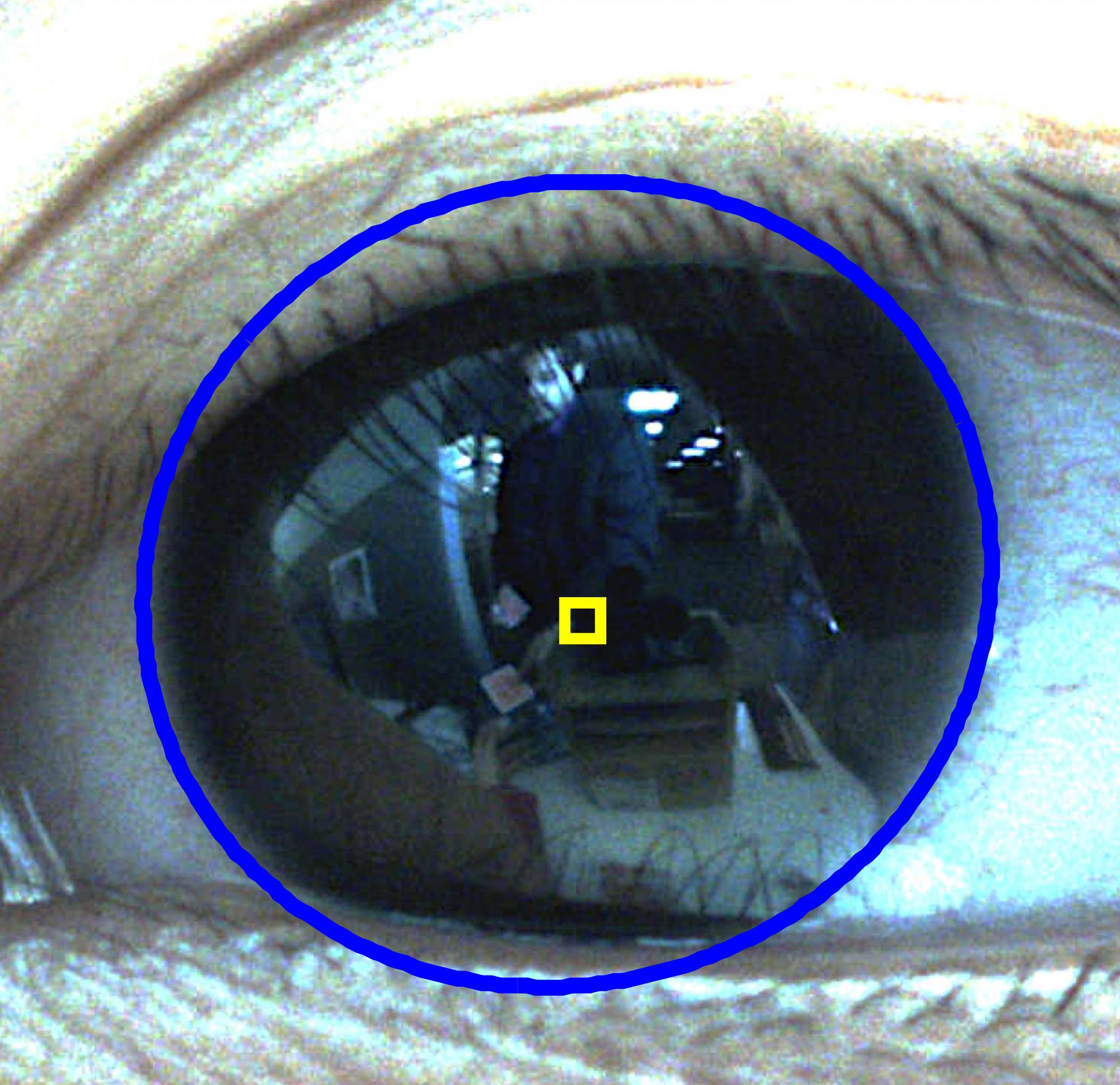} & \includegraphics[width=\linewidth]{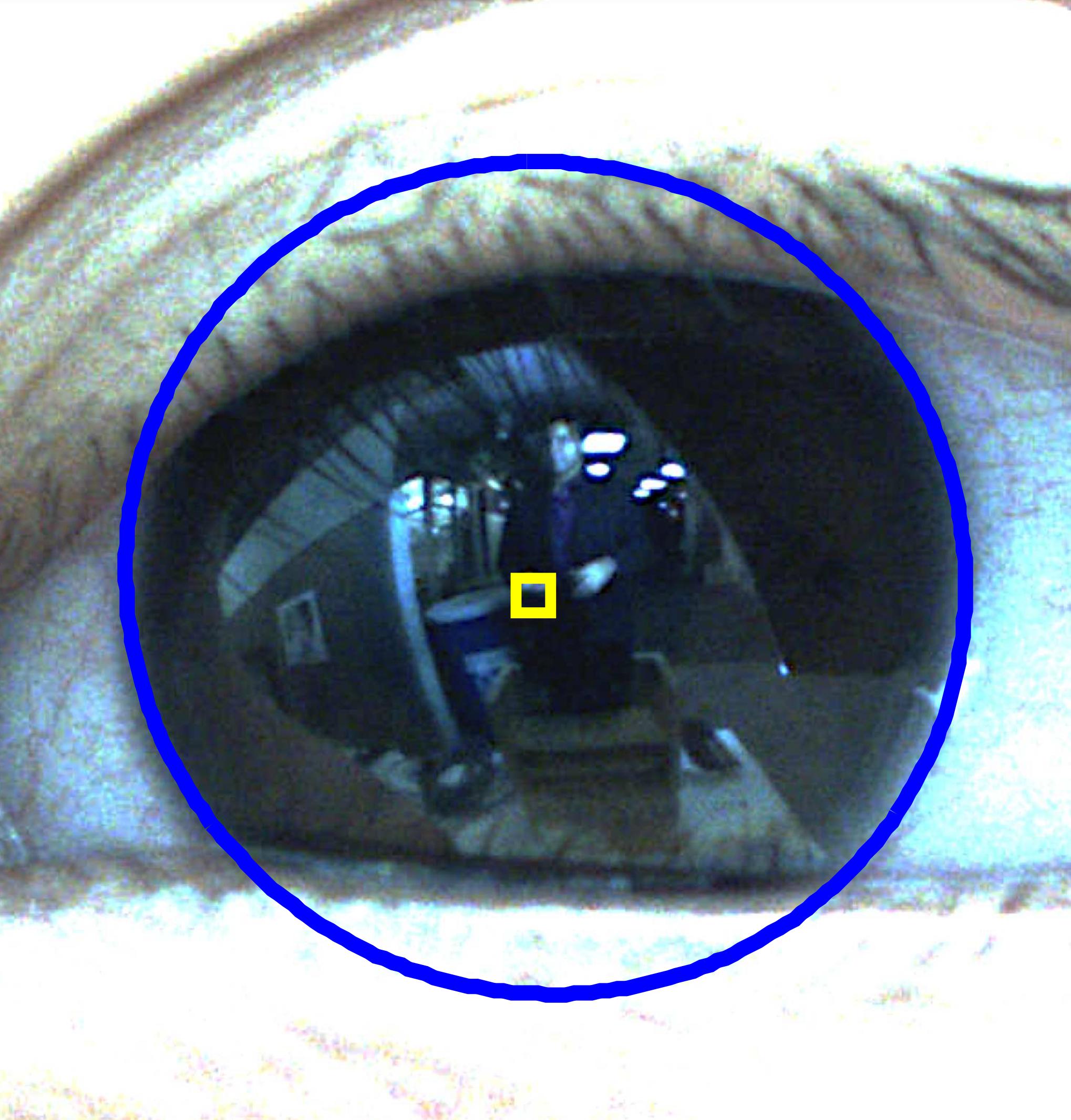} & \includegraphics[width=\linewidth]{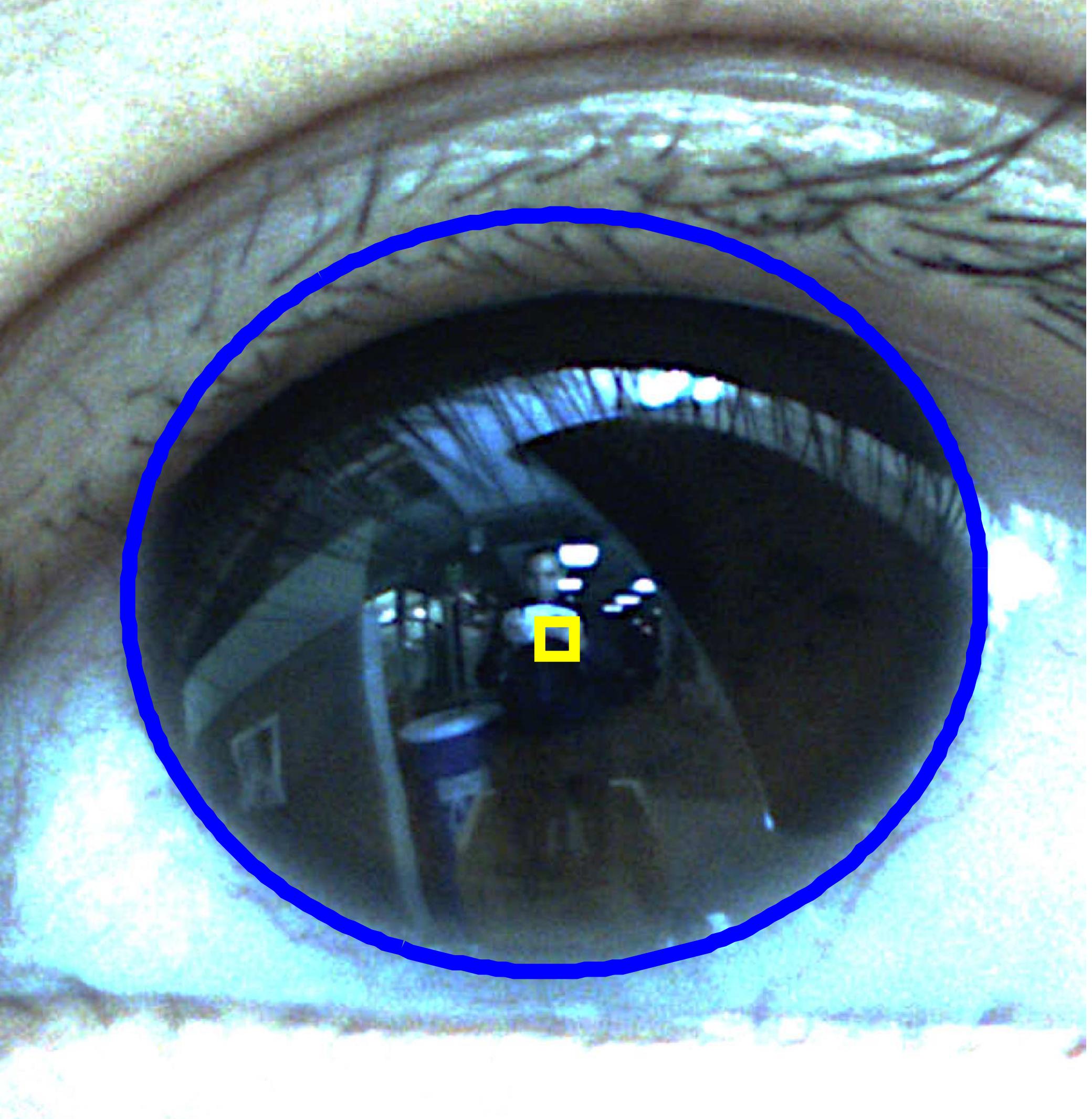}\\
(E) & \includegraphics[width=\linewidth]{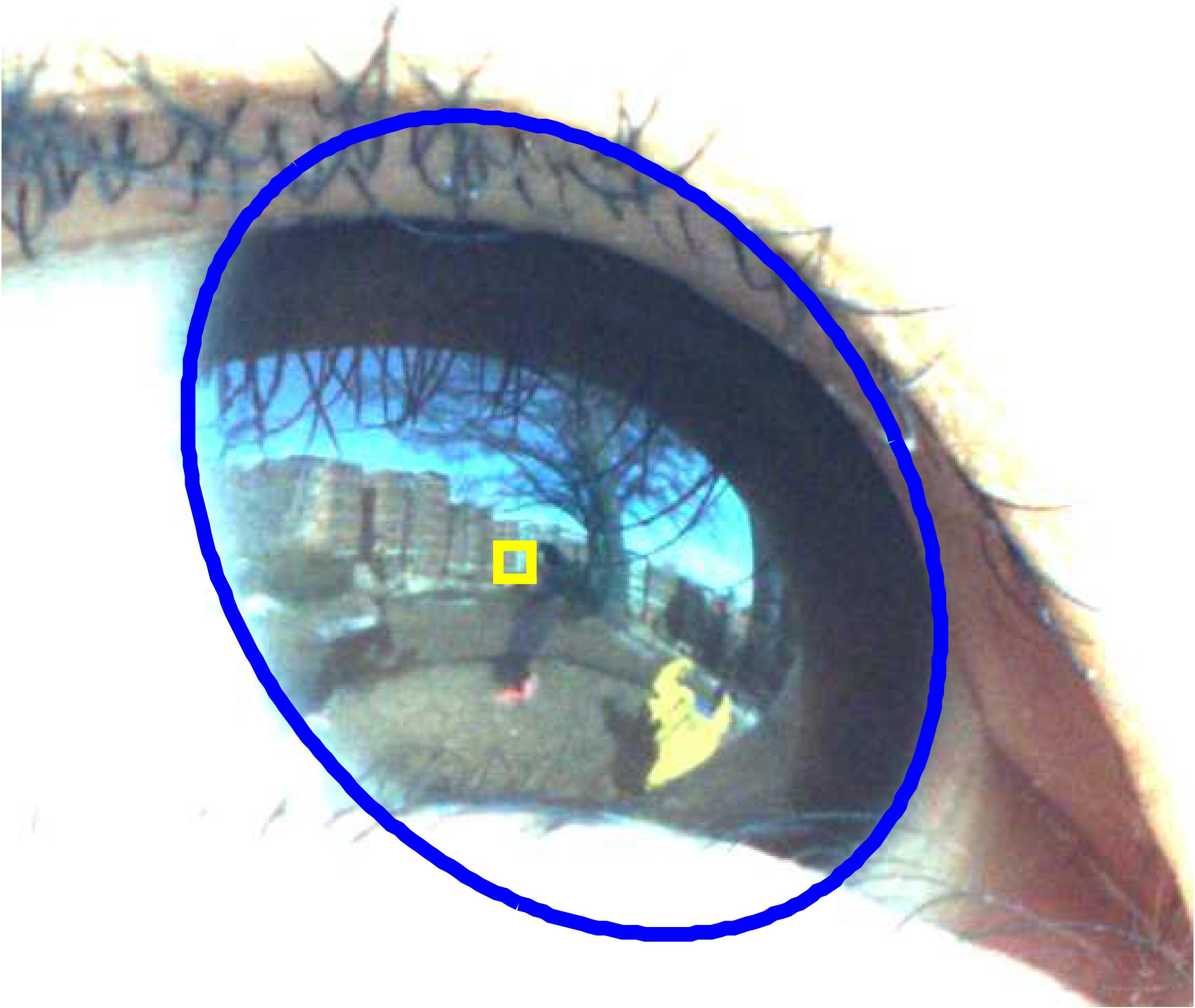} & \includegraphics[width=\linewidth]{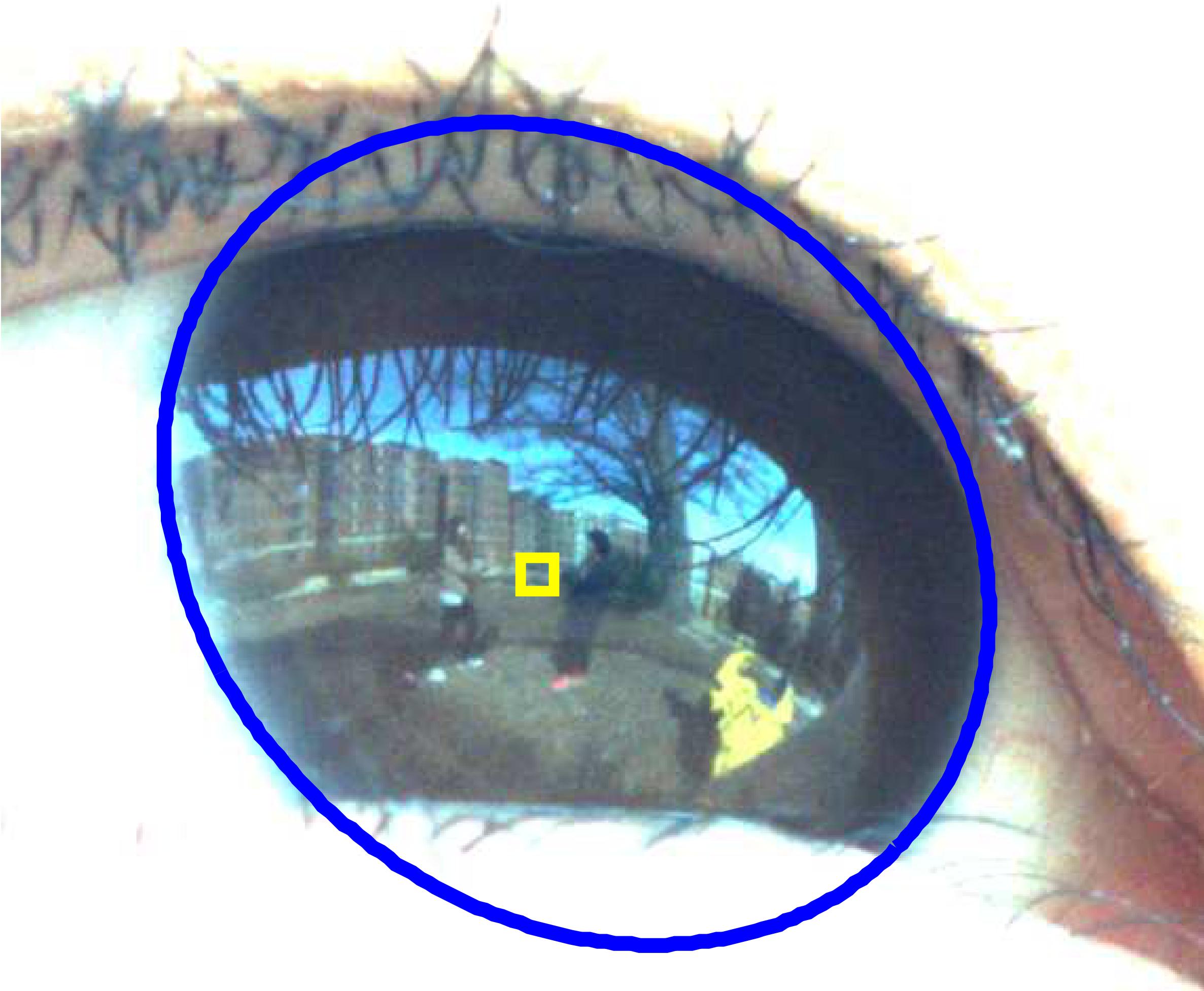} & \includegraphics[width=\linewidth]{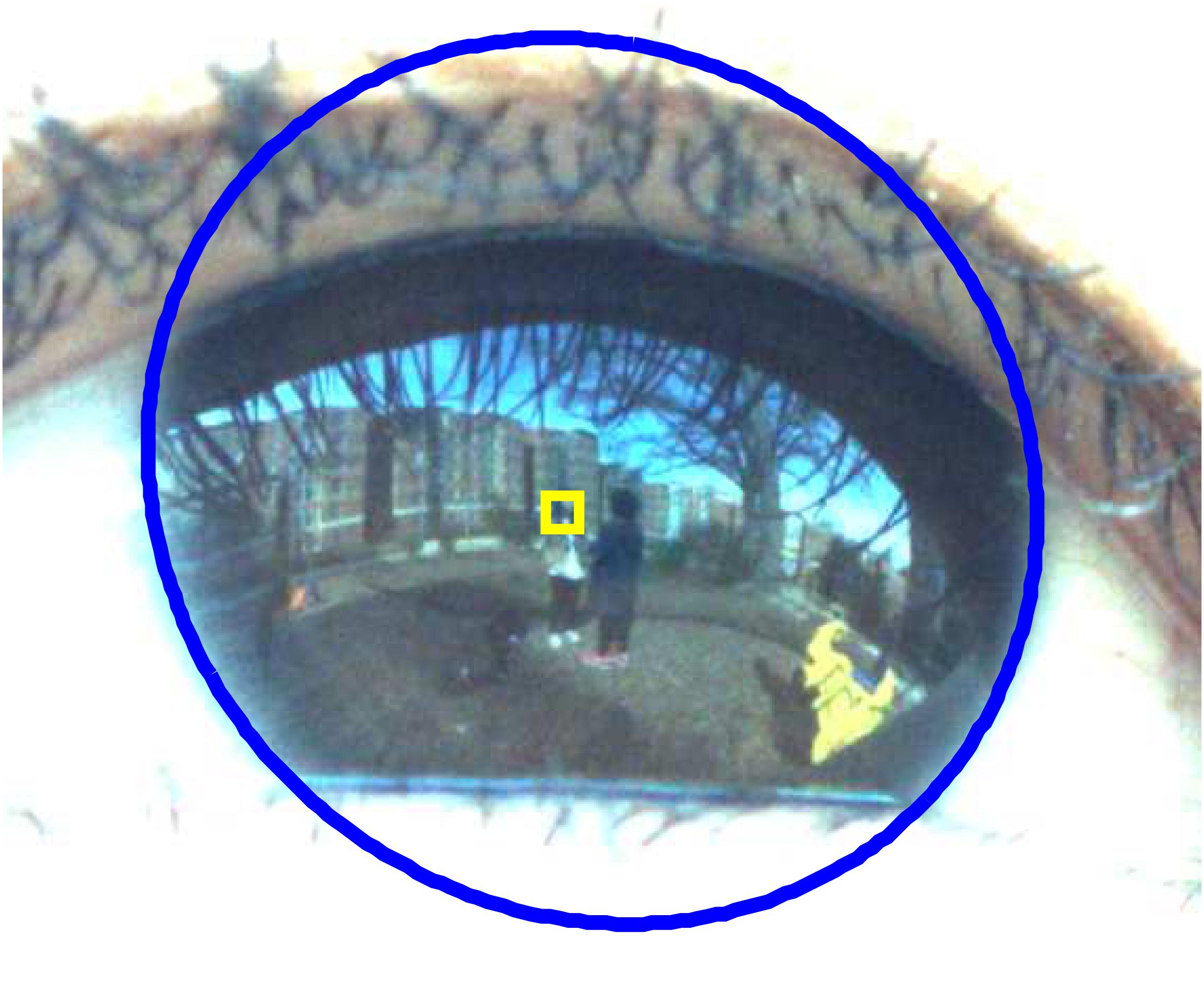}\\ 
\end{tabular}
\caption{Additional experiments; Each row shows gaze patterns of a user in a variety of naturalistic settings. (A) Conversation with two people. (B) Looking at a person swinging in the playground. (C) Looking at two people tossing a ball. (D) Magician performing a card trick for the user. (E) Looking at passersby in the street.}
\label{fig:app_more}
\end{figure*}

We conducted several additional experiments in various situations. This includes capturing an user (a) talking with two other people, (b) looking at a person swinging in the playground, (c) looking at two people tossing a ball, (d) interacting with a magician performing a card trick for the user, (e) looking at passersby in the street. Sample images are given in Figure~\ref{fig:app_more}. 

\section{Conclusion}
This paper introduced a noninvasive real-time camera-based gaze tracking system. This system is the first working real-time gaze measurement system based on the corneal imaging principle. It estimates the point of gaze from the reflected image in the cornea of a person looking at a scene. 
We describe a complete end-to-end system specification and design in the context of tabletop interaction. This approach could be easily adapted to other measurement scenarios. We describe the system components and their key properties which are needed for good performance. We present extensive evaluations of both end-to-end and absolute accuracies. We illustrate the use of our system in seven different experimental conditions. These results demonstrate the promise of our method to provide an alternative approach to gaze measurement with several advantages: it is noninvasive, requires minimal calibration and it can capture the subject's entire field of view. Moreover, its performance is comparable to SMI glasses, a state-of-the-art commercial gaze measurement system.

\setlength{\tabcolsep}{0pt}

\section*{Acknowledgment}
This research is supported by Samsung Scholarship awarded to Eunji Chong; JST Precursory Research for Embryonic Science and Technology (PRESTO) program and JSPS KAKENHI Grant Number 26280058 awarded to Atsushi Nakazawa; Simons Foundation Award 247332 and NSF Award IIS-1029679 awarded to James M. Rehg; and sponsored by Google, Inc.

\ifCLASSOPTIONcaptionsoff
  \newpage
\fi

\bibliographystyle{IEEEtran}
\bibliography{egbib}

\begin{IEEEbiography}[{\includegraphics[width=1in,height=1.25in,clip,keepaspectratio]{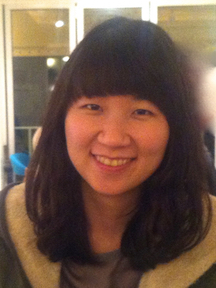}}]{Eunji Chong}
received the Bachelor of Science (B.S.) degree in Computer Science from Yonsei University, Korea in 2012. She is currently a PhD student in the School of Interactive Computing at the Georgia Institute of Technology. Her research interests include computer vision, machine learning, and computational behavioral science.
\end{IEEEbiography}
\begin{IEEEbiography}[{\includegraphics[width=1in,height=1.25in,clip,keepaspectratio]{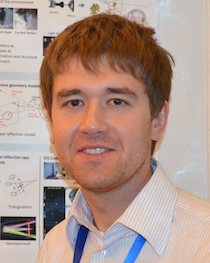}}]{Christian Nitschke}
received a Diplom (M.S.) in Media Systems from the Bauhaus Universita?t Weimar, Germany in 2006 and a Ph.D. in Engineering from Osaka University, Japan in 2011, where he continued to work as a Postdoctoral Researcher until 2013. He is currently an Assistant Professor at the Graduate School of Informatics, Kyoto University. His research interests include computer vision, computer graphics, visualization/display technologies, human-computer interfaces and eye gaze tracking.
\end{IEEEbiography}
\begin{IEEEbiography}[{\includegraphics[width=1in,height=1.25in,clip,keepaspectratio]{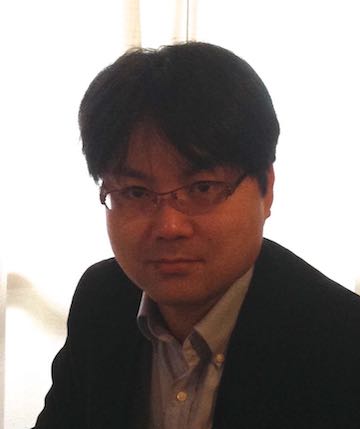}}]{Atsushi Nakazawa}
received a B.S. and Ph.D. in Engineering Science from Osaka University, Japan in 1997 and 2001 respectively. In 2001, he joined the Institute of Industrial Science, University of Tokyo as a postdoctoral researcher. From 2003 to 2013, he worked at Cybermedia Center, Osaka University as a lecturer. During the time, he was a visiting researcher at Georgia Institute of Technology, US from 2007 to 2008. Currently, he is an associate professor of Graduate School of Informatics, Kyoto University. His research interests include computer vision, robotics and human interfaces, in particular, estimating human posture from images, analysis of human motion using motion capture, humanoid robot control, image-based eye analysis, and eye gaze tracking. He is a member of IEEE, IPSJ, and RSJ.
\end{IEEEbiography}
\begin{IEEEbiography}[{\includegraphics[width=1in,height=1.25in,clip,keepaspectratio]{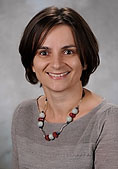}}]{Agata Rozga}
received her PhD degree in Developmental Psychology from the University of California, Los Angeles. She is a Research Scientist in the School of Interactive Computing at the Georgia Institute of Technology, where she is the director of the Georgia Tech Child Study Lab. She currently serves on the editorial board for the journal Focus on Autism and Other Developmental Disabilities, and has co-authored over 20 peer-reviewed publications across psychology and computing venues. Her research interests include autism, nonverbal communication, social development, and computational behavioral science.
\end{IEEEbiography}
\begin{IEEEbiography}[{\includegraphics[width=1in,height=1.25in,clip,keepaspectratio]{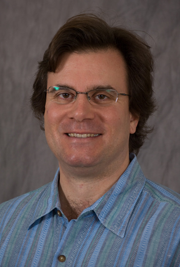}}]{James M. Rehg}
received the PhD degree in electrical and computer engineering from Carnegie Mellon University. He is a professor in the School of Interactive Computing at the Georgia Institute of Technology, where he is the director of the Center for Behavior Imaging, co-director of the Computation Perception Lab, and associate director of research in the Center for Robotics and Intelligent Ma- chines. From 1995-2001, he worked at the Cambridge Research Lab of DEC (and then Compaq), where he managed the computer vision research group. He received the US National Science Foundation CAREER award in 2001 and the Raytheon Faculty Fellowship from Georgia Tech in 2005. He and his students have received several best paper awards, including awards at ICML 2005 and BMVC 2010. Dr. Rehg served as the general cochair for IEEE CVPR 2009 and currently serves on the editorial board of the International Journal of Computer Vision. He has authored more than 100 peer-reviewed scientific papers and holds 23 issued US patents. His research interests include computer vision, robotics, and machine learning. He is a member of the IEEE.
\end{IEEEbiography}

\end{document}